\documentclass[times,sort&compress,3p]{elsarticle}
\journal{Journal of Multivariate Analysis}
\usepackage[labelfont=bf]{caption}

\usepackage{bm, txfonts, mathrsfs, multirow}
\usepackage[dvipsnames]{xcolor}
\usepackage[colorlinks = true]{hyperref}
\hypersetup{citecolor = blue,
	linkcolor = blue,
	urlcolor = blue}
\newcommand{\GG}[1]{}	

\usepackage{amsmath, amsfonts, amssymb, amsthm, booktabs, color, epsfig, graphicx, hyperref, url, mathtools}

\theoremstyle{plain}
\newtheorem{theorem}{Theorem}

\newtheorem{lemma}{Lemma}

\theoremstyle{definition}
\newtheorem{definition}{Definition}
\newtheorem{remark}{Remark}

\begin{document}

\begin{frontmatter}

\title{Hypothesis testing under uniform-block covariance structures}

\author[1]{Yifan Yang}
\author[2]{Shuo Chen \corref{mycorrespondingauthor}}
\author[1]{Ming Wang \corref{mycorrespondingauthor}}

\address[1]{Department of Population and Quantitative Health Sciences, Case Western Reserve University, Cleveland, Ohio, 44106 U.S.A.}
\address[2]{School of Medicine, University of Maryland, Baltimore, Maryland, 21201 U.S.A.}

\cortext[mycorrespondingauthor]{Corresponding authors. Email address: \url{shuochen@som.umaryland.edu} and \url{mxw827@case.edu}}

\begin{abstract}
	A block covariance structure is widely observed across large-scale and high-dimensional datasets in diverse fields such as biology, medicine, engineering, economics, and finance.
	This pattern entails partitioning a covariance matrix into uniform blocks, where each block exhibits equal variances and covariances.
	The importance of uniform-block structures lies in their ubiquity, interpretability, and ability to accommodate high dimensionality and data missingness.
	Despite their prevalence, statistical hypothesis testing under uniform-block covariance structures remains largely unexplored, and unknown statistical properties limit their application in research.
	To address this gap, we develop a comprehensive framework for joint hypothesis tests of both covariance and mean structures, leveraging a novel block Hadamard product representation of uniform-block matrices.
	Specifically, we derive closed-form likelihood ratio test statistics and information statistics, explicitly establishing their null distributions.
	Additionally, we perform simultaneous marginal mean tests under a procedure that controls the false discovery proportion (FDP). 
	Extensive simulations validate the consistency between theoretical and empirical distributions of the joint test statistics, assess the performance of the proposed FDP control procedure, and evaluate the robustness of the joint test statistics against structural disruptions and missing data. 
	Lastly, we apply our methodology to hypothesis testing in a high-dimensional imaging dataset.
\end{abstract}


\end{frontmatter}

\section{Introduction \label{Sec:introduction}}

	Covariance matrices with specific structures are fundamental to both the theoretical and practical aspects of multivariate analysis \citep{BilodeauBrenner1999, Anderson2003, Muirhead2005}.
	To address the challenges posed by large-scale and high-dimensional datasets resulting from technological advancements, various covariance structures have been developed.
	These customarily include bandability \citep{WuPourahmadi2003, BickelLevina2008a}, sparsity \citep{ElKaroui2008, BickelLevina2008b, CaiLiu2011}, and hybrid structures that combine sparsity with low-rank properties \citep{FanFanLv2008, FanLiaoMincheva2011}.
	An alternative strategy for mitigating high dimensionality in covariance structures is the use of block structures.
	For example, 
	\citet{Olkin1972} introduced block circular symmetry structures.
	\citet{Szatrowski1976} investigated covariance matrices with types I and II block compound symmetry structures \citep{Votaw1948}.
	Leiva and collaborators extensively studied covariance matrices with blocked compound symmetry or equicorrelation \citep{Leiva2007, RoyLeiva2011, RoyLeivaZezula2015, RoyZmyslonyFonseca2016, ZezulaKleinRoy2018}.
	In this paper, we examine a block covariance structure in the context of high-dimensional hypothesis testing, termed the \emph{uniform-block structure}, where each block contains identical block-wise entries.

	We focus on uniform-block structures because of their ubiquitous patterns, interpretability, and ability to handle high dimensionality. 
	First, uniform-block structures are important because they are prevalently observed in various high-throughput biomedical data, including genetics, proteomics, brain imaging, and RNA expression \citep{ChenKangXing2018, YangChenChen2024}, as well as in large-scale engineering studies \citep{RoustantDeville2017, RoustantPadonouDeville2020} and high-dimensional economic or financial data \citep{CadimaCalheirosPreto2010, ArchakovHansen2022}.
	Their widespread presence provides an opportunity to systematically investigate underlying scientific mechanisms within a unified framework. 
	Second, the importance of uniform-block structures lies in their exceptional interpretability. 
	Specifically, a covariance structure with uniform blocks represents a form of variable clustering that frequently appears in real-world applications, where variables are grouped such that both within- and between-group correlations are groupwise identical. 
	This offers more interpretability than the traditional clustering form, where between-group variables are assumed to be independent. 
	Third, uniform-block structures serve as a crucial link between existing covariance structures and high-dimensional statistical methods.  
	As a direct generalization of sphericity and intraclass symmetry structures \citep{Mauchly1940, Wilks1946} and a parallel version of blocked compound symmetry \citep{RoyLeiva2011, RoyLeivaZezula2015, RoyZmyslonyFonseca2016}, the parameterization of uniform-block structures depends on the number of blocks, significantly reducing the number of covariance parameters. 
	This leads to more efficient statistical inferences in high-dimensional settings, with additional advantages such as closed-form covariance estimators and effective missing value management.    
	These three aspects collectively highlight the importance of uniform-block structures in both theoretical research and real-world applications.

	Although the uniform-block structures play an important role in multivariate analysis, they remain largely unexplored, likely due to the lack of appropriate algebraic tools. 
	For example, while \citet{Geisser1963} was the first to explore the null distributions of information statistics for population means under uniform-block covariance structures, rigorous proofs were not provided.
	Another challenge arises from the covariance estimation.
	Specifically, when the population covariance matrix follows a uniform-block structure, traditional marginal mean statistics do not exactly follow $t$-distributions.
	This deviation occurs because the diagonal estimators of the sample covariance matrix do not follow a single scaled chi-squared distribution, but rather a linear combination of two independent weighted chi-squared distributions.
	Furthermore, when its dimension exceeds the sample size, 
	the sample covariance matrix (multiplied by the sample size minus one) follows a singular Wishart distribution, lacking a probability density with respect to the Lebesgue measure \citep{DiazCarciaJaimezMardia1997}.
	These limitations have led to the scarcity of research on uniform-block structures.

	Historically, structures with uniform blocks have been known by various names across different research fields but have not been comprehensively studied, particularly in large-scale statistical testing. 
	\citet{Geisser1963} referred to them as the uniform cases of certain orders and derived information statistics for testing population means for orders $1$ and $2$.  
	\citet{Morrison1972} later revisited Geisser's information statistics for general orders but did not provide new findings.
	\citet{HuangYang2010} investigated the random sampling issues involving uniform-block correlation matrices. 
	\citet{CadimaCalheirosPreto2010} described them as group block structures and explored their eigendecomposition.
	\citet{RoustantDeville2017} introduced the term parametric block correlation matrices and provided necessary and sufficient conditions for their positive definiteness. 
	\citet{RoustantPadonouDeville2020} investigated Gaussian process regression problems, referring to these matrices as generalized compound symmetry block covariance matrices. 
	More recently, \citet{ArchakovHansen2022} examined uniform-block structured matrices, established their canonical forms, and called them block matrices with block partitions.
	Despite these contributions, comprehensive research on the algebraic properties of uniform-block matrices remains limited, restricting their broader application in high-dimensional statistical inference and fields such as biometrics, economics, and finance.

	To bridge the gaps in high-dimensional statistical testing involving uniform-block structures, we propose likelihood ratio tests (LRTs) for population covariance structures, LRTs and Geisser's information tests for population mean structures, and a simultaneous multiple testing procedure for marginal mean components, all under covariance uniform-block structures, allowing the dimension of covariance matrices to exceed the sample size (except for the one-population covariance structure LRT).
	Our contributions are three-fold. 
	First, we introduce a novel block Hadamard product representation for uniform-block matrices, decomposing them into lower-dimensional diagonal and symmetric matrices. 
	Compared to the original uniform-block matrices, these diagonal and symmetric matrices, along with a predetermined vector, serve as their unique ``coordinate components’’, remarkably reducing the computational complexity of matrix operations such as taking the inverse or computing the square.
	Second, leveraging this decomposition, we derive explicit expressions for LRT statistics and Geisser's information statistics in both high-dimensional one- and multiple-population settings.
	Additionally, we rigorously establish their null distributions in closed forms.
	Third, we propose a multiple testing procedure that adjusts $p$-values for marginal mean tests, ensuring that the estimated false discovery proportion (FDP) remains below a predetermined significance level. 
	Both the block Hadamard product representation and high-dimensional hypothesis tests involving uniform-block structures will contribute to advancing both theoretical studies and practical applications across various fields.

	We structure the remainder of the paper as follows.
	Section~\ref{Sec:ub} presents the definitions and fundamental properties of uniform-block matrices.
	In Section~\ref{Sec:testing}, we derive the LRT statistics and Geisser's information statistics, and the FDP estimation procedure based on the theory of uniform-block matrices.
	Section~\ref{Sec:simulation} presents numerical demonstrations to illustrate the proposed testing procedures.
	In Section~\ref{Sec:real}, we apply the proposed statistical tests to a high-dimensional imaging dataset comprising measurements of $5$ neurometabolites across $89$ brain regions. 
	We conduct the hypothesis tests for the overall population, as well as for the gender-specific and age-specific subpopulations.
	Lastly, we summarize our findings and offer further remarks and discussions in Section~\ref{Sec:discussion}.
	Technical proofs and additional simulation results are provided in the \href{Supplementary Material.pdf}{Supplementary Material}.
	Throughout this paper, let $\top$ denote the transpose of a vector or matrix. 
	Let $\textbf{0}_{n \times m}$ and $\textbf{1}_{n \times m}$ denote the $n$ by $m$ all-zero matrix and all-one matrix, respectively.
	Additionally, let $\textbf{I}_n$ and $\textbf{J}_n \triangleq \textbf{1}_{n \times n}$ denote the $n$ by $n$ identify and all-one matrices.
	Let $\operatorname{diag}(\cdot)$ and $\operatorname {Bdiag}(\cdot)$ denote the diagonal matrix and the block-diagonal matrix.
	Let $\operatorname{tr}(\cdot)$ and $\det(\cdot)$ denote the trace and determinant of a square matrix, respectively. 
	Let $\operatorname{sum}(\cdot)$ denote the sum of all elements of a matrix.
	Let $\operatorname{corr}(\bm{\Sigma}) \triangleq \operatorname{diag}^{-1/2}\left(\sigma_{11}, \ldots, \sigma_{pp} \right) \bm{\Sigma} \operatorname{diag}^{-1/2}\left(\sigma_{11}, \ldots, \sigma_{pp} \right)$ denote the correlation matrix of covariance matrix $\bm{\Sigma}$ with diagonal elements $\sigma_{11}, \ldots, \sigma_{pp}$.
	We denote $\textbf{N} > 0$ if matrix $\textbf{N}$ is positive definite. 

\section{Uniform-block matrix and uniform-block structure \label{Sec:ub}}

	In this section, we begin by defining a uniform-block matrix and its underlying structure.
	We then introduce the block Hadamard product representation for uniform-block matrices, revealing their algebraic properties.
	
	\begin{definition}[partition-size vector and partitioned matrix by a partition-size vector]
		\label{Dfn:par}
		Let $\textbf{N} \in \mathbb{R} ^ {p \times p}$ be a real-valued matrix, and let $K \in \mathbb{Z}_+$ be a positive integer such that $K \leq p$. 
		
		(1) A column vector
		\begin{equation*}
			\bm{p} \triangleq \left(p_1, \ldots, p_K\right) ^ \top \in \mathbb{Z}_{+} ^ {K}
		\end{equation*}
		is said to be a \emph{partition-size vector}, 
		if $p = p_1 + \cdots + p_K$ and $p_k \geq 1$ for $k \in \{1, \ldots, K\}$.
		
		(2) The block matrix $\left(\textbf{N}_{kk'} \right) = \textbf{N}$ is said to be the \emph{partitioned matrix of $\textbf{N}$ (induced) by the partition-size vector $\bm{p} \triangleq \left(p_1, \ldots, p_K\right) ^ \top$}, 
		if the $(k,k')$th block $\textbf{N}_{kk'}$ has dimensions $p_k$ by $p_{k'}$ for $k, k' \in \{1, \ldots, K\}$.
	\end{definition}
	
	\begin{definition}[uniform-block matrix and structure]
		\label{Dfn:ub}
		Let $\left(\textbf{N}_{kk'}\right) = \textbf{N}$ be the partitioned matrix of $\textbf{N}$ by partition-size vector $\bm{p} \triangleq \left(p_1, \ldots, p_K \right) ^ \top$, 
		where $\textbf{N}$ is a symmetric matrix. 
		
		(1) The block matrix $\left(\textbf{N}_{kk'}\right) = \textbf{N}$ is said to be a \emph{uniform-block matrix}, 
		if there exist numbers $a_{kk} \in \mathbb{R}$ and $b_{kk'} \in \mathbb{R}$, for $k, k' \in \{1, \ldots, K\}$, satisfying that
		\begin{equation*}
			\textbf{N}_{kk} = a_{kk} \textbf{I}_{p_k} + b_{kk} \textbf{J}_{p_k}, \quad k = k'; \quad
			\textbf{N}_{kk'} = b_{kk'} \textbf{1}_{p_k \times p_{k'}}, \quad b_{k'k} = b_{kk'}, \quad k \neq k'.
		\end{equation*}
		We further denote this uniform-block matrix as 
		\begin{equation*}
			\textbf{N} \left[\textbf{A}, \textbf{B}, \bm{p} \right] \triangleq \left(\textbf{N}_{kk'}\right) = \textbf{N}, 
			\quad \text{ where } \quad 
			\textbf{A} \triangleq \operatorname{diag}\left(a_{11}, \ldots, a_{KK}\right) \in \mathbb{R} ^ {K \times K}, 
			\quad \textbf{B} \triangleq \left(b_{kk'}\right) \in \mathbb{R} ^ {K \times K}
		\end{equation*}
		are diagonal matrix and symmetric matrix, respectively.
		
		(2) The structure of a uniform-block matrix $\textbf{N} \left[\textbf{A}, \textbf{B}, \bm{p} \right]$ is said to be a \emph{uniform-block structure}.
	\end{definition}
	
	\begin{remark}[non-symmetric uniform-block matrix and structure]
		The definition of uniform-block matrices was proposed by \citet{YangChenChen2024}.
		In Definition~\ref{Dfn:ub}, we assume uniform-block matrices and structures are symmetric because our primary focus in this paper is on covariance matrices;
		however, a non-symmetric uniform-block matrix and structure can be defined by allowing $\textbf{B}$ to be a non-symmetric matrix.
		Unless explicitly specified otherwise, we consistently define uniform-block matrices and structures as symmetric.
	\end{remark}
	
	Following Definition~\ref{Dfn:ub}, we introduce two important instances of uniform-block matrices: the partitioned matrices of the identity matrix $\textbf{I}_{p}$ and the all-one matrix $\textbf{J}_{p}$ by a given partition-size vector $\bm{p} \triangleq \left(p_1, \ldots, p_K\right) ^ \top$.
	For simplicity, we denote these as $\textbf{I}\left[\bm{p}\right]$ and $\textbf{J}\left[\bm{p}\right]$, instead of $\textbf{I}\left[\textbf{I}_K, \textbf{0}_{K \times K}, \bm{p} \right]$ and $\textbf{J}\left[\textbf{0}_{K \times K}, \textbf{J}_{K}, \bm{p} \right]$, throughout the paper, i.e.,
	\begin{equation*}
		\textbf{I}\left[\bm{p}\right] \triangleq \textbf{I}\left[\textbf{I}_K, \textbf{0}_{K \times K}, \bm{p} \right] = \operatorname{Bdiag}\left(\textbf{I}_{p_1}, \ldots, \textbf{I}_{p_K}\right) = \textbf{I}_{p}, \quad 
		\textbf{J}\left[\bm{p}\right] \triangleq 
		\textbf{J}\left[\textbf{0}_{K \times K}, \textbf{J}_{K}, \bm{p} \right] = \left(\textbf{1}_{p_k \times p_{k'}}\right) = \textbf{J}_{p}.
	\end{equation*}
	By Definition~\ref{Dfn:ub}, we only know that a uniform-block matrix $\textbf{N}[\textbf{A}, \textbf{B}, \bm{p}]$ ``implicitly'' depends on matrices $\textbf{A}$, $\textbf{B}$, and vector $\bm{p}$.
	To make this dependence explicit and simplify algebraic calculations, we introduce the \emph{block Hadamard product representation} for uniform-block matrices utilizing the notations $\textbf{I}[\bm{p}]$ and $\textbf{J}[\bm{p}]$.
	
	\begin{lemma}[block Hadamard product representation]
		\label{Lem:ub_rep}
		Let $\textbf{N}\left[\textbf{A}, \textbf{B}, \bm{p} \right]$ be a uniform-block matrix with diagonal matrix $\textbf{A} \triangleq \text{diag}\left(a_{11}, \ldots, a_{KK}\right)$, 
		symmetric matrix $\textbf{B} \triangleq \left(b_{kk'}\right)$, 
		and partition-size vector $\bm{p} \triangleq \left(p_1, \ldots, p_K \right) ^ \top$.
		Suppose $p_k > 1$ for $k \in \{1, \ldots, K\}$, 
		then the following decomposition is unique:
		\begin{equation*}
			\textbf{N}\left[\textbf{A}, \textbf{B}, \bm{p} \right] 
			= \textbf{A} \circ \textbf{I} \left[\bm{p} \right] + \textbf{B} \circ \textbf{J} \left[\bm{p} \right],
		\end{equation*}
		where $\circ$ denotes the block Hadamard product satisfying that $\textbf{A} \circ \textbf{I} \left[\bm{p} \right]$ is the block-diagonal matrix $\operatorname{Bdiag}\left(a_{11} \textbf{I}_{p_1}, \ldots, a_{KK} \textbf{I}_{p_K}\right)$ and $\textbf{B} \circ \textbf{J} \left[\bm{p} \right]$ is the symmetric block matrix $\left(b_{kk'} \textbf{1}_{p_k \times p_{k'}}\right)$.
	\end{lemma}
	
	\begin{proof}[Proof Sketch of Lemma~\ref{Lem:ub_rep}]
		This can be readily verified from the definition. 
		A complete proof is provided in the \href{Supplementary Material.pdf}{Supplementary Material}. 
	\end{proof}
	
	\begin{remark}[block Hadamard product representation]
		The matrix operator $\circ$ can be regarded as a special form of the block Hadamard product \citep{HornMathiasNakamura1991, GuntherKlotz2012, YangChenChen2024}. 
		Compared to Definition~\ref{Dfn:ub}, the block Hadamard product representation is not valid if $p_k = 1$ for some $k$.
		Additionally, this representation is also applicable to non-symmetric uniform-block matrices.
	\end{remark}
	 	
	\begin{lemma}[properties of uniform-block matrices]
		\label{Lem:ops_all}
		Let $\textbf{N} \left[\textbf{A}, \textbf{B}, \bm{p} \right]$, $\textbf{N}_1 \left[ \textbf{A}_{1}, \textbf{B}_{1}, \bm{p} \right]$, 
		and $\textbf{N}_2 \left[ \textbf{A}_{2}, \textbf{B}_{2}, \bm{p} \right]$ be uniform-block matrices
		with diagonal matrices $\textbf{A}$, $\textbf{A}_{1}$, and $\textbf{A}_{2}$,
		symmetric matrices $\textbf{B}$, $\textbf{B}_{1}$, and $\textbf{B}_{2}$, 
		and the common partition-size vector $\bm{p} \triangleq \left(p_1, \ldots, p_K \right) ^ \top$ satisfying $p_k > 1$ for $k \in \{1, \ldots, K\}$.
		Throughout this paper, we denote 
		\begin{equation*}
			\textbf{P} \triangleq \operatorname{diag}(p_1, \ldots, p_K) \in \mathbb{R} ^ {K \times K}, \quad 
			\bm{\Delta} \triangleq \textbf{A} + \textbf{B} \textbf{P} \in \mathbb{R} ^ {K \times K}, \quad 
			\bar{p}_k \triangleq p_{1} + \cdots + p_{k}, \quad 
			k \in \{1, \ldots, K\}, \quad
			\bar{p}_0 \triangleq 0.
		\end{equation*}
				
		(1) (Addition/Subtraction) suppose $\textbf{N} ^ * = \textbf{N}_1 \pm \textbf{N}_2$; then the partitioned matrix of $\textbf{N} ^ *$ by $\bm{p}$ is still a uniform-block matrix, denoted as $\textbf{N} ^ *\left[ \textbf{A}^*, \textbf{B}^*, \bm{p} \right]$, where $\textbf{A}^* = \textbf{A}_1 \pm \textbf{A}_2$ and $\textbf{B}^* = \textbf{B}_1 \pm \textbf{B}_2$.
		
		(2) (Product) suppose $\textbf{N} ^ * = \textbf{N}_1 \textbf{N}_2$; in general, the partitioned matrix of $\textbf{N} ^ *$ by $\bm{p}$ is not a uniform-block matrix. 
		But if $\textbf{N}_1$ and $\textbf{N}_2$ are commute, i.e., $\textbf{N}_1 \textbf{N}_2 = \textbf{N}_2 \textbf{N}_1$, then it is still a uniform-block matrix, denoted as $\textbf{N} ^ * \left[ \textbf{A}^*, \textbf{B}^*, \bm{p} \right]$, where $\textbf{A} ^ * = \textbf{A}_1 \textbf{A}_2 = \textbf{A} ^ {*, \top}$ and $\textbf{B} ^ * = \textbf{A}_1 \textbf{B}_2 + \textbf{B}_1 \textbf{A}_2 + \textbf{B}_1 \textbf{P} \textbf{B}_2 = \textbf{B} ^ {*, \top}$. 
		
		(3) (Power) suppose $\textbf{N}^* = \textbf{N} ^ m$ with integer $m \geq 2$; then the partitioned matrix of $\textbf{N}^*$ by $\bm{p}$ is still a uniform-block matrix, denoted as $\textbf{N} ^ *\left[\textbf{A} ^ {(m)}, \textbf{B} ^ {(m)}, \bm{p} \right]$, 
		where $\textbf{A} ^ {(1)} = \textbf{A}$, $\textbf{B} ^ {(1)} = \textbf{B}$, $\textbf{A} ^ {(m')} = \textbf{A} ^ {(m' - 1)} \textbf{A}$ and $\textbf{B} ^ {(m')} = \textbf{A} ^ {(m' - 1)} \textbf{B} + \textbf{B} ^ {(m' - 1)} \textbf{A} + \textbf{B} ^ {(m' - 1)} \textbf{P} \textbf{B}$ for $m' \in \{2, \ldots, m\}$.
		In particular, when $m = 2$, the square formulas are $\textbf{A} ^ {(2)} = \textbf{A} ^ 2$ and $\textbf{B} ^ {(2)} = \textbf{A} \textbf{B} + \textbf{B} \textbf{A} + \textbf{B} \textbf{P} \textbf{B}$. 
		
		(4) (Eigenvalues) $\textbf{N}\left[ \textbf{A}, \textbf{B}, \bm{p} \right]$ has $p$ real-valued eigenvalues in total; 
		these are $a_{kk}$ with multiplicity $(p_k - 1)$ for $k \in \{1, \ldots, K\}$, and the remaining $K$ eigenvalues are identical to those of $\bm{\Delta}$.
		
		(5) (Determinant) $\textbf{N}\left[\textbf{A}, \textbf{B}, \bm{p} \right]$ has the determinant $\left(\prod_{k = 1}^{K} a_{kk} ^ {p_k - 1} \right) \det\left(\bm{\Delta} \right)$. 
		
		(6) (Inverse) suppose $\textbf{N}$ is invertible and $\textbf{N} ^ * = \textbf{N} ^ {-1}$; then the partitioned matrix of $\textbf{N} ^ *$ by $\bm{p}$ is still a uniform-block matrix, denoted as $\textbf{N} ^ *\left[\textbf{A} ^ *, \textbf{B} ^ *, \bm{p} \right]$, where $\textbf{A} ^ * = \textbf{A} ^ {- 1}$ and $\textbf{B} ^ * = - \bm{\Delta} ^ {- 1} \textbf{B} \textbf{A} ^ {- 1}$.
		
		(7) (Canonical Form) let $\lambda_j$ denote the $j$th eigenvalue of $\textbf{N}\left[\textbf{A}, \textbf{B}, \bm{p}\right]$ in the following order: $\lambda_1 = \cdots = \lambda_{\bar{p}_1 - 1} = a_{11}$, $\lambda_{\bar{p}_1 + 1} = \cdots = \lambda_{\bar{p}_2 - 1} = a_{22}$, $\ldots$, $\lambda_{\bar{p}_{K - 1} + 1} = \cdots = \lambda_{\bar{p}_K - 1} = a_{KK}$, and the remaining eigenvalues $\lambda_{\bar{p}_1}, \lambda_{\bar{p}_2}, \ldots, \lambda_{\bar{p}_K}$ are identical to the eigenvalues of $\bm{\Delta}$ (in the decreasing order).
		Thus, there exists an orthogonal matrix $\bm{\Gamma} \in \mathbb{R} ^ {p \times p}$ satisfying that $\bm{\Xi} \triangleq \bm{\Gamma} \textbf{N} \left[\textbf{A}, \textbf{B}, \bm{p}\right] \bm{\Gamma} ^ \top = \text{diag}\left(\lambda_1, \ldots, \lambda_p\right)$ is the canonical form of $\textbf{N} \left[\textbf{A}, \textbf{B}, \bm{p}\right]$.
		Additionally, $\bm{\Gamma}$ can be constructed by $K$ Helmert submatrices and $K$ row vectors as follows:
		\begin{equation}
			\label{Eq:Gamma}
			\bm{\Gamma} = \begin{pmatrix}
				\tilde{\textbf{H}}_1 & \textbf{0}_{(p_1 - 1) \times p_2} & \ldots & \textbf{0}_{(p_1 - 1) \times p_K} \\
				\xi_{1,1} \textbf{1}_{1 \times p_1} & \xi_{1,2} \textbf{1}_{1 \times p_2} & \ldots & \xi_{1,K} \textbf{1}_{1 \times p_K} \\
				\vdots & \vdots & \ddots & \vdots \\
				\textbf{0}_{(p_K - 1) \times p_1} & \textbf{0}_{(p_K - 1) \times p_2} & \ldots & \tilde{\textbf{H}}_K \\
				\xi_{K,1} \textbf{1}_{1 \times p_1} & \xi_{K,2} \textbf{1}_{1 \times p_2} & \ldots & \xi_{K,K} \textbf{1}_{1 \times p_K}
			\end{pmatrix},
		\end{equation}
		where $\tilde{\textbf{H}}_k \in \mathbb{R} ^ {(p_k - 1) \times p_k}$ is the submatrix of a standard Helmert matrix of order $p_k$ without the first row \citep{Lancaster1965},
		$\bm{\xi}_{k} \triangleq \left(\xi_{k,1}, \xi_{k,2}, \ldots, \xi_{k,K}\right) ^ \top \in \mathbb{R} ^ {K}$ is the eigenvector of $\bm{\Delta}$ corresponding to the eigenvalue $\lambda_{\bar{p}_k}$, 
		and the $\bar{p}_k$th row of $\bm{\Gamma}$ is normalized to the unit length for $k \in \{1, \ldots, K\}$.
	\end{lemma}
	
	\begin{proof}[Proof Sketch of Lemma~\ref{Lem:ops_all}]
		It is straightforward to verify properties (1), (2), (3), (6), and (7) by following the equalities 
		\begin{equation*}
			\begin{split}
				\left( \textbf{A}_1 \circ \textbf{I} \left[\bm{p} \right] \right)  \left( \textbf{A}_2 \circ \textbf{I} \left[\bm{p} \right] \right)
				& = \left( \textbf{A}_1 \textbf{A}_2 \right) \circ \textbf{I} \left[\bm{p} \right],
				\quad
				\left(\textbf{B}_1 \circ \textbf{J} \left[\bm{p} \right]\right)  \left(\textbf{B}_2 \circ \textbf{J} \left[\bm{p} \right]\right) 
				= \left(\textbf{B}_1  \textbf{P} \textbf{B}_2 \right) \circ \textbf{J} \left[\bm{p} \right], \\
				\left(\textbf{B}_1 \circ \textbf{J} \left[\bm{p} \right]\right)  \left( \textbf{A}_2 \circ \textbf{I} \left[\bm{p}\right] \right)
				& = \left( \textbf{B}_1  \textbf{A}_2 \right) \circ \textbf{J} \left[\bm{p} \right], 
				\quad
				\left( \textbf{A}_1 \circ \textbf{I}\left[\bm{p}\right] \right)  \left(\textbf{B}_2 \circ \textbf{J}\left[\bm{p} \right]\right)
				= \left( \textbf{A}_1  \textbf{B}_2 \right) \circ \textbf{J} \left[\bm{p}\right].
			\end{split}
		\end{equation*}
		Properties (4) and (5) can be confirmed using the induction method with respect to $K$.
		A complete proof is provided in the \href{Supplementary Material.pdf}{Supplementary Material}. 
	\end{proof}
		
	\begin{remark}[sufficient and necessary condition for positive definiteness]
		Although $\bm{\Delta}$ in Lemma~\ref{Lem:ops_all} is not symmetric in general, it possesses $K$ real-valued eigenvalues.
		This follows from the fact that $\bm{\Delta} = \left(\textbf{A} \textbf{P} ^ {-1} + \textbf{B} \right) \textbf{P}$ is similar, by definition, to a real-valued symmetric matrix $\textbf{P} ^ {1/2} \left(\textbf{A} \textbf{P} ^ {-1} + \textbf{B}\right) \textbf{P} ^ {1/2} $, which has $K$ real-valued eigenvalues. 
		Thus, $\textbf{N}\left[\textbf{A}, \textbf{B}, \bm{p}\right] > 0$ if and only if $\textbf{A} > 0$  (i.e., $a_{kk} > 0$ for $k \in \{1, \ldots, K\}$) and $\bm{\Delta} > 0$.
	\end{remark}
	
	\begin{remark}[properties for non-symmetric uniform-block matrices]
		For non-symmetric uniform-block matrices, the formulas for the sum, difference, product, determinant, and inverse (if it exists) remain the same expressions for $\textbf{A} ^ *$ and $\textbf{B} ^ *$ in Lemma~\ref{Lem:ops_all}.
		Although a non-symmetric uniform-block matrix $\textbf{N}\left[\textbf{A}, \textbf{B}, \bm{p}\right]$ still possesses $p$ eigenvalues, i.e., $a_{kk}$ with multiplicity $(p_k - 1)$ for $k \in \{1, \ldots, K\}$ and $K$ eigenvalues that are identical to those of $\bm{\Delta}$, some eigenvalues of $\bm{\Delta}$ may be complex valued.
	\end{remark}
	
	We summarize the advantages of the block Hadamard product representation as follows.
	First, 
	this representation provides an explicit expression for $\textbf{N}[\textbf{A}, \textbf{B}, \bm{p}]$ in terms of $\textbf{A}$, $\textbf{B}$, and $\bm{p}$.
	Second, 
	$\textbf{A}$, $\textbf{B}$, and $\bm{p}$ serve as (evaluable) ``coordinate components'' that determine the key properties of $\textbf{N}\left[\textbf{A}, \textbf{B}, \bm{p}\right]$, 
	including its power, inverse (if it exists), eigenvalues, and canonical form.
	Third, 
	computations involving the smaller $K$ by $K$ matrices $\textbf{A}$,  $\textbf{B}$, and $\textbf{P}$ can replace computations on the larger $p$ by $p$ matrix $\textbf{N}$, especially when $K$ is typically much smaller than $p$. 
	This reduction in matrix size substantially enhances computational efficiency.
	These advantages enhance the applicability of uniform-block matrices across various fields. 
	
	Before applying the theory of uniform-block matrices to hypothesis testing problems, we specify the relationships between a covariance matrix and its precision matrix, and correlation matrix.
	
	\begin{lemma}[uniform-block covariance matrices]
		\label{Lem:covariance}
		Let $\bm{\Sigma} \left[\textbf{A}, \textbf{B}, \bm{p} \right]$ be a uniform-block matrix with diagonal matrix $\textbf{A} \triangleq \operatorname{diag} \left(a_{11}, \ldots, a_{KK}\right)$, symmetric matrix $\textbf{B} \triangleq \left(b_{kk'}\right)$, 
		and partition-size vector $\bm{p} \triangleq \left(p_1, \ldots, p_K \right) ^ \top$ satisfying $p_k > 1$ for $k \in \{1, \ldots, K\}$.
		Furthermore, suppose $\textbf{A} > 0$ and $\bm{\Delta} \triangleq \textbf{A} + \textbf{B} \textbf{P} > 0$ with $\textbf{P} \triangleq \operatorname{diag}(p_1, \ldots, p_K)$ and ensure $\bm{\Sigma} \left[\textbf{A}, \textbf{B}, \bm{p} \right] > 0$.
		Then, the partitioned matrices of precision matrix $\bm{\Theta} \triangleq \bm{\Sigma} ^ {- 1}$ and correlation matrix $\bm{\Psi} \triangleq \operatorname{corr}\left(\bm{\Sigma}\right)$ by $\bm{p}$ are still uniform-block matrices, 
		denoted as $\bm{\Theta} \left[\textbf{A}_{\bm{\Theta}}, \textbf{B}_{\bm{\Theta}}, \bm{p}\right]$ and $\bm{\Psi} \left[\textbf{A}_{\bm{\Psi}}, \textbf{B}_{\bm{\Psi}}, \bm{p}\right]$, respectively, 
		where 
		\begin{equation}
			\label{Eq:correlation}
			\begin{cases}
				\textbf{A}_{\bm{\Theta}} = \textbf{A} ^ {- 1} \\
				\textbf{B}_{\bm{\Theta}} = - \bm{\Delta} ^ {-1} \textbf{B} \textbf{A} ^ {- 1}
			\end{cases}, 
			\begin{cases}
				\textbf{A}_{\bm{\Psi}} = \textbf{C} ^ {- 1/ 2} \textbf{A} \textbf{C} ^ {- 1/2} \\
				\textbf{B}_{\bm{\Psi}} = \textbf{C} ^ {- 1/ 2} \textbf{B} \textbf{C} ^ {- 1/2}
			\end{cases},
			\textbf{C} \triangleq \operatorname{diag}(c_{11}, \ldots, c_{KK}), \quad
			c_{kk} \triangleq a_{kk} + b_{kk},
			\quad k \in \{1, \ldots, K\}.
		\end{equation}
	\end{lemma}

	\begin{proof}[Proof of Lemma~\ref{Lem:covariance}]
		Use the results in Lemma~\ref{Lem:ops_all}. 
	\end{proof}

\section{Hypothesis testing under uniform-block structures \label{Sec:testing}} 

	In this section, we consider the following three hypothesis testing scenarios under the uniform-block structures.
	Given $M \geq 1$ populations,
	suppose we have $p$-dimensional normal vectors $\bm{X}_{1} ^ {(m)}, \ldots, \bm{X}_{n ^ {(m)}} ^ {(m)} \overset{\text{i.i.d.}}{\sim} N_p \left(\bm{\mu} ^ {(m)}, \bm{\Sigma} ^ {(m)}\right)$ with $\bm{\Sigma} ^ {(m)} > 0$ for $m \in \{1, \ldots, M\}$, 
	where $n ^ {(m)}$ denote the sample size from the $m$th population.
	
	\textbf{Scenario 1 (testing covariance structures)}: hypotheses for a single population to determine whether it follows a (unknown) uniform-block covariance structure, i.e., $M = 1$, 
	\begin{equation*}
		H_{1}: \bm{\Sigma} ^ {(1)} = \bm{\Sigma} ^ {(1)} \left[\textbf{A} ^ {(1)}, \textbf{B} ^ {(1)}, \bm{p} \right], 
		\quad 
		K_{1}: \bm{\Sigma} ^ {(1)} \text{ is unstructured}, 
	\end{equation*}
	where we consider unstructured $\bm{\mu} ^ {(1)}$ in both $H_1$ and $K_1$; 
	and for multiple populations to determine the equality of (unknown) uniform-block covariance structures, i.e., $M > 1$, 
	\begin{equation*}
		H_{3}: \bm{\Sigma} ^ {(1)} = \cdots = \bm{\Sigma} ^ {(M)} = \bm{\Sigma} \left[\textbf{A}, \textbf{B}, \bm{p} \right], 
		\quad 
		K_{3}: \bm{\Sigma} ^ {(m')} \neq \bm{\Sigma} ^ {(m)} \text{ for some $m' \neq m$, } 
	\end{equation*}
	where we consider unstructured $\bm{\mu} ^ {(m)}$ and structured $\bm{\Sigma} ^ {(m)} = \bm{\Sigma} ^ {(m)} \left[\textbf{A} ^ {(m)}, \textbf{B} ^ {(m)}, \bm{p} \right]$ for $m \in \{1, \ldots, M\}$ in both $H_3$ and $K_3$.
	
	\textbf{Scenario 2 (testing mean structures)}: hypotheses for a single population to determine whether the population mean vector equals a specified value $\bm{\mu}_{0} \in \mathbb{R} ^ {p}$, given a (unknown) uniform-block covariance structure, i.e., $M = 1$, 
	\begin{equation*}
		H_{2}: \bm{\mu} ^ {(1)} = \bm{\mu}_0, 
		\quad
		K_{2}: \bm{\mu} ^ {(1)} \neq \bm{\mu}_0, 
	\end{equation*}
	where we consider structured $\bm{\Sigma} ^ {(1)} = \bm{\Sigma} ^ {(1)} \left[\textbf{A} ^ {(1)}, \textbf{B} ^ {(1)}, \bm{p} \right]$ in both $H_2$ and $K_2$;
	and for multiple populations to determine the equality of (unknown) population mean vectors, given an equal (unknown) uniform-block covariance structure, i.e., $M > 1$, 
	\begin{equation*}
		H_{4}: \bm{\mu} ^ {(1)} = \cdots = \bm{\mu} ^ {(M)}, 
		\quad
		K_{4}: \bm{\mu} ^ {(m')} \neq \bm{\mu} ^ {(m)} \text{ for some $m' \neq m$},
	\end{equation*}
	where we consider structured $\bm{\Sigma} ^ {(1)} = \cdots = \bm{\Sigma} ^ {(M)} = \bm{\Sigma} \left[\textbf{A}, \textbf{B}, \bm{p} \right]$ in both $H_4$ and $K_4$.

	\textbf{Scenario 3 (simultaneously testing marginal means)}: simultaneous hypotheses for a single population to determine whether individual mean effects $\mu_j ^ {(1)}$ in $\bm{\mu} ^ {(1)} \triangleq \left(\mu_1 ^ {(1)}, \ldots, \mu_p ^ {(1)} \right) ^ \top$ for $j \in \{1, \ldots p\}$ are null,  given a (unknown) uniform-block covariance structure, i.e., $M = 1$, 
	\begin{equation*}
		H_{5, j}: \mu_j ^ {(1)} = 0, 
		\quad
		K_{5, j}: \mu_j ^ {(1)} \neq 0,
		\quad j \in \{1, \ldots, p\}.
	\end{equation*}
	where we consider structured $\bm{\Sigma} ^ {(1)} = \bm{\Sigma} ^ {(1)} \left[\textbf{A} ^ {(1)}, \textbf{B} ^ {(1)}, \bm{p} \right]$ in all $H_{5, j}$ and $K_{5, j}$ for $j \in \{1, \ldots, p\}$.
	
	\textbf{Conditions for positive definiteness in the above scenarios}:
	we assume that $\textbf{A} ^ {(m)} > 0$, $\textbf{A} > 0$,
	$\bm{\Delta} ^ {(m)} \triangleq \textbf{A} ^ {(m)} + \textbf{B} ^ {(m)} \textbf{P} > 0$, $\bm{\Delta} \triangleq \textbf{A} + \textbf{B} \textbf{P} > 0$ to ensure that $\bm{\Sigma} ^ {(m)} > 0$ and $\bm{\Sigma} > 0$ for $m \in \{1, \ldots, M\}$,
	where $\textbf{A} ^ {(m)}$ and $\textbf{A}$ are diagonal matrices, 
	$\textbf{B} ^ {(m)}$ and $\textbf{B}$ are symmetric matrices, 
	$\bm{p} \triangleq \left(p_1, \ldots, p_K\right) ^ \top$ is a common partition-size vector satisfying $p_k > 1$ for $k \in \{1, \ldots, K\}$, 
	and 
	$\textbf{P} \triangleq \operatorname{diag}\left(p_1, \ldots, p_K\right)$.
	
	Before proceeding with the calculation of the test statistics, 
	we introduce the maximum likelihood estimators for the uniform-block covariance matrices.
	
	\textbf{Maximum likelihood estimators for the $m$th population}:
	for each $m$, we denote 
	\begin{equation*}
		\bar{\bm{X}} ^ {(m)} \triangleq \left(\bm{X}_1 ^ {(m)} + \cdots + \bm{X}_{n ^ {(m)}} ^ {(m)} \right) / n ^ {(m)}, \quad
		\textbf{W} ^ {(m)} \triangleq \sum_{i = 1}^{n ^ {(m)}} \left(\bm{X}_i ^ {(m)} - \bar{\bm{X}} ^ {(m)}\right)\left(\bm{X}_i ^ {(m)} - \bar{\bm{X}} ^ {(m)}\right) ^ \top, \quad 
		\textbf{S} ^ {(m)} \triangleq \textbf{W} ^ {(m)} / \left(n ^ {(m)} - 1\right), 
	\end{equation*}
	and partition $\textbf{S} ^ {(m)} \triangleq \left(\textbf{S}_{kk'} ^ {(m)}\right)$ by $\bm{p}$, satisfying that blocks $\textbf{S}_{kk'} ^ {(m)}$ have dimensions $p_k$ by $p_{k'}$ for $k, k' \in \{1, \ldots, K\}$.
	Thus, 
	if $n ^ {(m)} > K + (K + 1) K / 2$ (i.e., the sample size exceeds the number of unknown covariance parameters),
	the following maximum likelihood estimators, for $k, k' \in \{1, \ldots, K\}$,  
	\begin{equation}
		\label{Eq:ls}
		\widehat{a}_{kk} ^ {(m)} = \frac{\operatorname{tr}\left(\textbf{S}_{kk} ^ {(m)}\right)}{p_k} - \widehat{b}_{kk} ^ {(m)}, \quad
		\widehat{b}_{kk'} ^ {(m)} = \begin{cases}
			\dfrac{\operatorname{sum}\left( \textbf{S}_{kk'} ^ {(m)} \right)}{p_k p_{k'}}, & k \neq k' \\
			\dfrac{\operatorname{sum}\left( \textbf{S}_{kk'} ^ {(m)} \right) - \operatorname{tr}\left( \textbf{S}_{kk'} ^ {(m)} \right)}{p_k (p_{k'} - 1)}, & k = k'
		\end{cases}, \quad
		\widehat{\textbf{A}} ^ {(m)} \triangleq \operatorname{diag} \left(\widehat{a}_{11} ^ {(m)}, \ldots, \widehat{a}_{KK} ^ {(m)}\right), \quad
		\widehat{\textbf{B}} ^ {(m)} \triangleq \left(\widehat{b}_{kk'} ^ {(m)} \right)
	\end{equation}
	with $\widehat{b}_{kk'} ^ {(m)} = \widehat{b}_{k'k} ^ {(m)}$ for $k \neq k'$, are the uniformly minimum-variance unbiased estimators \citep{YangChenChen2024} for $\textbf{A} ^ {(m)}$ and $\textbf{B} ^ {(m)}$ in $H_1$, $H_2$, $K_2$, $K_3$, $H_{5j}$, and $K_{5j}$ for $j \in \{1, \ldots, p\}$.
	
	Moreover, we assume that $\widehat{\textbf{A}} ^ {(m)}$ and $\widehat{\bm{\Delta}} ^ {(m)} \triangleq \widehat{\textbf{A}} ^ {(m)} + \widehat{\textbf{B}} ^ {(m)} \textbf{P}$ are positive definite with probability $1$ for $m \in \{1, \ldots, M\}$. 
	Additionally, let 
	\begin{equation}
		\label{Eq:ls2}
		\widehat{\textbf{A}}_{\bm{\Theta}} ^ {(m)} = \widehat{\textbf{A}} ^ {(m), -1}, \quad
		\widehat{\textbf{B}}_{\bm{\Theta}} ^ {(m)} = - \widehat{\bm{\Delta}} ^ {(m), - 1} \widehat{\textbf{B}} ^ {(m)} \widehat{\textbf{A}} ^ {(m), - 1}, \quad 
		m \in \{1, \ldots, M\}. 
	\end{equation}
	Then, according to Lemma~\ref{Lem:covariance},
	$m \in \{1, \ldots, M\}$, 
	\begin{equation*}
		\widehat{\bm{\Sigma}} ^ {(m)} \left[\widehat{\textbf{A}} ^ {(m)}, \widehat{\textbf{B}} ^ {(m)}, \bm{p} \right] = \widehat{\textbf{A}} ^ {(m)} \circ \textbf{I}[\bm{p}] + \widehat{\textbf{B}} ^ {(m)} \circ \textbf{J}[\bm{p}], \quad
		\widehat{\bm{\Theta}} ^ {(m)} \left[\widehat{\textbf{A}}_{\bm{\Theta}} ^ {(m)}, \widehat{\textbf{B}}_{\bm{\Theta}} ^ {(m)}, \bm{p} \right] = \widehat{\textbf{A}}_{\bm{\Theta}} ^ {(m)} \circ \textbf{I}[\bm{p}] + \widehat{\textbf{B}}_{\bm{\Theta}} ^ {(m)} \circ \textbf{J}[\bm{p}]
	\end{equation*}
	are the plug-in estimators for covariance matrices $\bm{\Sigma} ^ {(m)} \left[\textbf{A} ^ {(m)}, \textbf{B} ^ {(m)}, \bm{p} \right]$ and precision matrices $\bm{\Theta} ^ {(m)} \left[\textbf{A}_{\bm{\Theta}} ^ {(m)}, \textbf{B}_{\bm{\Theta}} ^ {(m)}, \bm{p} \right]$, respectively.
	 
	\textbf{Maximum likelihood estimators for the pooled population}:
	we denote
	\begin{equation*}
		n \triangleq n ^ {(1)} + \cdots + n ^ {(M)}, \quad 
		f ^ {(m)} \triangleq n ^ {(m)} / n, \quad
		\textbf{S} \triangleq \left(\textbf{W} ^ {(1)} + \cdots + \textbf{W} ^ {(M)}\right) / (n - M).
	\end{equation*}
	We partition $\textbf{S} \triangleq \left(\textbf{S}_{kk'} \right)$ by $\bm{p}$ such that blocks $\textbf{S}_{kk'}$ have dimensions $p_k$ by $p_{k'}$ for $k, k' \in \{1, \ldots, K\}$.
	Replacing $\textbf{S}_{kk'} ^ {(m)}$ in~\eqref{Eq:ls} with $\textbf{S}_{kk'}$, 
	we obtain 
	\begin{equation*}
		\widehat{a}_{kk} = \frac{\operatorname{tr}\left(\textbf{S}_{kk} \right)}{p_k} - \widehat{b}_{kk}, \quad
		\widehat{b}_{kk'} = \begin{cases}
			\dfrac{\operatorname{sum}\left( \textbf{S}_{kk'} \right)}{p_k p_{k'}}, & k \neq k' \\
			\dfrac{\operatorname{sum}\left( \textbf{S}_{kk'} \right) - \operatorname{tr}\left( \textbf{S}_{kk'} \right)}{p_k (p_{k'} - 1)}, & k = k'
		\end{cases}, \quad
		\widehat{\textbf{A}} \triangleq \operatorname{diag} \left(\widehat{a}_{11}, \ldots, \widehat{a}_{KK}\right), \quad
		\widehat{\textbf{B}} \triangleq \left(\widehat{b}_{kk'} \right)
	\end{equation*}
	with $\widehat{b}_{k'k} = \widehat{b}_{kk'}$ for $k \neq k'$. 
	If $n > K + (K + 1) K / 2$, 
	they are the uniformly minimum-variance unbiased estimators for $\textbf{A}$ and $\textbf{B}$ in $H_3$, $H_4$, and $K_4$.
	Additionally, we assume that $\widehat{\textbf{A}}$ and $\widehat{\bm{\Delta}} \triangleq \widehat{\textbf{A}} + \widehat{\textbf{B}} \textbf{P}$ are positive definite with probability $1$ and denote
	\begin{equation*}
		\widehat{\textbf{A}}_{\bm{\Theta}} = \widehat{\textbf{A}} ^ { -1}, \quad
		\widehat{\textbf{B}}_{\bm{\Theta}} = - \widehat{\bm{\Delta}} ^ {- 1} \widehat{\textbf{B}} \widehat{\textbf{A}} ^ {- 1}.
	\end{equation*}
	Similarly,  
	\begin{equation*}
		\widehat{\bm{\Sigma}} \left[\widehat{\textbf{A}}, \widehat{\textbf{B}}, \bm{p} \right] = \widehat{\textbf{A}} \circ \textbf{I}[\bm{p}] + \widehat{\textbf{B}} \circ \textbf{J}[\bm{p}], \quad 
		\widehat{\bm{\Theta}} \left[\widehat{\textbf{A}}_{\bm{\Theta}}, \widehat{\textbf{B}}_{\bm{\Theta}}, \bm{p} \right] = \widehat{\textbf{A}}_{\bm{\Theta}} \circ \textbf{I}[\bm{p}] + \widehat{\textbf{B}}_{\bm{\Theta}} \circ \textbf{J}[\bm{p}]
	\end{equation*}
	are the plug-in estimators for the covariance matrix $\bm{\Sigma} \left[\textbf{A}, \textbf{B}, \bm{p} \right]$ and the precision matrix $\bm{\Theta} \left[\textbf{A}_{\bm{\Theta}}, \textbf{B}_{\bm{\Theta}}, \bm{p} \right]$, respectively.

\subsection{Joint hypothesis tests for covariance structures in Scenario 1 \label{Subsec:covariance}}	
	
	\begin{theorem}[LRT statistics related to covariance structures]
		\label{Thm:covariance}
		(1) Given that $n > p$ and $p_k > 2$ for $k \in \{1, \ldots, K\}$, 
		the LRT statistic $\Lambda_1$ for testing $H_1$ versus $K_1$, raised to the power of $2 / n$, is computed as
		\begin{equation*}
			\Lambda_1 ^ {2 / n}
			= \left\{\prod_{k = 1}^{K}\left(p_k - 1\right) ^ {p_k - 1}\right\}
			\frac{\det\left(\textbf{V} ^ {(1)} \right)}{\prod_{k = 1}^{K}\left\{\textbf{V}_{\bar{p}_k, \bar{p}_k} ^ {(1)} \left(\sum_{j = \bar{p}_{k - 1} + 1}^{\bar{p}_k - 1} \textbf{V}_{j, j} ^ {(1)} \right) ^ {p_k - 1} \right\}}, 
		\end{equation*}
		where $\textbf{V} ^ {(1)} \triangleq \bm{\Gamma}_1 \textbf{W} ^ {(1)} \bm{\Gamma}_1 ^ \top$,
		and $\bm{\Gamma}_1$ is determined by~\eqref{Eq:Gamma}, i.e.,  $\bm{\Gamma}_1 \bm{\Sigma} ^ {(1)}\left[\textbf{A} ^ {(1)}, \textbf{B} ^ {(1)}, \bm{p}\right] \bm{\Gamma}_1 ^ \top$ is the canonical form of $\bm{\Sigma} ^ {(1)}\left[\textbf{A} ^ {(1)}, \textbf{B} ^ {(1)}, \bm{p}\right]$ presented in $H_1$.
		Under the null hypothesis $H_1$, 
		$\Lambda_1$ follows a product of mutually independent beta variates, i.e., 
		\begin{equation*}
			\Lambda_1 \sim
			\left\{\prod_{j = 1}^{p - 1} \beta_j ^ {n / 2} \left(\frac{n - 1}{2} - \frac{p - j}{2}, \frac{p - j}{2}\right) \right\}
			\prod_{k = 1}^{K}
			\left\{\prod_{j = 2}^{p_k - 1} \beta_{j, k} ^ {n / 2} \left(\frac{n - 1}{2}, \frac{j - 1}{p_k - 1}\right) \right\},
		\end{equation*}
		where $\beta_j(\cdot, \cdot)$ and $\beta_{j, k} (\cdot, \cdot)$ denote the beta variates with shape parameters for all $j$ and $k$.
		
		(2) Given that $\widehat{\textbf{A}} ^ {(m)} > 0$, $\widehat{\textbf{A}} > 0$, $\widehat{\bm{\Delta}} ^ {(m)} > 0$, $\widehat{\bm{\Delta}} > 0$ hold with probability $1$, $n - n ^ {(m)} > M - 1$, $n ^ {(m)} > 1$ for $m \in \{1, \ldots, M\}$, $n > M$, and $p_k > 1$ for $k \in \{1, \ldots, K\}$, 
		the LRT statistic $\Lambda_3$ for testing $H_3$ versus $K_3$, raised to the power of $2 / n$, is computed as 
		\begin{equation*}
			\Lambda_3 ^ {2 / n}
			= \left(\prod_{m = 1}^{M} f ^ {(m), f ^ {(m)}}\right) ^ {- p}
			\prod_{m = 1}^{M}
			\left\{\prod_{k = 1}^{K} \frac{\textbf{V}_{\bar{p}_k, \bar{p}_k} ^ {(m)}}{\sum_{m' = 1}^{M} \textbf{V}_{\bar{p}_k, \bar{p}_k} ^ {(m')}} \left(\frac{\sum_{j = \bar{p}_{k - 1} + 1}^{\bar{p}_k - 1} \textbf{V}_{j, j} ^ {(m)}}{\sum_{j = \bar{p}_{k - 1} + 1} ^ {\bar{p}_k - 1} \sum_{m' = 1} ^ {M} \textbf{V}_{j, j} ^ {(m')}} \right) ^ {p_k - 1} \right\} ^ {f ^ {(m)}},
		\end{equation*}
		where $\textbf{V} ^ {(m)} \triangleq \bm{\Gamma}_3 \textbf{W} ^ {(m)} \bm{\Gamma}_3 ^ \top$,
		and $\bm{\Gamma}_3$ is determined by~\eqref{Eq:Gamma}, i.e.,
		$\bm{\Gamma}_3 \bm{\Sigma} \left[\textbf{A}, \textbf{B}, \bm{p}\right] \bm{\Gamma}_3 ^ \top$ is the canonical form of $\bm{\Sigma} \left[\textbf{A}, \textbf{B}, \bm{p}\right]$ presented in $H_3$.
		Under the null hypothesis $H_3$, 
		$\Lambda_3$ follows a product of distributions of Dirichlet components, i.e., 
		\begin{equation*}
			\Lambda_3 \sim \left(\prod_{m = 1}^{M} f ^ {(m), f ^ {(m)}} \right) ^ {- p n / 2}
			\prod_{k = 1}^{K}
			\left\{\prod_{m = 1}^{M} D_{1, m, k} ^ {n ^ {(m)} / 2} \right\}
			\left\{\prod_{m = 1}^{M} D_{2, m, k} ^ {(p_k - 1) n ^ {(m)} / 2} \right\},
		\end{equation*}
		where $\left(D_{1, 1, k}, \ldots, D_{1, M, k}\right) \triangleq \bm{D}_{1, k}\left(\bm{\alpha}\right)$ and $\left(D_{2, 1, k}, \ldots, D_{2, M, k} \right) \triangleq \bm{D}_{2, k} \left(\bm{\alpha}_{k}\right)$ are $M$-dimensional Dirichlet variates with parameters $\bm{\alpha} \triangleq \left(n ^ {(1)} - 1, \ldots, n ^ {(M)} - 1\right) / 2$ and $\bm{\alpha}_{k} \triangleq \left(p_k - 1\right)\bm{\alpha}$ for $k \in \{1, \ldots, K\}$.
	\end{theorem}
	
	\begin{proof}[Proof of Theorem~\ref{Thm:covariance}]
		A complete proof is provided in the \href{Supplementary Material.pdf}{Supplementary Material}. 
	\end{proof}
	
	\begin{remark}[high-dimensionality in Scenario 1]
		The LRT statistic $\Lambda_1$ requires $n > p$ to ensure that $\textbf{V} ^ {(1)} > 0$ with probability $1$.
		However, the condition $n > p$ can be relaxed if $\bm{\Sigma} ^ {(1)}$ in $K_1$ follows a structured pattern rather than being arbitrarily unstructured.
		Furthermore, there is a normal approximation to the null distribution related to $\Lambda_1$ when both $p$ and $n$ go to infinity with $p / n \to c \in (0, 1]$, where $c$ is a constant.
		In contrast, 
		the LRT statistic $\Lambda_3$ does not impose this restriction $n > p$.
		Specifically, the second part of Theorem~\ref{Thm:covariance} remains valid even when $n ^ {(m)} < p$ and $n < p$, as long as $K$ by $K$ matrices $\widehat{\textbf{A}} ^ {(m)}$, $\widehat{\textbf{A}}$, $\widehat{\bm{\Delta}} ^ {(m)}$, and $\widehat{\bm{\Delta}}$ are positive definite with probability $1$ (and $n > \max \{n ^ {(m)} + M - 1, K + (K + 1) K / 2\}$).
		We emphasize that when $n ^ {(m)} < p$ and $n < p$, 
		$\left(n ^ {(m)} - 1\right)\textbf{S} ^ {(m)}$ and $(n - M)\textbf{S}$ follow singular Wishart distributions lacking probability density functions with respect to Lebesgue measure \citep{DiazCarciaJaimezMardia1997}.
	\end{remark}

\subsection{Joint hypothesis tests for mean structures in Scenario 2 \label{Subsec:mean}}	
	
	\begin{theorem}[LRT statistics related to mean structures]
		\label{Thm:mean}
		(1) Given that $\widehat{\textbf{A}} ^ {(1)} > 0$, $\widehat{\bm{\Delta}} ^ {(1)} > 0$ hold with probability $1$ and $p_k > 2$ for $k \in \{1, \ldots, K\}$,
		the LRT statistic $\Lambda_2$ for testing $H_2$ versus $K_2$, raised to the power of $2 / n$, is computed as
		\begin{equation*}
			\Lambda_2 ^ {2 / n} = 
			\prod_{k = 1}^{K} 
			\frac{\textbf{V}_{\bar{p}_k, \bar{p}_k} ^ {(1)}}{\textbf{H}_{\bar{p}_k, \bar{p}_k} ^ {(1)} + \textbf{V}_{\bar{p}_k, \bar{p}_k} ^ {(1)}}
			\left\{\frac{\sum_{j = \bar{p}_{k - 1} + 1}^{\bar{p}_k - 1} \textbf{V}_{j, j} ^ {(1)}}{\sum_{j = \bar{p}_{k - 1} + 1}^{\bar{p}_k - 1} \left(\textbf{H}_{j, j} + \textbf{V}_{j, j} ^ {(1)}\right)}\right\} ^ {p_k - 1},
		\end{equation*}
		where $\textbf{V} ^ {(1)} \triangleq \bm{\Gamma}_2 \textbf{W} ^ {(1)} \bm{\Gamma}_2 ^ \top$, 
		$\textbf{H} ^ {(1)} \triangleq \left(\bm{Y} ^ {(1)} - \bm{\nu}_0 \right)\left(\bm{Y} ^ {(1)} - \bm{\nu}_0 \right) ^ \top$, 
		$\bm{Y} ^ {(1)} \triangleq \sqrt{n} \bm{\Gamma}_2 \bar{\bm{X}} ^ {(1)}$, 
		$\bm{\nu}_0 \triangleq \sqrt{n} \bm{\Gamma}_2 \bm{\mu}_0$, 
		and $\bm{\Gamma}_2$ is determined by~\eqref{Eq:Gamma}, i.e.,
		$\bm{\Xi} ^ {(1)} \triangleq \bm{\Gamma}_2 \bm{\Sigma} ^ {(1)}\left[\textbf{A} ^ {(1)}, \textbf{B} ^ {(1)}, \bm{p}\right] \bm{\Gamma}_2 ^ \top$ is the canonical form of $\bm{\Sigma} ^ {(1)}\left[\textbf{A} ^ {(1)}, \textbf{B} ^ {(1)}, \bm{p}\right]$ presented in $H_2$.
		Under the alternative hypothesis $K_2$ with a general $\bm{\nu}_0$, we denote $\bm{\nu} ^ {(1)} \triangleq \sqrt{n}\bm{\Gamma}_2\bm{\mu} ^ {(1)}$ and $\bm{\Omega} ^ {(1)} \triangleq \left(\bm{\nu} ^ {(1)} - \bm{\nu}_0\right)\left(\bm{\nu} ^ {(1)} - \bm{\nu}_0\right) ^ \top$.
		Then, 
		$\Lambda_2$ follows a product of mutually independent noncentral type II beta variates, i.e., 
		\begin{equation*}
			\Lambda_2 \sim 
			\prod_{k = 1}^{K}
			\beta_{1, k} ^ {n / 2} \left(\frac{n - 1}{2}, \frac{1}{2}; \left(\bm{\Xi} ^ {(1), - 1/2} \bm{\Omega} ^ {(1)} \bm{\Xi} ^ {(1), - 1/2}\right)_{\bar{p}_k, \bar{p}_k}\right) 
			\beta_{2, k} ^ {(p_k - 1) n / 2} \left(\frac{(n - 1)(p_k - 1)}{2}, \frac{p_k - 1}{2}; \sum_{j = \bar{p}_{k - 1} + 1}^{\bar{p}_{k} - 1} \left(\bm{\Xi} ^ {(1), - 1/2} \bm{\Omega} ^ {(1)} \bm{\Xi} ^ {(1), - 1/2}\right)_{j, j} \right),
		\end{equation*}
		where $\beta_{1,k} (\cdot, \cdot; \cdot)$ and $\beta_{2, k} \left(\cdot, \cdot; \cdot\right)$ denote the type II beta variates with shape parameters and noncentrality parameters for $k \in \{1, \ldots, K\}$.
		Alternatively, 
		$\Lambda_2 ^ {- 2 / n}$ follows a product of mutually independent noncentral $F$ variates, i.e.,
		\begin{equation*}
			\begin{split}
				\Lambda_2 ^ {- 2 / n}
				& \sim \prod_{k = 1}^{K}
				\left\{1 + \frac{1}{n - 1} F_{1, k} \left(1, n - 1; \left(\bm{\Xi} ^ {(1), - 1/2} \bm{\Omega} ^ {(1)} \bm{\Xi} ^ {(1), - 1/2}\right)_{\bar{p}_k, \bar{p}_k}\right) \right\} \\
				& \times
				\left\{1 + \frac{1}{n - 1} F_{2, k} \left(p_k - 1, \left(p_k - 1\right)\left(n - 1\right); \sum_{j = \bar{p}_{k - 1} + 1}^{\bar{p}_{k} - 1} \left(\bm{\Xi} ^ {(1), - 1/2} \bm{\Omega} ^ {(1)} \bm{\Xi} ^ {(1), - 1/2}\right)_{j, j} \right) \right\} ^ {p_k - 1},
			\end{split} 
		\end{equation*}
		where $F_{1, k} (\cdot, \cdot; \cdot)$ and $F_{2, k} (\cdot, \cdot; \cdot)$ denote the $F$ variates with degrees of freedoms and noncentrality parameters for $k \in \{1, \ldots, K\}$.
		Under the null hypothesis $H_2$ with $\bm{\nu}_0 = \bm{\nu} ^ {(1)}$, 
		$\bm{\Omega} ^ {(1)} = \textbf{0}_{p \times p}$, 
		and the above noncentrality parameters are zeros.
		Thus, $\Lambda_2$ and $\Lambda_2 ^ {- 2 / n}$ follow a product of mutually independent central beta variates and central $F$ variates, respectively.
		
		(2) Given that $\widehat{\textbf{A}} > 0$, $\widehat{\bm{\Delta}} > 0$ hold with probability $1$, $n > M$, and $p_k > 1$ for $k \in \{1, \ldots, K\}$,
		the LRT statistic $\Lambda_4$ for testing $H_4$ versus $K_4$, raised to the power of $2 / n$, is computed as
		\begin{equation*}
			\Lambda_4 ^ {2 / n}
			= \prod_{k = 1} ^ {K}
			\frac{\sum_{m = 1}^{M} \textbf{V}_{\bar{p}_k, \bar{p}_k} ^ {(m)}}{\widetilde{\textbf{H}}_{\bar{p}_k, \bar{p}_k} + \sum_{m = 1}^{M} \textbf{V}_{\bar{p}_k, \bar{p}_k} ^ {(m)}}
			\left\{\frac{\sum_{j = \bar{p}_{k - 1} + 1}^{\bar{p}_k - 1} \sum_{m = 1}^{M} \textbf{V}_{\bar{p}_k, \bar{p}_k} ^ {(m)}}{\sum_{j = \bar{p}_{k - 1} + 1}^{\bar{p}_k - 1} \left(\widetilde{\textbf{H}}_{j, j} + \sum_{m = 1}^{M} \textbf{V}_{\bar{p}_k, \bar{p}_k} ^ {(m)}\right)}\right\} ^ {p_k - 1},
		\end{equation*}
		where $\textbf{V} ^ {(m)} \triangleq \bm{\Gamma}_4 \textbf{W} ^ {(m)} \bm{\Gamma}_4 ^ \top$, 
		$\widetilde{\textbf{H}} \triangleq \sum_{m = 1}^{M} \left(\bm{Y} ^ {(m)} - \sqrt{n ^ {(m)}} \bar{\bm{Y}} \right)\left(\bm{Y} ^ {(m)} - \sqrt{n ^ {(m)}} \bar{\bm{Y}} \right) ^ \top$, 
		$\bar{\bm{Y}} \triangleq \sum_{m = 1}^{M} \sqrt{n ^ {(m)}} \bm{Y} ^ {(m)} / n$, 
		$\bm{Y} ^ {(m)} \triangleq \sqrt{n ^ {(m)}} \bm{\Gamma}_4 \bar{\bm{X}} ^ {(m)}$, 
		and $\bm{\Gamma}_4$ is determined by~\eqref{Eq:Gamma}, i.e.,
		$\bm{\Xi} \triangleq \bm{\Gamma}_4 \bm{\Sigma} \left[\textbf{A}, \textbf{B}, \bm{p}\right] \bm{\Gamma}_4 ^ \top$ is the canonical form of $\bm{\Sigma}\left[\textbf{A}, \textbf{B}, \bm{p}\right]$ presented in $H_4$.
		Under the alternative hypothesis $K_4$,
		we denote $\bm{\nu} ^ {(m)} \triangleq \sqrt{n ^ {(m)}} \bm{\Gamma}_4 \bm{\mu} ^ {(m)}$ for $m \in \{1, \ldots, M\}$ and noncentrality parameter $\bm{\Omega} \triangleq \left(\bm{\nu} ^ {(1)}, \ldots, \bm{\nu} ^ {(M)}\right) \left\{\textbf{I}_M - \left(\sqrt{f ^ {(1)}}, \ldots, \sqrt{f ^ {(M)}}\right) ^ \top \left(\sqrt{f ^ {(1)}}, \ldots, \sqrt{f ^ {(M)}}\right)\right\} \left(\bm{\nu} ^ {(1)}, \ldots, \bm{\nu} ^ {(M)}\right) ^ \top$.
		Then, 
		$\Lambda_4$ follows a product of mutually independent noncentral type II beta variates, i.e., 
		\begin{equation*}
			\begin{split}
				\Lambda_4 \sim 
				\prod_{k = 1}^{K}
				\beta_{1, k} ^ {n / 2} \left(\frac{n - M}{2}, \frac{M - 1}{2}; \left(\bm{\Xi} ^ {- 1/2} \bm{\Omega} \bm{\Xi} ^ {- 1/2} \right)_{\bar{p}_k, \bar{p}_k}\right) 
				\beta_{2, k} ^ {(p_k - 1) n / 2}
				\left(\frac{(n - M)(p_k - 1)}{2}, \frac{(M - 1)(p_k - 1)}{2}; \sum_{j = \bar{p}_{k - 1} + 1}^{\bar{p}_k - 1} \left(\bm{\Xi} ^ {- 1/2} \bm{\Omega} \bm{\Xi} ^ {- 1/2} \right)_{j, j}\right),
			\end{split}
		\end{equation*}
		where $\beta_{1, k} (\cdot, \cdot; \cdot)$ and $\beta_{2, k} (\cdot, \cdot; \cdot)$ denote the type II beta variates with shape parameters and noncentrality parameters for $k \in \{1, \ldots, K\}$.
		Alternatively, 
		$\Lambda_4 ^ {- 2 / n}$ follows a product of mutually independent noncentral $F$ variates, i.e.,
		\begin{equation*}
			\begin{split}
				\Lambda_4 ^ {- 2 / n}
				& \sim \prod_{k = 1}^{K}
				\left\{1 + \frac{M - 1}{n - M} F_{1, k} \left(M - 1, n - M; \left(\bm{\Xi} ^ {- 1/2} \bm{\Omega} \bm{\Xi} ^ {- 1/2} \right)_{\bar{p}_k, \bar{p}_k} \right) \right\} \\
				& \times \left\{1 + \frac{M - 1}{n - M} F_{2, k} \left(\left(M - 1\right)\left(p_k - 1\right), \left(p_k - 1\right)\left(n - M\right); \sum_{j = \bar{p}_{k - 1} + 1}^{\bar{p}_k - 1} \left(\bm{\Xi} ^ {- 1/2} \bm{\Omega} \bm{\Xi} ^ {- 1/2} \right)_{j, j}\right) \right\} ^ {p_k - 1},
			\end{split} 
		\end{equation*}
		where $F_{1, k}(\cdot, \cdot; \cdot)$ and $F_{2, k}(\cdot, \cdot; \cdot)$ denote the $F$ variates with degrees of freedoms and noncentrality parameters for $k \in \{1, \ldots, K\}$.
		Under the null hypothesis $H_4$, $\bm{\Omega} = \textbf{0}_{p \times p}$, and the above noncentrality parameters are zeros.
		Thus, $\Lambda_4$ and $\Lambda_4 ^ {- 2 / n}$ follow a product of mutually independent central beta variates and central $F$ variates, respectively.
	\end{theorem}
	
	\begin{proof}[Proof of Theorem~\ref{Thm:mean}]
		A complete proof is provided in the \href{Supplementary Material.pdf}{Supplementary Material}. 
	\end{proof}
	
	In the calculation of $\Lambda_2$ or $\Lambda_4$, 
	we transform the matrix $\textbf{W} ^ {(m)}$ into the matrix $\textbf{V} ^ {(m)}$ using the (estimated) $\bm{\Gamma}_2$ or $\bm{\Gamma}_4$.
	Consequently, if $\textbf{W} ^ {(m)}$ contains missing values, 
	$\Lambda_2$ or $\Lambda_4$ requires a data imputation process. 
	However, the following information statistics leverage the uniform-block structures, eliminating the need for imputation of missing data.
	
	When $K = 1$ or $K = 2$, \citet{Geisser1963} introduced an information statistic for testing $H_2$ versus $K_2$,
	and derived its exact null distribution in closed form.
	However, for $K > 2$, \citet{Geisser1963} only outlined an algorithm for computing the information statistic, without providing formal proofs. 
	In the rest of this section, 
	we elucidate the exact null distributions of one-population Geisser's information statistic with a rigorous proof.
	We then extend Geisser's information statistic to the multiple-population case.
	Our proofs confirm that Geisser's null distributions are equivalent to a linear combination of mutually independent $F$-variates, further augmented by either Hotelling's $T ^ 2$ statistic (if $M = 1$) or Hotelling-Lawley $T_0 ^ 2$ statistic (if $M > 1$).
	
	\begin{theorem}[Geisser's information statistics related to mean structures]
		\label{Thm:information}
		(1) Given that $\widehat{\textbf{A}} ^ {(1)} > 0$, $\widehat{\bm{\Delta}} ^ {(1)} > 0$ hold with probability $1$, $n > K$, and $p_k > 1$ for $k \in \{1, \ldots, K\}$,
		the Geisser's information statistic $G_2$ for testing $H_2$ versus $K_2$ is computed as 
		\begin{equation*}
			G_2 = n \left(\bar{\bm{X}} ^ {(1)} - \bm{\mu}_0 \right) ^ \top 
			\widehat{\bm{\Theta}} ^ {(1)} \left[\widehat{\textbf{A}}_{\bm{\Theta}} ^ {(1)}, \widehat{\textbf{B}}_{\bm{\Theta}} ^ {(1)}, \bm{p}\right]
			\left(\bar{\bm{X}} ^ {(1)} - \bm{\mu}_0 \right).
		\end{equation*}
		Under the alternative hypothesis $K_2$ with a general $\bm{\mu}_0$, 
		we denote $K$ by $p$ matrix $\bm{C}_{\bm{p}} \triangleq \operatorname{Bdiag}\left(p_1 ^ {- 1}\textbf{1}_{1 \times p_1}, \ldots, p_K ^ {- 1} \textbf{1}_{1 \times p_K}\right)$, 
		and we partition $\bm{\mu} ^ {(1)} = \left(\bm{\mu}_1 ^ {(1), \top}, \ldots, \bm{\mu}_K ^ {(1), \top}\right) ^ \top$ and $\bm{\mu}_0 = \left(\bm{\mu}_{0, 1} ^ \top, \ldots, \bm{\mu}_{0, K} ^ \top \right) ^ \top$ satisfying that $\bm{\mu}_k ^ {(1)}$ and $\bm{\mu}_{0, k}$ have dimension $p_k$ for $k = 1, \ldots, K$.
		Then, 
		$G_2$ follows a sum of mutually independent noncentral $F$ variates plus the noncentral Hotelling's $T ^ 2$ statistic, i.e., 
		\begin{equation*}
			G_2 \sim \sum_{k = 1}^{K}
			(p_k - 1) F_{k}\left(\left(p_k - 1\right), \left(p_k - 1 \right)\left(n - 1\right); \Omega_k\right) + T ^ 2(\Omega_{K + 1}),
		\end{equation*}
		where 
		$T ^ 2(\Omega_{K + 1}) \triangleq K(n - 1)(n - K) ^ {- 1} F_{k + 1}\left(K, n - K; \Omega_{K + 1} \right)$ is the noncentral Hotelling's $T ^ 2$ statistic, 
		and $F_{k}\left(\cdot, \cdot; \cdot \right)$ denotes the $F$ variate with degrees of freedoms and noncentrality parameters, and $\Omega_k \triangleq \left(\bm{\mu}_k ^ {(1)} - \bm{\mu}_{0, k}\right) ^ \top \left(a_{kk} ^ {- 1} \textbf{I}_{p_k} - a _{kk} ^ {-1} p_{k} ^ {- 1}\textbf{J}_{p_k}\right) \left(\bm{\mu}_k ^ {(1)} - \bm{\mu}_{0, k}\right)$ for $k \in \{1, \ldots, K\}$, 
		$\Omega_{K + 1} \triangleq n \left(\bm{\mu} ^ {(1)} - \bm{\mu}_0 \right) ^ \top \bm{C}_{\bm{p}} ^ \top \left(\textbf{P} \bm{\Delta} ^ {- 1} \right)\bm{C}_{\bm{p}} \left(\bm{\mu} ^ {(1)} - \bm{\mu}_0 \right)$.
		Under the null hypothesis $H_2$ with $\bm{\mu} ^ {(1)} = \bm{\mu}_0$, $\Omega_1 = \cdots = \Omega_{K + 1} = 0$. 
		Thus, $G_2$ follows a sum of mutually independent central $F$ variates plus the central Hotelling's $T ^ 2$ statistic.
		
		(2) Given that $\widehat{\textbf{A}} > 0$, $\widehat{\bm{\Delta}} > 0$ hold with probability $1$, $n > M + K$, and $p_k > 1$ for $k \in \{1, \ldots, K\}$,
		the Geisser's information statistic $G_4$ for testing $H_4$ versus $K_4$ is computed as 
		\begin{equation*}
			G_4 = \sum_{m = 1}^{M} n ^ {(m)} \left(\bar{\bm{X}} ^ {(m)} - \bar{\bm{X}}\right) ^ \top 
			\widehat{\bm{\Theta}}\left[\widehat{\textbf{A}}_{\bm{\Theta}}, \widehat{\textbf{B}}_{\bm{\Theta}}, \bm{p}\right]
			\left(\bar{\bm{X}} ^ {(m)} - \bar{\bm{X}}\right).
		\end{equation*}
		Under the alternative hypothesis $K_4$, 
		we denote $\bar{\bm{\mu}} \triangleq \sum_{m = 1}^{M} f ^ {(m)} \bm{\mu} ^ {(m)}$, 
		$M$ by $M$ matrices $\textbf{N} \triangleq \operatorname{diag}\left(n ^ {(1)}, \ldots, n ^ {(M)}\right)$ and $\textbf{M} \triangleq \textbf{N} ^ {- 1/ 2} \left(\textbf{I}_M - n ^ {- 1} \textbf{J}_M \textbf{N}\right)$.
		Then, 
		$G_4$ follows a sum of mutually independent noncentral $F$ variates plus the noncentral Hotelling-Lawley $T_0^2$ statistic, i.e., 
		\begin{equation*}
			G_4 \sim \sum_{k = 1}^{K} (M - 1) \left(p_k - 1\right)
			F_{k}\left(\left(M - 1\right)\left(p_k - 1\right), \left(n - M\right)\left(p_k - 1\right); \Omega_k\right) + T_0 ^ 2(\bm{\Omega}),
		\end{equation*}
		where $T_0 ^ 2(\cdot)$ is the noncentral Hotelling-Lawley $T_0 ^ 2$ statistic with $\bm{\Omega} \triangleq \sum_{m = 1}^{M} n ^ {(m)} \bm{C}_{\bm{p}} \left(\bm{\mu} ^ {(m)} - \bar{\bm{\mu}}\right) \left(\bm{\mu} ^ {(m)} - \bar{\bm{\mu}}\right) ^ \top \bm{C}_{\bm{p}} ^ \top$, 
		$F_{k}(\cdot, \cdot; \cdot)$ denotes the $F$ variate with degrees of freedoms and noncentral parameters,
		and let $\otimes$ denote the Kronecker product, 
		$\Omega_k \triangleq \left(\bm{\mu} ^ {(1), \top}, \ldots, \bm{\mu} ^ {(M), \top} \right) ^ \top \left\{(\textbf{M} ^ \top \textbf{M}) \otimes \operatorname{Bdiag}\left(\textbf{0}_{p_1 \times p_1}, \cdots, a_{kk} ^ {-1} \textbf{I}_{p_k} - a_{kk} ^ {- 1} p_k ^ {- 1} \textbf{J}_{p_k}, \cdots, \textbf{0}_{p_K \times p_K}\right) \right\} \left(\bm{\mu} ^ {(1), \top}, \ldots, \bm{\mu} ^ {(M), \top} \right)$ for $k \in \{1, \ldots, K\}$.
		Under the null hypothesis $H_4$, $\Omega_k = 0$ and $\bm{\Omega} = \textbf{0}_{p \times p}$.
		Thus, $G_4$ follows a sum of mutually independent central $F$ variates plus the central Hotelling-Lawley $T_0 ^ 2$ statistic.
	\end{theorem}
	
	\begin{proof}[Proof Sketch of Theorem~\ref{Thm:information}]
		The proof can be carried out in three steps.
		First, decompose $G_2$ and $G_4$ into a sum of several components.
		Second, examine the independence among these components.
		Third, determine the distribution of each component.
		A complete proof is provided in the \href{Supplementary Material.pdf}{Supplementary Material}. 
	\end{proof}
	
	\begin{remark}[approximate distributions of the Hotelling's $T_0 ^ 2$ statistic]
		The Hotelling-Lawley $T_0 ^ 2$ statistic, also known as the Hotelling-Lawley trace, is widely employed for testing the equality of multiple population means \citep{Lawley1938, Hotelling1947, Hotelling1951}. 
		However, it is intractable to derive the exact null or non-null distribution of the Hotelling-Lawley $T_0 ^ 2$ statistic, which has prompted the development of various approximations in the literature \citep{Ito1956, Ito1960, PillaiYoung1971, Siotani1971, McKeon1974}.
		We explore $F$-type distributions in the simulation studies and real data analysis as their approximations \citep{McKeon1974, Betz1987}.
	\end{remark}
	
	\begin{remark}[handling missing data]
		Compared to the LRT statistics $\Lambda_2$ and $\Lambda_4$, 
		the information statistics $G_2$ and $G_4$ rely solely on the estimates of $\textbf{A} ^ {(m)}$ and $\textbf{B} ^ {(m)}$, 
		without requiring data transformations such as $\textbf{V} ^ {(m)}$.
		The expressions in~\eqref{Eq:ls} demonstrate that $\widehat{a}_{kk} ^ {(m)} + \widehat{b}_{kk} ^ {(m)}$ represents the average of the diagonal elements of $\textbf{S}_{kk} ^ {(m)}$, 
		$\widehat{b}_{kk} ^ {(m)}$ represents the average of the off-diagonal elements of $\textbf{S}_{kk} ^ {(m)}$, 
		and $\widehat{b}_{kk'} ^ {(m)}$ represents the average of all elements of $\textbf{S}_{kk'} ^ {(m)}$.
		While conventional sample covariance matrices may be undefined in the presence of missing values, the proposed estimators remain valid when omitting the missing values from $\textbf{S} ^ {(m)}$ (or $\textbf{S}$), as long as $\textbf{S}_{kk'} ^ {(m)}$ (or $\textbf{S}_{kk'}$) includes at least one observation.
	\end{remark}
	
	\begin{remark}[high-dimensionality in Scenario 2]
		Compared to classical LRT statistics and Hotelling's $T ^ 2$ statistic, 
		the LRT statistics $\Lambda_2$ and $\Lambda_4$, as well as Geisser's information statistics $G_2$ and $G_4$, 
		are applicable even when $n ^ {(m)} < p$ and $n < p$, as long as  $\widehat{\textbf{A}} ^ {(m)}$, $\widehat{\textbf{A}}$, $\widehat{\bm{\Delta}} ^ {(m)}$, and $\widehat{\bm{\Delta}}$ are positive definite with probability $1$ (and $n > \max\left\{M + K, K + (K + 1)K / 2\right\}$).
		When $n ^ {(m)} < p$ and $n < p$,
		we remark that $\left(n ^ {(m)} - 1\right)\textbf{S} ^ {(m)}$ and $(n - M) \textbf{S}$ follow singular Wishart distributions lacking probability density functions with respect to Lebesgue measure \citep{DiazCarciaJaimezMardia1997}.
	\end{remark}

\subsection{Simultaneous hypothesis tests for individual means in Scenario 3 \label{Subsec:FDP}}	
	
	We first simplify our notation due to the single population. 
	For ease of presentation, we suppress the superscript $^ {(1)}$ throughout this section.
	Specifically, we consider 
	$\bm{X}_1, \ldots, \bm{X}_n \overset{\text{i.i.d.}}{\sim} N_p\left(\bm{\mu}, \bm{\Sigma}\left[\textbf{A}, \textbf{B}, \bm{p}\right]\right)$ with $\bm{\mu} \triangleq \left(\mu_1, \ldots, \mu_p\right) ^ \top$, $\textbf{A} > 0$ and $\bm{\Delta} \triangleq \textbf{A} + \textbf{B} \textbf{P} > 0$.
	The sample mean and the sample covariance matrix are written as $\bar{\bm{X}} \triangleq \left(\bar{X}_1, \ldots, \bar{X}_p\right) ^ \top$ and $\textbf{S}$, respectively. 
	Additionally, 
	we define $\textbf{C} \triangleq \operatorname{diag}\left(c_{11}, \ldots, c_{KK}\right)$ and $\textbf{D} \triangleq \operatorname{Bdiag}\left(\sqrt{c_{11}}\textbf{I}_{p_1}, \ldots, \sqrt{c_{KK}} \textbf{I}_{p_K}\right)$, 
	where $c_{kk} \triangleq a_{kk} + b_{kk}$ for $k \in \{1, \ldots, K\}$.
	Thus, $\textbf{D}$ contains the square-rooted diagonal elements of $\bm{\Sigma}\left[\textbf{A}, \textbf{B}, \bm{p}\right]$.
	Using the formulas in~\eqref{Eq:ls} and omitting the superscript $^ {(m)}$ since $m = 1$,
	we denote the estimators of $\textbf{A}$, $\textbf{B}$, $\textbf{C}$, $\textbf{D}$, and $\bm{\Sigma}\left[\textbf{A}, \textbf{B}, \bm{p}\right]$ as follows: 
	$\widehat{\textbf{A}} \triangleq \operatorname{diag}\left(\widehat{a}_{11}, \ldots, \widehat{a}_{KK}\right)$, 
	$\widehat{\textbf{B}} \triangleq \left(\widehat{b}_{kk'}\right)$,
	$\widehat{\textbf{C}} \triangleq \operatorname{diag}\left(\widehat{c}_{11}, \ldots, \widehat{c}_{KK}\right)$, 
	$\widehat{\textbf{D}} \triangleq \operatorname{Bdiag}\left(\sqrt{\widehat{c}_{11}}\textbf{I}_{p_1}, \ldots, \sqrt{\widehat{c}_{KK}} \textbf{I}_{p_K}\right)$,
	where $\widehat{c}_{kk} \triangleq \widehat{a}_{kk} + \widehat{b}_{kk}$ for $k \in \{1, \ldots, K\}$, 
	and $\widehat{\bm{\Sigma}}\left[\widehat{\textbf{A}}, \widehat{\textbf{B}}, \bm{p}\right] = \widehat{\textbf{A}} \circ \textbf{I}[\bm{p}] + \widehat{\textbf{B}} \circ \textbf{J}[\bm{p}]$.
	
	To conduct simultaneous tests for $H_{5, j}: \mu_j = 0$ versus $K_{5, j}: \mu_j \neq 0$, $j \in \{1, \ldots, p\}$, 
	given unknown $\bm{\Sigma}\left[\textbf{A}, \textbf{B}, \bm{p}\right]$, 
	we consider the transformed data $\sqrt{n} \widehat{\textbf{D}} ^ {-1} \bar{\bm{X}} \sim N_p\left(\widetilde{\bm{\mu}}, \widetilde{\bm{\Sigma}}\right) \approx N_p\left(\widetilde{\bm{\mu}}, \operatorname{corr}\left(\bm{\Sigma}\left[\textbf{A}, \textbf{B}, \bm{p}\right]\right)\right)$,
	where $\widetilde{\bm{\mu}} \triangleq \sqrt{n} \widehat{\textbf{D}} ^ {- 1} \bm{\mu}$, 
	$\widetilde{\bm{\Sigma}} \triangleq \widehat{\textbf{D}} ^ {- 1} \bm{\Sigma}\left[\textbf{A}, \textbf{B}, \bm{p}\right]\widehat{\textbf{D}} ^ {- 1}$.
	The approximation $\approx$ holds because $\widehat{c}_{kk}$ uniformly converges to $c_{kk}$ for $k \in \{1, \ldots, K\}$ under normality \citep{BickelLevina2008a, FanHan2017}.
	This implies that $\mu_j = 0$ for all $j$ is equivalent to $\widetilde{\mu}_j = 0$ for all $j$. 
	Thus, 
	we test $\widetilde{\mu}_j = 0$ for $j \in \{1, \ldots, p\}$ using the marginal test statistic vector
	\begin{equation}
		\label{Eq:T}
		\bm{T} = \left(T_1, \ldots, T_p\right) ^ \top \triangleq \sqrt{n} \widehat{\textbf{D}} ^ {-1} \bar{\bm{X}},  
		\text{ where }  
		T_j \triangleq \frac{\sqrt{n} \bar{X}_j}{\sqrt{\widehat{c}_{kk}}},  
		k \in \{1, \ldots, K\} \text{ such that }
		\bar{p}_{k - 1} + 1 \leq j \leq \bar{p}_k.
	\end{equation}
	However, two challenges arise in multiple hypothesis testing under a uniform-block covariance structure.
	
	First, the marginal test statistic $T_j$ in~\eqref{Eq:T} does not exactly follow $t$-distributions with degrees of freedom parameter $(n - 1)$.
	More precisely, 
	the $j$th marginal test statistic $T_j \triangleq \sqrt{n} \bar{X}_j / \sqrt{\widehat{c}_{kk}}$, 
	where $k \in \{1, \ldots, K\}$ satisfying $\bar{p}_{k - 1} + 1 \leq j \leq \bar{p}_k$, 
	does not follow $t$-distributions with degrees of freedom parameter $(n - 1)$, 
	because $\widehat{c}_{kk}$ does not follow a chi-square distribution; 
	instead, it follows a linear combination of two independent weighted chi-square distributions, i.e., $(n - 1)p_k \widehat{c}_{kk} \sim \left(a_{kk} + p_{k} b_{kk}\right) \chi ^ 2 (n - 1) + a_{kk} \chi ^ 2\left((n - 1)(p_k - 1)\right)$ (see the proof of Theorem~\ref{Thm:information}).
	Therefore,
	applying the Welch-Satterthwaite method \citep{Satterthwaite1946, Welch1947}, 
	we approximate $(n - 1) p_k \widehat{c}_{kk} \overset{\text{Welch-Satterthwaite}}{\sim} d_k \chi ^ 2 (v_k)$, 
	where $d_k\triangleq (a_{kk} ^ 2 + p_k b_{kk} ^ 2 + 2 a_{kk} b_{kk}) / (a_{kk} + b_{kk})$ and $v_k \triangleq (n - 1) p_{k} \left(a_{kk} + b_{kk}\right) ^ 2 / (a_{kk} ^ 2 + 2 a_{kk} b_{kk} + p_k b_{kk} ^ 2)$ are obtained by matching their means and variances.
	This approximation remains valid in high-dimensional settings \citep{ZhangGuoZhou2020}: under regularity conditions, $(n - 1) p_k \widehat{c}_{kk}$ and $d_k \chi ^ 2 (v_k)$ have the same asymptotic distribution.
	Thus, under $\widetilde{\mu}_j = 0$, ~\eqref{Eq:T} becomes, for $j \in \{1, \ldots, p\}$, 
	\begin{equation}
		\label{Eq:Tj}
		T_j \triangleq \sqrt{n} \bar{X}_j / \sqrt{\widehat{c}_{kk}}
		\overset{\text{Welch-Satterthwaite}}{\sim} t(v_k), 
		\quad 
		v_k \triangleq (n - 1) p_{k} \left(a_{kk} + b_{kk}\right) ^ 2 / (a_{kk} ^ 2 + 2 a_{kk} b_{kk} + p_k b_{kk} ^ 2),
	\end{equation} 
	where $k \in \left\{1, \ldots, K\right\}$ satisfies that $\bar{p}_{k - 1} + 1 \leq j \leq \bar{p}_k$ and $t(\cdot)$ denotes the $t$-variate with a degrees of freedom parameter.
	The (unadjusted) $p$-values of marginal tests are calculated as 
	\begin{equation}
		\label{Eq:pvalue}
		P_j \triangleq 2 \Phi_{v_k}\left(- \vert T_j \vert\right), \quad
		j \in \{1, \ldots, p\}, 
	\end{equation}
	where $\Phi_{v_k}(\cdot)$ is the cumulative distribution function of $t(v_k)$.
	
	Second, an appropriate measure for type I errors is required due to the dependence among marginal test statistics.
	Conventionally, 
	when the test statistics are independent or their joint distribution exhibits positive regression dependence on subsets \citep{BenjaminiYekutieli2001}, 
	the standard Benjamini--Hochberg procedure can be applied to control the false discovery rate (FDR) \citep{BenjaminiHochberg1995}. 
	However, when the simultaneous test statistics are correlated, e.g., in the presence of uniform-block structure, 
	the expectation and variance of the number of false discoveries can be strongly influenced by these non-negligible dependencies \citep{FinnerRoters2002, SchwartzmanLin2011}.
	Thus, incorporating the dependence structure among test statistics
	is essential to avoid misleading statistical inferences, 
	such as loss of statistical power and distortion in the ranking of $p$-values \citep{LeekStorey2008}.
	In settings with strong correlations, \citet{FanHanGu2012} and \citet{FanHan2017} suggest directly estimating the false discovery proportion (FDP), 
	which is more relevant to the present study.
	Unlike FDR, which provides an expectation-based measure of error rate, FDP offers a direct assessment of false discoveries, i.e., $\operatorname{FDR} = \operatorname{E}\left(\operatorname{FDP}(T)\right)$, 
	where $T$ is a test threshold.
	Below, we describe how to control FDP in our setting following the approach of \citet{FanHanGu2012} and \citet{FanHan2017}.
	
	Given a threshold value $t \in [0, 1]$, 
	we define the number of false discoveries $V(t)$, 
	the number of true discoveries $S(t)$, 
	and the total number of discoveries $R(t)$ as the following empirical processes:
	\begin{equation*}
		V(t) \triangleq \# \left\{\text{ true null } P_j: P_j \leq t \right\}, \quad
		S(t) \triangleq \# \left\{\text{ false null } P_j: P_j \leq t \right\}, \quad
		R(t) \triangleq \# \left\{P_j: P_j \leq t\right\}.
	\end{equation*}
	Additionally, 
	we define $\operatorname{FDP}(t) \triangleq V(t) / R(t)$, 
	where the convention $0 / 0 = 0$ is used.
	
	Let $\bm{\Psi} \triangleq \operatorname{corr}\left(\bm{\Sigma}\left[\textbf{A}, \textbf{B}, \bm{p}\right]\right) > 0$ denote the population correlation matrix.
	By Lemma~\ref{Lem:covariance}, 
	$\bm{\Psi}$ follows a uniform-block structure, expressed as $\bm{\Psi} = \bm{\Psi}\left[\textbf{A}_{\bm{\Psi}}, \textbf{B}_{\bm{\Psi}}, \bm{p}\right]$, 
	where $\textbf{A}_{\bm{\Psi}} = \textbf{C} ^ {-1 / 2}\textbf{A} \textbf{C} ^ {- 1 / 2}$ and $\textbf{B}_{\bm{\Psi}} = \textbf{C} ^ {-1 / 2}\textbf{B} \textbf{C} ^ {- 1 / 2}$, as provided in~\eqref{Eq:correlation}.
	Let $\lambda_j ^ {(\bm{\Psi})} > 0$ and $\bm{\xi}_j ^ {(\bm{\Psi})} \in \mathbb{R} ^ {p}$ be the $j$th eigenvalue and its corresponding eigenvector (normalized to the unit length) of $\bm{\Psi}\left[\textbf{A}_{\bm{\Psi}}, \textbf{B}_{\bm{\Psi}}, \bm{p}\right]$ in decreasing order.
	Using Lemma~\ref{Lem:ops_all}, 
	these can be computed and then reordered.
	
	For a fixed constant $\kappa_0 \in \{1, \ldots, p - 1\}$, $\bm{\Psi}\left[\textbf{A}_{\bm{\Psi}}, \textbf{B}_{\bm{\Psi}}, \bm{p}\right]$ can be decomposed as
	\begin{equation*}
		\bm{\Psi}\left[\textbf{A}_{\bm{\Psi}}, \textbf{B}_{\bm{\Psi}}, \bm{p}\right]
		= \sum_{j = 1}^{\kappa_0} \lambda_j ^ {(\bm{\Psi})} 
		\bm{\xi}_j ^ {(\bm{\Psi})} \bm{\xi}_j ^ {(\bm{\Psi}), \top} + \sum_{j' = \kappa_0 + 1}^{p} \lambda_{j'} ^ {(\bm{\Psi})} 
		\bm{\xi}_{j'} ^ {(\bm{\Psi})} \bm{\xi}_{j'} ^ {(\bm{\Psi}), \top}
		\triangleq \textbf{L}_{p \times \kappa_0} \textbf{L} ^ \top + \textbf{Q}_{p \times p},
	\end{equation*}
	where $\textbf{L}_{p \times \kappa_0} \triangleq \left(\bm{l}_1, \ldots, \bm{l}_{\kappa_0}\right)$ contains the first $\kappa_0$ principal components, with $\bm{l}_j \triangleq \sqrt{\lambda_j ^ {(\bm{\Psi})}} \bm{\xi}_j ^ {(\bm{\Psi})}$ as its $j$th column for $j = 1, \ldots, \kappa_0$.
	Define $\bm{Z} \triangleq \left(Z_1, \ldots, Z_p\right) ^ \top \triangleq \sqrt{n} \textbf{D} ^ {- 1} \bar{\bm{X}}$ and $\breve{\bm{\mu}} \triangleq \left(\breve{\mu}_1, \ldots, \breve{\mu}_p \right) ^ \top \triangleq \sqrt{n} \textbf{D} ^ {- 1} \bm{\mu}$.
	Then, $\bm{Z} \sim N_p \left(\breve{\bm{\mu}} , \textbf{D} ^ {- 1} \bm{\Sigma}\left[\textbf{A}, \textbf{B}, \bm{p} \right] \textbf{D} ^ {- 1}\right) = N_p \left(\breve{\bm{\mu}}, \bm{\Psi}\left[\textbf{A}_{\bm{\Psi}}, \textbf{B}_{\bm{\Psi}}, \bm{p}\right]\right)$.
	Accordingly, the $j$th component of $\bm{Z}$ decomposes as 
	\begin{equation*}
		Z_j = \breve{\mu}_j + \bm{l} ^ {(j), \top} \bm{w} + q_j, \quad j = 1, \ldots p,
	\end{equation*}
	where $\bm{l} ^ {(j), \top}$ is the $j$th row of $\textbf{L}$, $\bm{w} \sim N_K \left(\textbf{0}_{K \times 1}, \textbf{I}_K\right)$ and $\bm{q} \triangleq \left(q_1, \ldots, q_p\right) ^ \top \sim N_p \left(\textbf{0}_{p \times 1}, \textbf{Q}\right)$ are independent random vectors.
	The oracle FDP and its approximation are defined as:
	\begin{equation*}
		\begin{split}
			\operatorname{FDP} ^ {(\text{O})} (t)
			\triangleq \sum_{j \in \{\text{true nulls}\}} 
			\frac{\Phi\left\{\ell_j\left(z_{t / 2} + \tau_j\right)\right\} + \Phi\left\{\ell_j\left(z_{t / 2} - \tau_j\right)\right\}}{R(t)}, \quad
			\operatorname{FDP} ^ {(\text{A})} (t)
			\triangleq \sum_{j = 1}^{p} 
			\frac{\Phi\left\{\ell_j\left(z_{t / 2} + \tau_j\right)\right\} + \Phi\left\{\ell_j\left(z_{t / 2} - \tau_j\right)\right\}}{R(t)},
		\end{split}
	\end{equation*}
	where $0 / 0 = 0$ is used conventionally. 
	Here, $\ell_j \triangleq \left(1 - \sum_{j = 1}^{\kappa_0} l_{j', j} ^ 2\right) ^ {- 1 / 2}$, 
	$l_{j', j}$ is the $j$th component of $\bm{l} ^ {(j'), \top}$,
	$\tau_j \triangleq \bm{l} ^ {(j), \top} \bm{w}$ for $j = 1, \ldots, p$,
	and $z_{t / 2} \triangleq \Phi ^ {- 1}(t / 2)$ is the lower $t / 2$ quantile of a standard normal distribution.
	
	Given the estimator $\widehat{\bm{\Psi}}\left[\widehat{\textbf{A}}_{\bm{\Psi}}, \widehat{\textbf{B}}_{\bm{\Psi}}, \bm{p}\right]$ of the population correlation matrix, where
	\begin{equation}
		\label{Eq:ls3}
		\widehat{\textbf{A}}_{\bm{\Psi}}
		= \widehat{\textbf{C}} ^ {- 1/2} \widehat{\textbf{A}} \widehat{\textbf{C}} ^ {- 1/2}, \quad
		\widehat{\textbf{B}}_{\bm{\Psi}}
		= \widehat{\textbf{C}} ^ {- 1/2} \widehat{\textbf{B}} \widehat{\textbf{C}} ^ {- 1/2}
	\end{equation}
	provided in~\eqref{Eq:correlation}.
	Let $\widehat{\textbf{L}}_{p \times \kappa_0}$ denote the first $\kappa_0$ principal components of $\widehat{\bm{\Psi}}\left[\widehat{\textbf{A}}_{\bm{\Psi}}, \widehat{\textbf{B}}_{\bm{\Psi}}, \bm{p}\right]$, 
	and $\widehat{\bm{w}} \triangleq \left(\widehat{\textbf{L}} ^ \top \widehat{\textbf{L}}\right) ^ {- 1} \widehat{\textbf{L}} ^ \top \bm{T}$, 
	where the marginal test statistic vector $\bm{T}$ is defined in~\eqref{Eq:T}.
	We define the adjusted $p$-values as
	\begin{equation}
		\label{Eq:pvalue_hat}
		\widehat{P}_j \triangleq 2 \Phi\left(- \left \vert \widehat{\ell}_j \left(T_j - \widehat{\tau}_j \right)\right \vert\right), \quad j \in \{1, \ldots, p\},
	\end{equation}
	and the FDP estimator as
	\begin{equation}
		\label{Eq:FDP_hat}
		\widehat{\operatorname{FDP}}(t)
		\triangleq \sum_{j = 1}^{p} 
		\frac{\Phi\left\{\widehat{\ell}_j\left(z_{t / 2} + \widehat{\tau}_j\right)\right\} + \Phi\left\{\widehat{\ell}_j\left(z_{t / 2} - \widehat{\tau}_j\right)\right\}}{R(t)},
	\end{equation}
	where the convention $0 / 0 = 0$ is used, 
	$\widehat{\ell}_j \triangleq \left(1 - \sum_{j = 1}^{\kappa_0} \widehat{l}_{j', j} ^ 2\right) ^ {- 1 / 2}$, 
	$\widehat{l}_{j', j}$ is the $j$th component of $\widehat{\bm{l}} ^ {(j'), \top}$,
	and $\widehat{\bm{l}} ^ {(j), \top}$ is the $j$th row vector of $\widehat{\textbf{L}}$, 
	$\widehat{\tau}_j \triangleq \widehat{\bm{l}} ^ {(j), \top} \widehat{\bm{w}}$ for $j \in \{1, \ldots, p\}$.
	
	\begin{remark}[steps for estimating FDP under uniform-block structures]
		Determine and fix $\alpha \in (0, 1)$ and select $\kappa_0 \in \{1, \ldots, p - 1\}$, 
		where $\kappa_0$ can be chosen based on suggestions in \citet{FanHan2017}.
		
		\textbf{Step 1}: Compute the estimates $\widehat{\textbf{A}}$, $\widehat{\textbf{B}}$, $\widehat{\textbf{C}}$, and $\widehat{\textbf{D}}$ using~\eqref{Eq:ls}.
		
		\textbf{Step 2}: Calculate $\bm{T}$ using~\eqref{Eq:T} and the corresponding unadjusted $p$-values $P_j$ for $j \in \{1, \ldots, p\}$ using~\eqref{Eq:pvalue}, 
		where the degrees of freedom parameters are given in~\eqref{Eq:Tj}.
		
		\textbf{Step 3}: Set candidate threshold values: $t_1, \ldots, t_S \in [0, 1]$, where $S$ is pre-determined, say $S = 200$.
		
		\textbf{Step 4}: Compute the numbers of total discoveries $R(t_s) \triangleq \# \left\{P_j: P_j \leq t_s\right\}$ using the unadjusted $p$-values $P_j$ from Step 2 for each $s \in \{1, \ldots, S\}$. 
		
		\textbf{Step 5}: Compute the estimates $\widehat{\bm{\Psi}}\left[\widehat{\textbf{A}}_{\bm{\Psi}}, \widehat{\textbf{B}}_{\bm{\Psi}}, \bm{p}\right]$, 
		$\widehat{\textbf{A}}_{\bm{\Psi}}$, $\widehat{\textbf{B}}_{\bm{\Psi}}$ using~\eqref{Eq:ls3} and results from Step 1; 
		then, calculate its eigenvalues and eigenvectors using the canonical form result of Lemma~\ref{Lem:ops_all};
		thus, construct $\widehat{\textbf{L}}$ from the first $\kappa_0$ principal components.
		
		\textbf{Step 6}: Calculate $\widehat{\bm{w}} \triangleq \left(\widehat{\textbf{L}} ^ \top \widehat{\textbf{L}}\right) ^ {- 1} \widehat{\textbf{L}} ^ \top \bm{T}$, 
		alternatively, $\widehat{\bm{w}}$ can be obtained using the quantile regression or other robust regression approaches \citep{VutovDickhaus2022, VutovDickhaus2023};
		then compute $\widehat{\tau}_j \triangleq \widehat{\bm{l}} ^ {(j), \top} \widehat{\bm{w}}$ for $j \in \{1, \ldots, p\}$.
		
		\textbf{Step 7}: Calculate the adjusted $p$-values $\widehat{P}_j$ using~\eqref{Eq:pvalue_hat} for $j \in \{1, \ldots, p\}$ and the $\operatorname{FDP}(t)$ estimator $\widehat{\operatorname{FDP}}(t)$ using~\eqref{Eq:FDP_hat}.
		
		\textbf{Step 8}: Determine $t_{s_{\alpha}} \triangleq \max \left\{t_{s}: \widehat{\operatorname{FDP}}\left(t_s\right) \leq \alpha, s \in \left\{1, \ldots, S \right\} \right\}$, 
		i.e., the largest threshold $t_{s}$ satisfying $\widehat{\operatorname{FDP}}\left(t_s\right) \leq \alpha$.
		
		\textbf{Step 9}: Reject the null hypotheses $H_{5j}$ if the adjusted $p$-values $\widehat{P}_j \leq t_{s_{\alpha}}$ for $j \in \{1, \ldots, p\}$.
	\end{remark}
	
	The following theorem, adapted from \citet{FanHan2017}, establishes the approximation bounds for the proposed FDP estimator.
	
	\begin{theorem}[approximation bounds for the FDP estimator]
		\label{Thm:FDP}
		For the population correlation matrix $\bm{\Psi}\left[\textbf{A}_{\bm{\Psi}}, \textbf{B}_{\bm{\Psi}}, \bm{p}\right]$, 
		suppose there exists $\delta_1 > 0$, $\delta_2 \geq 0$, and $\delta_3 < \infty$ satisfying that $\sqrt{\lambda_{\kappa_0 + 1} ^ {(\bm{\Psi}), 2} + \cdots + \lambda_{p} ^ {(\bm{\Psi}), 2}} / p = O\left(p ^ {- \delta_1}\right)$, $R ^ {-1}(t) = O\left(p ^ {- (1 - \delta_2)}\right)$, and $\ell_j \leq \delta_3$ for $j \in \{1, \ldots, p\}$, respectively.
		Additionally, suppose $\lambda_j ^ {(\bm{\Psi})} - \lambda_{j + 1} ^ {(\bm{\Psi})} \geq g_p$ for a positive sequence $g_p$ for $j \in \{1, \ldots, \kappa_0\}$. 
		Then, given the proposed estimator in~\eqref{Eq:ls3}, 
		for sufficiently large $p$ and some $\delta_4 > 0$, 
		\begin{equation*}
			\begin{split}
				\left \vert \operatorname{FDP} ^ {(\text{O})} (t) - \operatorname{FDP} (t) \right \vert = O_{P}
				\left\{p ^ {\delta_2} \left(p ^ {- \delta_1 / 2} + n ^ {- 1/ 2}\right)\right\}, \quad
				\left \vert \widehat{\operatorname{FDP}}(t) - \operatorname{FDP} ^ {(\text{A})} (t) \right \vert
				= O_P \left[
				p ^ {\delta_2} \left\{\kappa_0 p ^ {- \delta_4} g_p / p + (\kappa_0 + 1) p ^ {- \delta_4} + \left \Vert \breve{\bm{\mu}} \right \Vert p ^ {-1 / 2} \right\}
				\right].
			\end{split}
		\end{equation*}
		
		
		
	\end{theorem}
	
	\begin{proof}[Proof Sketch of Theorem~\ref{Thm:FDP}]
		We verify the conditions of Theorems $2$ and $3$ in \citet{FanHan2017}, 
		as well as the arguments of the modified hard-threshold estimators in \citet{YangChenChen2024}.
		A complete proof is provided in the \href{Supplementary Material.pdf}{Supplementary Material}. 
	\end{proof}
	
	\begin{remark}[high-dimensionality in Scenario 3]
		Note that the estimators~\eqref{Eq:ls} or~\eqref{Eq:ls3} require that $n > K + (K + 1) K / 2$.
		However, when $K > n$ and $K$ grows with $n$, 
		\citet{YangChenChen2024} proposed a modified hard-thresholding estimators for $\textbf{A}$ and $\textbf{B}$ (or $\textbf{A}_{\bm{\Psi}}$ and $\textbf{B}_{\bm{\Psi}}$) that achieve consistency in both Frobenius and spectral norms.
		Consequently,
		if $K > n$ and $\log(K) / n \to 0$ as $n \to \infty$.
		Theorem~\ref{Thm:FDP} remains valid if we replace the estimators~\eqref{Eq:ls} or~\eqref{Eq:ls3} with the modified hard-thresholding estimators.
	\end{remark}
		
\section{Simulation studies \label{Sec:simulation}}	
	
	\subsection{Study 1: LRT statistics in Scenario 1 \label{Subsec:study1}}

	This study numerically compares the empirical distributions of $- \log \Lambda_1$ and $- \log \Lambda_3$ with their exact null distributions, as given in Theorem~\ref{Thm:covariance},
	across various sample sizes, covariance matrix dimensions, 
	and population numbers using Monte Carlo methods.
	 
	For $M = 1$,
	the	true population mean vector and covariance matrix are given by  $\bm{\mu} ^ {(1)} = \bm{\mu}_{*}$ and $\bm{\Sigma} ^ {(1)}\left[\textbf{A} ^ {(1)}, \textbf{B} ^ {(1)}, \bm{p}\right]$, respectively, where $\bm{p} = \left(p_1, p_2, p_3, p_4, p_5 \right) ^ \top = \left(\lfloor 3\rho \rfloor, \lfloor 4\rho \rfloor, \lfloor 5\rho \rfloor, \lfloor 6\rho \rfloor, \lfloor 7\rho \rfloor \right) ^ \top$.
	Here, $\rho \in \mathbb{R} ^ +$ controls the dimension $p$, 
	$\lfloor \cdot \rfloor$ denotes the floor function, and $K = 5$,
	where $p = \sum_{k = 1}^{5} \lfloor k \rho \rfloor$.
	Additionally, $\bm{\mu}_* = \textbf{1}_{p \times 1}$, 
	$\textbf{A} ^ {(1)} = \textbf{A}_{*}$, 
	$\textbf{B} ^ {(1)} = \textbf{B}_{*}$, 
	and 
	\begin{align*}
		\textbf{A}_{*} & = \begin{pmatrix*}[r]
			0.016 &  &  &  &  \\
			& 0.214 &  &  &  \\
			&  & 0.749 &  &  \\
			&  &  & 0.068 &  \\
			&  &  & & 0.100
		\end{pmatrix*}, \quad \quad \quad \quad \quad 
		\textbf{B}_{*} && =  \begin{pmatrix*}[r]
			 6.731 & -1.690 &  0.696 & -2.936 &  1.913 \\
			-1.690 &  5.215 &  3.815 & -1.010 &  0.703 \\
			 0.696 &  3.815 &  4.328 & -3.357 & -0.269 \\
			-2.936 & -1.010 & -3.357 &  6.788 &  0.000 \\
			 1.913 &  0.703 & -0.269 &  0.000 &  3.954
		\end{pmatrix*}, \\
		\textbf{P} & = \begin{pmatrix*}[r]
			\lfloor 3\rho \rfloor &  &  &  &  \\
			& \lfloor 4\rho \rfloor &  &  &  \\
			&  & \lfloor 5\rho \rfloor &  &  \\
			&  &  & \lfloor 6\rho \rfloor &  \\
			&  &  & & \lfloor 7\rho \rfloor
		\end{pmatrix*},  \quad
		\bm{\Delta} ^ {(1)} = \textbf{A} ^ {(1)} + \textbf{B} ^ {(1)} \textbf{P} && = 
		\begin{pmatrix*}[r]
			 20.21 & -6.761 &  3.480 & -17.62 &  13.39 \\
			-5.071 &  21.07 &  19.08 & -6.060 &  4.921 \\
			 2.088 &  15.26 &  22.39 & -20.14 & -1.882 \\
			-8.809 & -4.040 & -16.79 &  40.80 &  0.001 \\
			 5.739 &  2.812 & -1.345 &  0.001 &  27.78
		\end{pmatrix*}.
	\end{align*}
	Because $\bm{\Delta} ^ {(1)}$ has eigenvalues 
	$\lambda_{\bar{p}_1} = 59.61$, 
	$\lambda_{\bar{p}_{2}} = 34.87$, 
	$\lambda_{\bar{p}_{3}} = 28.94$, 
	$\lambda_{\bar{p}_{4}} = 8.34$, 
	and $\lambda_{\bar{p}_5} = 0.48$, 
	there exists a $p$ by $p$ orthogonal matrix $\bm{\Gamma}_1$, determined by Lemma~\ref{Lem:ops_all}, such that $\bm{\Xi} \triangleq \bm{\Gamma}_1 \bm{\Sigma} ^ {(1)}\left[\textbf{A} ^ {(1)}, \textbf{B} ^ {(1)}, \bm{p}\right] \bm{\Gamma}_1 ^ \top = \operatorname{diag}(0.016 \times \textbf{1}_{1 \times (p_1 - 1)}$, $59.61$, $0.214 \times \textbf{1}_{1 \times (p_2 - 1)}$, $34.87$, $0.749 \times \textbf{1}_{1 \times (p_3 - 1)}$, $28.94$, $0.068 \times \textbf{1}_{1 \times (p_4 - 1)}$, $8.34$, $0.100 \times \textbf{1}_{1 \times (p_5 - 1)}$, $0.48)$.
	
	Monte Carlo simulations with $10^5$ replicates are used to compare the empirical and theoretical distributions of the negative logarithm of the LRT statistic, $- \log \Lambda_1$. 
	Specifically, 
	for each replicate, 
	we adjust $\bm{p}$ by varying the values of $\rho$ and generate an i.i.d. sample of size $n \in \{50, 100\}$ from $N_p\left(\bm{\mu} ^ {(1)}, \bm{\Sigma} ^ {(1)}\left[\textbf{A} ^ {(1)}, \textbf{B} ^ {(1)}, \bm{p}\right]\right)$.
	Because $\bm{\Gamma}_1$ is unknown,
	we consider two scenarios for comparison: one where the true $\bm{\Gamma}_1$ is known and another where it is estimated.
	To estimate $\bm{\Gamma}_1$,  
	we first compute $\widehat{\textbf{A}} ^ {(1)}$ and $\widehat{\textbf{B}} ^ {(1)}$ using~\eqref{Eq:ls}. 
	Then, using Lemma~\ref{Lem:ops_all}, 
	we obtain the consistent estimate $\widehat{\bm{\Gamma}}_1$ based on the eigenvalues of $\widehat{\textbf{A}} ^ {(1)}$ and $\widehat{\bm{\Delta}} ^ {(1)} \triangleq \widehat{\textbf{A}} ^ {(1)} + \widehat{\textbf{B}} ^ {(1)} \textbf{P}$.
	Using both the truth $\bm{\Gamma}_1$ and its estimate $\widehat{\bm{\Gamma}}_1$, 
	we determine $- \log \Lambda_1$ based on the result in the first part of Theorem~\ref{Thm:covariance}.
	This yields the empirical distributions of $- \log \Lambda_1$ under the true and estimated $\bm{\Gamma}_1$, respectively, based on $10^5$ replicates.
	Additionally, in each replicate, we generate a theoretical version of $\Lambda_1$ as a product of mutually independent beta distributions, as described in the first part of Theorem~\ref{Thm:covariance}.
	This allows us to construct the theoretical distribution of $- \log \Lambda_1$ based on $10^5$ replicates.
	The quantiles of these distributions are summarized in Table~\ref{Tab:study1}.
	
	For $M > 1$, we consider the case of $M = 3$ populations.
	The partition-size vector $\bm{p}$, true population mean vectors, and covariance matrices are identical to those in the case of $M = 1$.
	Specifically, $\bm{p} = \left(\lfloor 3\rho \rfloor, \lfloor 4\rho \rfloor, \lfloor 5\rho \rfloor, \lfloor 6\rho \rfloor, \lfloor 7\rho \rfloor \right) ^ \top$, where $\rho$ varies across different values. 
	The mean vectors are given by $\bm{\mu} ^ {(m)} = m \bm{\mu}_{*}$, 
	and the covariance matrices take the form $\bm{\Sigma} ^ {(m)}\left[\textbf{A} ^ {(m)}, \textbf{B} ^ {(m)}, \bm{p}\right]$, 
	where $\textbf{A} ^ {(m)} = \textbf{A}_{*}$ and $\textbf{B} ^ {(m)} = \textbf{B}_{*}$ for $m \in \{1, \ldots, M\}$.
	Additionally,
	in hypothesis $H_3$,
	the common covariance matrix $\bm{\Sigma}\left[\textbf{A}, \textbf{B}, \bm{p}\right]$ has $\textbf{A} = \textbf{A}_{*}$ and $\textbf{B} = \textbf{B}_{*}$. 
	Consequently, its canonical form remains the same as in the case of $M = 1$, satisfying that $\bm{\Xi} = \bm{\Gamma}_3 \bm{\Sigma}\left[\textbf{A}, \textbf{B}, \bm{p}\right] \bm{\Gamma}_3 ^ \top$ with $\bm{\Gamma}_3 = \bm{\Gamma}_1$.
	
	Similarly, 
	we compare the empirical and theoretical distributions of the negative logarithm LRT statistic, $- \log \Lambda_3$, using Monte Carlo simulations with $10^5$ replicates.
	Specifically, 
	for each replicate, we generate i.i.d. samples from $N_p\left(\bm{\mu} ^ {(m)}, \bm{\Sigma} ^ {(m)}\left[\textbf{A} ^ {(m)}, \textbf{B} ^ {(m)}, \bm{p}\right]\right)$ for $m \in \{1, 2, 3\}$ with sizes $\left(n ^ {(1)}, n ^ {(2)}, n ^ {(3)} \right) \in \{(50, 50, 50), (50, 75, 100)\}$.
	The matrix $\bm{\Gamma}_3$ is estimated similarly, based on $\widehat{\textbf{A}}$ and $\widehat{\textbf{B}}$, 
	where we apply formula~\eqref{Eq:ls} to the pooled data matrix $\textbf{S}$.
	Using both the true $\bm{\Gamma}_3$ and its estimate $\widehat{\bm{\Gamma}}_3$, 
	we compute $- \log \Lambda_3$ according to the second part of Theorem~\ref{Thm:covariance}.
	Based on $10^5$ replicates, we obtain the empirical distributions of $- \log \Lambda_3$ corresponding to both the true $\bm{\Gamma}_3$ and its estimate $\widehat{\bm{\Gamma}}_3$.
	Additionally, we derive the theoretical distribution of $- \log \Lambda_3$ using the product of Dirichlet components, also based on $10^5$ replicates.
	The quantiles of these distributions are summarized in Table~\ref{Tab:study1}.
	
	Table~\ref{Tab:study1} demonstrates the consistency between the empirical distributions obtained using the true $\bm{\Gamma}_1$ and $\bm{\Gamma}_3$ and those obtained using their estimates, as evidenced by the small differences in quantiles.
	Specifically, we observe that the quantiles of $- \log \Lambda_3$ are similar across various sample sizes and dimensions.
	It is because the parameters for the Dirichlet variates are relatively large, depending on the sample sizes.  
	Furthermore, 
	the empirical null distributions based on the true $\bm{\Gamma}_1$ and $\bm{\Gamma}_3$ closely align with their theoretical null distributions, as indicated by the remarkably similar quantiles.
	
	\begin{table}[!htb]
		\scriptsize
		\centering
		\caption{
			Quantiles of theoretical and empirical distributions related to $- \log \Lambda_1$ and $- \log \Lambda_3$ under various sample sizes $n$ or $(n ^ {(1)}, n ^ {(2)}, n ^ {(3)})$ and covariance matrix dimensions $p$. 
			Here ``Theory'', ``Emp-True'', and ``Emp-Estd'' denote the theoretical distribution and the empirical distributions computed using the true and estimated $\bm{\Gamma}_1$ or $\bm{\Gamma}_3$, respectively}
		\begin{tabular}{ccccccccccc}
			\hline
			\multirow{2}{*}{Statistic} & \multirow{2}{*}{Sample Size} & \multirow{2}{*}{Dimension} & \multirow{2}{*}{Distribution} & \multicolumn{7}{c}{Quantile} \\ \cline{5-11} 
			&  &  &  & \multicolumn{1}{c}{0.01} & \multicolumn{1}{c}{0.05} & \multicolumn{1}{c}{0.10} & \multicolumn{1}{c}{0.50} & \multicolumn{1}{c}{0.90} & \multicolumn{1}{c}{0.95} & \multicolumn{1}{c}{0.99} \\ \hline
			\multirow{12}{*}{$- \log \Lambda_1$} & \multirow{6}{*}{50} & \multirow{3}{*}{25} & Theory & 164.841 & 174.749 & 180.089 & 200.082 & 221.767 & 228.006 & 240.459 \\
			&  &  & Emp-True & 164.771 & 174.782 & 180.234 & 199.994 & 221.517 & 227.729 & 240.034 \\
			&  &  & Emp-Estd & 159.805 & 169.696 & 175.078 & 194.658 & 215.903 & 222.128 & 234.256 \\ \cline{4-4}
			&  & \multirow{3}{*}{45} & Theory & 797.822 & 826.438 & 841.832 & 898.889 & 959.700 & 977.789 & 1012.022 \\
			&  &  & Emp-True & 799.655 & 826.727 & 842.174 & 898.844 & 959.791 & 977.214 & 1010.907 \\
			&  &  & Emp-Estd & 794.448 & 821.397 & 836.837 & 893.492 & 954.313 & 971.805 & 1005.664 \\ \cline{4-4}
			& \multirow{6}{*}{100} & \multirow{3}{*}{75} & Theory & 1947.604 & 1985.120 & 2005.652 & 2078.955 & 2154.591 & 2175.725 & 2217.200 \\
			&  &  & Emp-True & 1947.094 & 1985.197 & 2005.283 & 2078.937 & 2153.998 & 2176.062 & 2217.139 \\
			&  &  & Emp-Estd & 1941.953 & 1980.155 & 2000.129 & 2073.893 & 2148.817 & 2170.911 & 2211.889 \\ \cline{4-4}
			&  & \multirow{3}{*}{95} & Theory & 3938.323 & 4007.708 & 4044.784 & 4180.730 & 4322.452 & 4363.505 & 4445.066 \\
			&  &  & Emp-True & 3936.660 & 4006.240 & 4044.432 & 4179.979 & 4321.377 & 4362.547 & 4442.011 \\
			&  &  & Emp-Estd & 3931.817 & 4001.301 & 4039.282 & 4174.853 & 4316.309 & 4357.202 & 4436.587 \\ \cline{4-4}
			\multirow{12}{*}{$- \log \Lambda_3$} & \multirow{6}{*}{50, 50, 50} & \multirow{3}{*}{200} & Theory & 4.292 & 5.587 & 6.382 & 9.928 & 14.551 & 16.074 & 19.262 \\
			&  &  & Emp-True & 4.244 & 5.548 & 6.375 & 9.915 & 14.569 & 16.127 & 19.300 \\
			&  &  & Emp-Estd & 4.211 & 5.540 & 6.359 & 9.901 & 14.549 & 16.098 & 19.299 \\ \cline{4-4}
			&  & \multirow{3}{*}{400} & Theory & 4.213 & 5.540 & 6.362 & 9.903 & 14.586 & 16.104 & 19.205 \\
			&  &  & Emp-True & 4.220 & 5.557 & 6.378 & 9.903 & 14.537 & 16.081 & 19.159 \\
			&  &  & Emp-Estd & 4.227 & 5.564 & 6.380 & 9.904 & 14.551 & 16.050 & 19.225 \\ \cline{4-4}
			& \multirow{6}{*}{50, 75, 100} & \multirow{3}{*}{200} & Theory & 4.290 & 5.636 & 6.452 & 10.011 & 14.768 & 16.297 & 19.431 \\
			&  &  & Emp-True & 4.282 & 5.608 & 6.448 & 10.020 & 14.733 & 16.279 & 19.358 \\
			&  &  & Emp-Estd & 4.258 & 5.606 & 6.454 & 10.007 & 14.726 & 16.259 & 19.363 \\ \cline{4-4}
			&  & \multirow{3}{*}{400} & Theory & 4.372 & 5.720 & 6.547 & 10.172 & 14.942 & 16.550 & 19.817 \\
			&  &  & Emp-True & 4.353 & 5.737 & 6.558 & 10.191 & 14.979 & 16.575 & 19.810 \\
			&  &  & Emp-Estd & 4.355 & 5.734 & 6.576 & 10.208 & 14.960 & 16.559 & 19.859 \\ \hline
		\end{tabular}
		\label{Tab:study1}
	\end{table}

\subsection{Study 2: LRT and Geisser's information statistics in Scenario 2 \label{Subsec:study2}}
	
	We employ Monte Carlo simulations to numerically compare the empirical and theoretical distributions of $- \log \Lambda_2$, $G_2$, $- \log \Lambda_4$, and $G_4$.
	Various sample sizes, covariance matrix dimensions, and numbers of populations are considered.
	The statistics and their related distributions are provided in Theorem~\ref{Thm:mean} and Theorem~\ref{Thm:information}.
	
	For $M = 1$, we set $\bm{\mu}_0 = 2 \bm{\mu}_*$, 
	while keeping the remaining configurations the same as in Study 1.
	As a result, $\bm{\mu} ^ {(1)} \neq \bm{\mu}_0$, implying non-zero noncentrality parameters.
	Following a procedure similar to that in Study 1, 
	we compare the empirical theoretical distributions of $- \log \Lambda_2$ and $G_2$ with their corresponding theoretical (noncentral) distributions.
	For $M > 1$, we set $M = 3$ while maintaining the same configurations as in Study 1.
	This ensures that $\bm{\mu} ^ {(1)} \neq \cdots \neq \bm{\mu} ^ {(M)}$ for $m \in \{1, \ldots, M\}$, 
	leading to non-zero noncentrality parameters.
	We then compare the empirical theoretical distributions of $- \log \Lambda_4$ and $G_4$ with their respective theoretical (noncentral) distributions.
	
	The quantiles of these distributions are summarized in Table~\ref{Tab:study2}.
	The results indicate that $\widehat{\bm{\Gamma}}_2$ and $\widehat{\bm{\Gamma}}_4$ provide reasonable and reliable estimates of $\bm{\Gamma}_2$ and $\bm{\Gamma}_4$ in hypothesis testing, as evidenced by small differences in quantiles.
	Specifically, compared to the other test statistics,
	the differences in $G_4$'s quantiles are slightly larger.
	This is because we use the $F$-approximation.
	As the sample size and dimension increase,  
	these differences tend to decrease.
	Moreover, the proposed theoretical distributions demonstrate high accuracy, as the quantile gaps between them and the empirical distributions remain small. 
	
	\begin{table}[!htb]
		\scriptsize
		\centering
		\caption{
			Quantiles of theoretical and empirical distributions related to $- \log \Lambda_2$, $G_2$, $- \log \Lambda_4$, and $G_4$ under various sample sizes $n$ or $(n ^ {(1)}, n ^ {(2)}, n ^ {(3)})$ and covariance matrix dimensions $p$. 
			Here ``Theory-$\beta$'', ``Theory-$F$'',  ``Emp-True'', and ``Emp-Estd'' denote the theoretical distributions computed using $\beta$ and $F$ variates and the empirical distributions computed using the true and estimated $\bm{\Gamma}_2$ or $\bm{\Gamma}_4$, respectively}
		\begin{tabular}{ccccccccccc}
			\hline
			\multirow{2}{*}{Statistic} & \multirow{2}{*}{Sample Size} & \multirow{2}{*}{Dimension} & \multirow{2}{*}{Distribution} & \multicolumn{7}{c}{Quantile} \\ \cline{5-11} 
			&  &  &  & \multicolumn{1}{c}{0.01} & \multicolumn{1}{c}{0.05} & \multicolumn{1}{c}{0.10} & \multicolumn{1}{c}{0.50} & \multicolumn{1}{c}{0.90} & \multicolumn{1}{c}{0.95} & \multicolumn{1}{c}{0.99} \\ \hline
			\multirow{16}{*}{$- \log \Lambda_2$} & \multirow{8}{*}{50} & \multirow{4}{*}{200} & Theory-$\beta$ & 141.251 & 149.082 & 153.390 & 169.314 & 185.950 & 190.838 & 200.346 \\
			&  &  & Theory-$F$ & 141.232 & 149.074 & 153.427 & 169.126 & 185.929 & 190.851 & 200.176 \\
			&  &  & Emp-True & 141.330 & 149.235 & 153.468 & 169.244 & 185.882 & 190.947 & 200.519 \\
			&  &  & Emp-Estd & 142.169 & 150.761 & 155.281 & 172.237 & 190.531 & 195.979 & 206.606 \\ \cline{4-4}
			&  & \multirow{4}{*}{400} & Theory-$\beta$ & 240.485 & 250.604 & 256.179 & 276.576 & 297.696 & 304.039 & 316.169 \\
			&  &  & Theory-$F$ & 240.645 & 250.602 & 256.218 & 276.427 & 297.699 & 303.933 & 315.920 \\
			&  &  & Emp-True & 240.485 & 250.798 & 256.382 & 276.535 & 297.929 & 304.087 & 315.916 \\
			&  &  & Emp-Estd & 242.320 & 252.803 & 258.549 & 279.740 & 302.159 & 308.726 & 321.250 \\ \cline{4-4}
			& \multirow{8}{*}{100} & \multirow{4}{*}{200} & Theory-$\beta$ & 203.917 & 212.998 & 218.026 & 236.496 & 255.845 & 261.310 & 271.826 \\
			&  &  & Theory-$F$ & 203.964 & 213.028 & 218.057 & 236.434 & 255.883 & 261.534 & 272.480 \\
			&  &  & Emp-True & 203.419 & 212.936 & 218.011 & 236.519 & 255.759 & 261.482 & 272.441 \\
			&  &  & Emp-Estd & 203.745 & 213.989 & 219.497 & 239.774 & 261.184 & 267.463 & 279.897 \\ \cline{4-4}
			&  & \multirow{4}{*}{400} & Theory-$\beta$ & 309.673 & 321.169 & 327.288 & 349.587 & 372.762 & 379.465 & 392.705 \\
			&  &  & Theory-$F$ & 309.749 & 321.128 & 327.301 & 349.843 & 373.174 & 380.149 & 393.137 \\
			&  &  & Emp-True & 309.801 & 320.930 & 327.081 & 349.632 & 373.216 & 379.888 & 393.181 \\
			&  &  & Emp-Estd & 310.765 & 322.692 & 329.334 & 353.099 & 378.136 & 385.286 & 399.467 \\ \cline{4-4}
			\multirow{8}{*}{$G_2$} & \multirow{4}{*}{50} & \multirow{2}{*}{200} & Theory-$F$ & 363.040 & 397.348 & 417.770 & 502.838 & 620.680 & 662.726 & 757.247 \\
			&  &  & Emp-Estd & 364.627 & 398.060 & 418.429 & 503.089 & 622.276 & 665.764 & 758.529 \\ \cline{4-4}
			&  & \multirow{2}{*}{400} & Theory-$F$ & 609.109 & 651.250 & 676.122 & 782.336 & 928.565 & 980.169 & 1093.770 \\
			&  &  & Emp-Estd & 608.117 & 651.133 & 676.163 & 782.550 & 928.270 & 980.103 & 1094.986 \\ \cline{4-4}
			& \multirow{4}{*}{100} & \multirow{2}{*}{200} & Theory-$F$ & 585.508 & 632.703 & 660.821 & 775.082 & 919.698 & 968.126 & 1068.022 \\
			&  &  & Emp-Estd & 583.128 & 631.706 & 660.026 & 774.562 & 919.049 & 968.409 & 1068.667 \\ \cline{4-4}
			&  & \multirow{2}{*}{400} & Theory-$F$ & 886.849 & 948.495 & 983.724 & 1124.459 & 1303.677 & 1363.759 & 1486.688 \\
			&  &  & Emp-Estd & 888.460 & 949.555 & 984.760 & 1125.190 & 1304.122 & 1361.374 & 1482.571 \\ \cline{4-4}
			\multirow{16}{*}{$- \log \Lambda_4$} & \multirow{8}{*}{50, 50, 50} & \multirow{4}{*}{200} & Theory-$\beta$ & 318.833 & 330.654 & 337.112 & 360.896 & 385.889 & 393.131 & 406.487 \\
			&  &  & Theory-$F$ & 318.758 & 330.604 & 337.218 & 361.161 & 386.017 & 393.432 & 407.226 \\
			&  &  & Emp-True & 318.547 & 330.604 & 337.213 & 360.887 & 385.598 & 392.883 & 406.585 \\
			&  &  & Emp-Estd & 318.752 & 331.444 & 338.383 & 363.642 & 390.240 & 398.192 & 413.042 \\ \cline{4-4}
			&  & \multirow{4}{*}{400} & Theory-$\beta$ & 527.626 & 543.009 & 551.193 & 581.187 & 612.214 & 620.945 & 638.257 \\
			&  &  & Theory-$F$ & 527.241 & 543.005 & 551.174 & 581.233 & 612.424 & 621.375 & 637.926 \\
			&  &  & Emp-True & 527.004 & 542.688 & 551.046 & 580.863 & 612.186 & 621.099 & 638.219 \\
			&  &  & Emp-Estd & 528.160 & 544.062 & 552.845 & 583.951 & 616.337 & 625.781 & 644.197 \\ \cline{4-4}
			& \multirow{8}{*}{50, 75, 100} & \multirow{4}{*}{200} & Theory-$\beta$ & 380.547 & 393.829 & 400.905 & 427.004 & 453.997 & 461.785 & 476.855 \\
			&  &  & Theory-$F$ & 380.306 & 393.849 & 401.053 & 427.098 & 453.996 & 461.771 & 476.698 \\
			&  &  & Emp-True & 380.461 & 393.802 & 401.042 & 426.983 & 453.941 & 461.803 & 476.572 \\
			&  &  & Emp-Estd & 379.912 & 394.231 & 401.888 & 429.748 & 458.874 & 467.515 & 483.679 \\ \cline{4-4}
			&  & \multirow{4}{*}{400} & Theory-$\beta$ & 597.175 & 613.469 & 622.394 & 654.403 & 687.196 & 696.672 & 714.654 \\
			&  &  & Theory-$F$ & 597.091 & 613.805 & 622.701 & 654.431 & 687.281 & 696.683 & 714.633 \\
			&  &  & Emp-True & 597.171 & 613.365 & 622.422 & 654.352 & 687.043 & 696.476 & 714.608 \\
			&  &  & Emp-Estd & 597.227 & 614.384 & 623.867 & 657.275 & 691.627 & 701.368 & 720.579 \\ \cline{4-4}
			\multirow{8}{*}{$G_4$} & \multirow{4}{*}{50} & \multirow{2}{*}{200} & Theory-$F$ & 829.206 & 868.154 & 889.722 & 969.491 & 1057.849 & 1083.678 & 1134.088 \\
			&  &  & Emp-Estd & 787.685 & 835.267 & 862.303 & 965.388 & 1088.123 & 1127.383 & 1208.735 \\ \cline{4-4}
			&  & \multirow{2}{*}{400} & Theory-$F$ & 1347.150 & 1395.346 & 1421.539 & 1518.941 & 1624.758 & 1656.679 & 1716.816 \\
			&  &  & Emp-Estd & 1291.751 & 1352.865 & 1385.726 & 1514.615 & 1665.485 & 1713.819 & 1813.891 \\ \cline{4-4}
			& \multirow{4}{*}{100} & \multirow{2}{*}{200} & Theory-$F$ & 1022.731 & 1066.383 & 1090.676 & 1180.361 & 1277.379 & 1306.107 & 1361.526 \\
			&  &  & Emp-Estd & 975.909 & 1029.930 & 1060.191 & 1177.642 & 1312.849 & 1355.319 & 1441.427 \\ \cline{4-4}
			&  & \multirow{2}{*}{400} & Theory-$F$ & 1594.203 & 1646.432 & 1677.057 & 1785.145 & 1899.771 & 1934.086 & 2001.236 \\
			&  &  & Emp-Estd & 1528.828 & 1597.220 & 1635.584 & 1780.382 & 1946.916 & 1999.018 & 2102.660 \\ \hline
		\end{tabular}
		\label{Tab:study2}
	\end{table}

\subsection{Study 3: simultaneous marginal tests in Scenario 3 \label{Subsec:study3}}	
	
	We evaluate the proposed FDP estimation procedure for multiple marginal tests through Monte Carlo approaches.
	
	Because of the single population, we fix $n = 100$ and set $\bm{p} = \left(\lfloor 3\rho \rfloor, \lfloor 4\rho \rfloor, \lfloor 5\rho \rfloor, \lfloor 6\rho \rfloor, \lfloor 7\rho \rfloor \right) ^ \top$, varying $\rho$ over a predefined set of values $\rho \in \{12, 24\}$.
	The true population covariance matrix is defined as $\bm{\Sigma} ^ {(1)}\left[\textbf{A} ^ {(1)}, \textbf{B} ^ {(1)}, \bm{p}\right]$, where $\textbf{A} ^ {(1)} = \textbf{A}_*$ and $\textbf{B} ^ {(1)} = \textbf{B}_*$, consistent with Study 1.
	The population mean is designed as $\bm{\mu} ^ {(1)} = \bm{\mu}_{\star}$, 
	where $\bm{\mu}_{\star}$ is constructed as follows. 
	First, we define the effect size $\kappa_{\bm{\mu}} \in \{0.5, 1.0, 1.5\}$ as the mean over its standard deviation.
	Second, for each value of $\kappa_{\bm{\mu}}$, 
	we set $\mu_{\star, \bar{p}_{k - 1} + 1} = \mu_{\star, \bar{p}_{k - 1} + 2} = \mu_{\star, \bar{p}_{k - 1} + 3} = \kappa_{\bm{\mu}} \times \sqrt{c_{kk}} / \sqrt{n}$ for $k \in \{1, \ldots, K\}$ and $\mu_{\star, j} = 0$ otherwise. 
	That is, only the first three mean components are non-zero in each sub-block of the covariance structure.
	Thus, the total number of non-zero mean components is $3K = 15$, implying there are $15$ false null hypotheses among all $p$ hypotheses.
	
	For Study 3, we conduct $10^3$ Monte Carlo replicates.
	In each replicate, 
	we generate $n = 100$ normal observations from $N_p\left(\bm{\mu} ^ {(1)}, \bm{\Sigma} ^ {(1)}\left[\textbf{A} ^ {(1)}, \textbf{B} ^ {(1)}, \bm{p}\right]\right)$ for each combination of $\rho$ and $\kappa_{\bm{\mu}}$.
	We then apply the proposed FDP estimation procedure as described in the previous section.
	Specifically, 
	we fix $\alpha = 0.05$ and select $\kappa_0$ using \citet{FanHan2017}'s approach.
	The estimates $\widehat{\textbf{A}} ^ {(1)}$, $\widehat{\textbf{B}} ^ {(1)}$, $\widehat{\textbf{C}} ^ {(1)}$, and $\widehat{\textbf{D}} ^ {(1)}$ are computed using~\eqref{Eq:ls}.
	Next, we construct the marginal test vector $\bm{T} \triangleq \sqrt{n} \widehat{\textbf{D}} ^ {(1), - 1} \bar{\bm{X}}$ and compute the unadjusted $p$-values $P_j$ using~\eqref{Eq:pvalue}.
	The estimated correlation matrix $\widehat{\bm{\Psi}} ^ {(1)}\left[\widehat{\textbf{A}}_{\bm{\Psi}} ^ {(1)}, \widehat{\textbf{B}}_{\bm{\Psi}} ^ {(1)}, \bm{p}\right]$ is also determined.
	We consider a candidate set of $200$ threshold values, ranging from $\exp(- 2)$ to $\exp(- 7)$.
	For each threshold $t_s$, we estimate the realized $\operatorname{FDP}\left(t_s\right)$ and compute the adjusted $p$-values $\widehat{P}_j$ for $j \in \{1, \ldots, p\}$ using the build-in functions of the R package \texttt{pfa} \citep{FanHanGu2012}.
	The FDP estimate at $t_s$ is denoted as ``PFA-UB''.
	For comparison, 
	we also evaluate two additional covariance matrix estimators for the marginal test statistic $\sqrt{n}\bar{\bm{X}}$: ``PFA-True'', the true covariance matrix $\bm{\Sigma} ^ {(1)}\left[\textbf{A} ^ {(1)}, \textbf{B} ^ {(1)}, \bm{p}\right]$, and ``PFA-POET'', the principal orthogonal complement thresholding (POET) estimator from the R package \texttt{POET} \citep{FanLiaoMincheva2011}. 
	For each $t_s$, 
	we compute the number of true discoveries, $\widehat{S}(t_s)$, 
	modified from the build-in functions from \citet{VutovDickhaus2022} and \citet{VutovDickhaus2023},
	for each estimator when the corresponding adjusted $p$-values $\widehat{P}_j$ satisfy $\widehat{P}_j \leq t_s$. 
	Across $1000$ replicates, 
	we calculate the median and standard error of
	$\widehat{\operatorname{FDP}}(t_s)$ and the average and standard error of $\widehat{S}(t_s)$ for ``PFA-True'', ``PFA-UB'', and ``PFA-POET''. 
	Given $\alpha = 0.05$, 
	we report $t_{\alpha}$ in Table~\ref{Tab:study3} for each estimator. 
	The complete results for all $200$ threshold values are provided in the \href{Supplementary Material.pdf}{Supplementary Material}.
	
	Based on the results in Table~\ref{Tab:study3}, 
	``PFA-UB'' exhibits greater similarity to ``PFA-True'' than
	``PFA-POET'', as it better captures the underlying covariance structure.
	Specifically, 
	``PFA-POET'' yields a relatively higher estimated FDP than ``PFA-True'', resulting in a smaller $t_{\alpha}$ based on the same candidate threshold values under the same nominal level $\alpha$.	
	As the number of tests $p$ and effect sizes $\kappa_{\bm{\mu}}$ increase, 
	the performances of ``PFA-UB'' and ``PFA-POET'' become increasingly similar.
	Notably, the standard error of $\widehat{\operatorname{FDP}}(t)$ is relatively large compared to its median, suggesting that FDP values are not tightly concentrated around the $\operatorname{FDR}(t)$ due to strong correlation.
	
	\begin{table}[!htb]
		\scriptsize
		\centering
		\caption{
			Results of the estimated FDP and $t_{\alpha}$ for various methods, 
			where ``PFA-POET'', ``PFA-True'', and ``PFA-UB'' refer to methods utilizing the POET covariance matrix estimator, the true covariance matrix, and the proposed correlation matrix estimator, respectively.
			``Dimension'' $p$ represents the dimension of the covariance matrix.
			``Effect size'' $\kappa_{\bm{\mu}}$ is defined as the mean divided by its standard deviation.
			``$t_{\alpha}$'' denotes the largest threshold satisfying $\widehat{\operatorname{FDP}}(t_{\alpha}) \leq \alpha$.
			``S.E.'' refers to the standard error. 
		}
		\begin{tabular}{cccccccccc}
			\hline
			\multirow{2}{*}{Dimension} & \multirow{2}{*}{\begin{tabular}[c]{@{}c@{}}Effect\\ Size\end{tabular}} & \multirow{2}{*}{Method} & \multirow{2}{*}{$t_{\alpha}$} & \multicolumn{2}{c}{$R(t)$} & \multicolumn{2}{c}{$S(t)$} & \multicolumn{2}{c}{$\widehat{\operatorname{FDP}}(t)$} \\ \cline{5-10} 
			&  &  &  & Mean & S.E. & Mean & S.E. & Median & S.E. \\ \hline
			\multirow{9}{*}{300} & \multirow{3}{*}{0.5} & PFA-POET & 0.00163 & 55.132 & 54.622 & 4.320 & 3.207 & 0.049 & 0.348 \\
			&  & PFA-True & 0.05477 & 17.998 & 32.675 & 11.510 & 1.240 & 0.043 & 0.441 \\
			&  & PFA-UB & 0.05209 & 17.141 & 31.771 & 9.539 & 2.784 & 0.035 & 0.451 \\ \cline{3-3}
			& \multirow{3}{*}{1.0} & PFA-POET & 0.00184 & 58.731 & 54.045 & 8.151 & 3.098 & 0.050 & 0.340 \\
			&  & PFA-True & 0.04954 & 17.737 & 30.679 & 14.141 & 0.774 & 0.044 & 0.418 \\
			&  & PFA-UB & 0.04052 & 14.622 & 27.458 & 13.292 & 1.753 & 0.048 & 0.422 \\ \cline{3-3}
			& \multirow{3}{*}{1.5} & PFA-POET & 0.00328 & 69.177 & 55.035 & 11.422 & 2.609 & 0.050 & 0.349 \\
			&  & PFA-True & 0.05209 & 20.799 & 31.303 & 14.912 & 0.294 & 0.044 & 0.407 \\
			&  & PFA-UB & 0.04155 & 16.976 & 27.661 & 14.517 & 1.129 & 0.048 & 0.410 \\ \cline{3-3}
			\multirow{9}{*}{600} & \multirow{3}{*}{0.5} & PFA-POET & 0.00111 & 94.387 & 101.395 & 3.844 & 3.015 & 0.050 & 0.357 \\
			&  & PFA-True & 0.05342 & 29.598 & 58.958 & 11.493 & 1.241 & 0.029 & 0.447 \\
			&  & PFA-UB & 0.05080 & 28.077 & 56.539 & 9.556 & 2.851 & 0.039 & 0.465 \\ \cline{3-3}
			& \multirow{3}{*}{1.0} & PFA-POET & 0.00103 & 94.953 & 99.578 & 7.760 & 3.102 & 0.050 & 0.337 \\
			&  & PFA-True & 0.04831 & 28.081 & 55.590 & 14.160 & 0.792 & 0.045 & 0.421 \\
			&  & PFA-UB & 0.03758 & 22.134 & 48.373 & 13.298 & 1.772 & 0.039 & 0.431 \\ \cline{3-3}
			& \multirow{3}{*}{1.5} & PFA-POET & 0.00184 & 111.170 & 104.344 & 11.064 & 2.639 & 0.049 & 0.348 \\
			&  & PFA-True & 0.05080 & 31.859 & 56.696 & 14.919 & 0.301 & 0.048 & 0.408 \\
			&  & PFA-UB & 0.03485 & 22.601 & 46.063 & 14.533 & 1.136 & 0.048 & 0.409 \\ \hline
		\end{tabular}
		\label{Tab:study3}
	\end{table}

\subsection{Study 4: robustness analysis for the proposed joint tests \label{Subsec:study4}}		
	
	In this section, we conduct a robustness analysis for the proposed joint test statistics.
	Specifically, 
	we access their robustness in terms of empirical statistical powers and sizes under two scenarios: (1) random noise that disrupts the uniform-block structures for $- \log \Lambda_1$, $- \log \Lambda_2$, $G_2$, $- \log \Lambda_3$, $- \log \Lambda_4$, and $G_4$; 
	(2) random missingness that results in incomplete sample covariance matrices for $G_2$ and $G_4$.
	
	We begin with the same simulation setup as in Study 1 for both $M = 1$ and $M = 3$, except for the designs of population means and covariance matrices.
	Specifically, 
	we fix $n = 50$ and $\left(n ^ {(1)}, n ^ {(2)}, n ^ {(3)}\right) = \left(50, 50, 50\right)$ for $M = 1$ and $M = 3$, respectively, 
	while setting $\bm{p} = \left(\lfloor 3\rho \rfloor, \lfloor 4\rho \rfloor, \lfloor 5\rho \rfloor, \lfloor 6\rho \rfloor, \lfloor 7\rho \rfloor \right) ^ \top$ with $\rho = 4$.
	
	To generate distinct mean structures, 
	we follow the population mean designs proposed by \citet{BenjaminiHochberg1995} and \citet{ChenQin2010}.
	For $M = 1$, 
	let $\bm{\mu}_0 = \textbf{0}_{p \times 1}$ denote the fixed hypothesized mean.
	We specify the components of $\bm{\mu} ^ {(1)} \triangleq \left(\mu_1 ^ {(1)}, \ldots, \mu_p ^ {(1)} \right) ^ \top$ as follows. 
	Let $\ell \in \{0.00, 0.10, 0.20, 0.50, 0.80, 0.90, 1.00\}$ represent the proportion of true null hypotheses among all $p$ null hypotheses. 
	At a fixed value of $\ell$, 
	three configurations are considered for the false null hypotheses, i.e., $\{\mu_{j_{l}} ^ {(1)} \neq 0: j_{l} \in \{1, \ldots, p\}, 1 \leq l \leq \lfloor (1 - \ell) p \rfloor \}$ if $\lfloor (1 - \ell) p \rfloor \geq 1$ and $\varnothing$ if otherwise. 
	These configurations include the equal configuration, where $\mu_{j_1} ^ {(1)} = \cdots = \mu_{j_{\lfloor (1 - \ell) p \rfloor}} ^ {(1)} = L$; the linearly increasing configuration, where $\mu_{j_1} ^ {(1)} = L / (\lfloor (1 - \ell) p \rfloor), \mu_{j_2} ^ {(1)} = 2 L / (\lfloor (1 - \ell) p \rfloor), \ldots, \mu_{j_{\lfloor (1 - \ell) p \rfloor }} ^ {(1)} = L$; and the linear decreasing configuration, where $\mu_{j_1} ^ {(1)} = - L / (\lfloor (1 - \ell) p \rfloor), \mu_{j_2} ^ {(1)} = - 2 L / (\lfloor (1 - \ell) p \rfloor), \ldots, \mu_{j_{\lfloor (1 - \ell) p \rfloor }} ^ {(1)} = - L$.
	Here, $L$ is determined by solving $\Vert \bm{\mu} ^ {(1)} - \bm{\mu}_0 \Vert ^ 2 / \operatorname{tr} ^ {1/2}(\bm{\Sigma}_{\epsilon} ^ {(1), 2}) = 0.02$, 
	where $\bm{\Sigma}_{\epsilon} ^ {(1)}$ is the population covariance matrix discussed later.
	For $M = 3$, 
	we set $\bm{\mu} ^ {(2)} = 0.75 \times \bm{\mu} ^ {(1)}$ and $\bm{\mu} ^ {(3)} = 0.50 \times \bm{\mu} ^ {(1)}$.
	Now $L$ is determined by solving $\Vert \bm{\mu} ^ {(1)} - \bm{\mu}_0 \Vert ^ 2 / \operatorname{tr} ^ {1/2}(\bm{\Sigma}_{\epsilon} ^ {(1), 2}) = 0.5$.
	
	To evaluate robustness against random noise, 
	we define $\bm{\Sigma}_{\epsilon} ^ {(m)} \triangleq \bm{\Sigma}_{\epsilon} ^ {(m)} \left[\textbf{A} ^ {(m)}, \textbf{B} ^ {(m)}, \bm{p} \right] + \textbf{E}_{\epsilon}$, 
	where $m = 1$ for $M = 1$ and $m \in \{1, 2, 3\}$ for $M = 3$.
	The matrices $\textbf{A} ^ {(m)} = \textbf{A}_*$ and $\textbf{B} ^ {(m)} = \textbf{B}_*$ are the same as in Study 1, 
	while $\textbf{E}_{\epsilon} \sim W_p(\epsilon \textbf{I}_p, p)$ represents random noise at levels $\epsilon \in \{0, 1, 5\} \times 10^{-2}$.
	When $\epsilon = 0$, $\bm{\Sigma}_{\epsilon} ^ {(m)}$ reduces to the uniform-block matrix in Study 1.
	Using Monte Carlo simulations with $10 ^ 4$ replicates, 
	we compute the empirical statistical power and size for each mean structure and $\epsilon$ value. 
	Specifically, 
	for each replicate,
	we generate a random sample from $N_p\left(\bm{\mu} ^ {(m)}, \bm{\Sigma}_{\epsilon} ^ {(m)}\right)$, 
 	compute the test statistics, 
 	and compare them with the theoretical critical values (i.e., $(1 - \alpha)$th percentiles of theoretical distributions) at a predetermined significance level of $\alpha = 0.05$.
	Rejections are counted when the test statistic exceeds the critical value.
	The empirical statistical powers (when $\ell \neq 1$) and sizes (when $\ell = 1$) of the proposed joint statistics are reported in the \href{Supplementary Material.pdf}{Supplementary Material}. 
	
	To assess robustness against random missingness, 
	we focus on the population covariance matrices $\bm{\Sigma}_{\epsilon} ^ {(m)}$ with $\epsilon = 0$, 
	considering $m = 1$ for $M = 1$ and $m \in \{1, 2, 3\}$ for $M = 3$.
	Monte Carlo simulations with $10 ^ 4$ replicates are conducted to evaluate empirical statistical powers and sizes under different levels of missingness in the sample covariance matrices. 
	Specifically, 
	for each replicate,
	we generate a random sample from $N_p\left(\bm{\mu} ^ {(m)}, \bm{\Sigma}_{\epsilon} ^ {(m)}\right)$ with $\epsilon = 0$,
	then introduce a proportion of $\{0.00, 0.05, 0.10, 0.15\}$ of missing values in $\textbf{S} ^ {(m)}$ for $m \in \{1, \ldots, M\}$.
	Without imputing the missing values, 
	we compute the test statistics $G_2$ and $G_4$, 
	counting rejections when the test statistic exceeds the critical value.
	The results for empirical statistical powers (when $\ell \neq 1$) and sizes (when $\ell = 1$) of the proposed $G_2$ and $G_4$ are summarized in the \href{Supplementary Material.pdf}{Supplementary Material}.
	
	The above simulation results indicate that when the uniform-block covariance structure is slightly or moderately disrupted, the proposed test statistics exhibit only minor losses in empirical statistical powers and slight deviations in sizes from the nominal level. 
	Furthermore, even with moderate levels of random missingness in the sample covariance structure, the proposed test statistics maintain reasonable and reliable empirical performance. 
	These findings suggest that our tests enhance robustness in joint hypothesis testing by effectively leveraging the covariance structure.

\section{Real data analysis \label{Sec:real}}
	
	Our motivating dataset originates from a high-dimensional echo-planar spectroscopic imaging (EPSI) study.
	In previous research, \citet{ChiappelliRowlandWijtenburg2019} investigated the associations between regional neurometabolite levels and cardiovascular risk factors, suggesting that neurometabolite levels could serve as early biomarkers for cognitive decline. 
	They employed linear regression models to estimate the associations between Framingham cardiovascular risk scores \citep{DAgostinoVasanPencina2008} and regional neurometabolite levels, adjusting for covariates such as gender and age due to their potential relationships with neurometabolites. 
	For comprehensive details on the study design, data preprocessing, and statistical analysis, please refer to \citet{ChiappelliRowlandWijtenburg2019}.
	
	\textbf{Objectives.}
	Leveraging the study protocol and data \citep{ChiappelliRowlandWijtenburg2019}, 
	the current analysis aimed to examine the interdependence of regional neurometabolite levels and to validate the influence of gender and age on these levels.
	Our data analysis evaluated model specifications and provided statistical insights for future research.
	
\begin{figure}[!htb]
		\centering
		\includegraphics[width = 0.6 \linewidth]{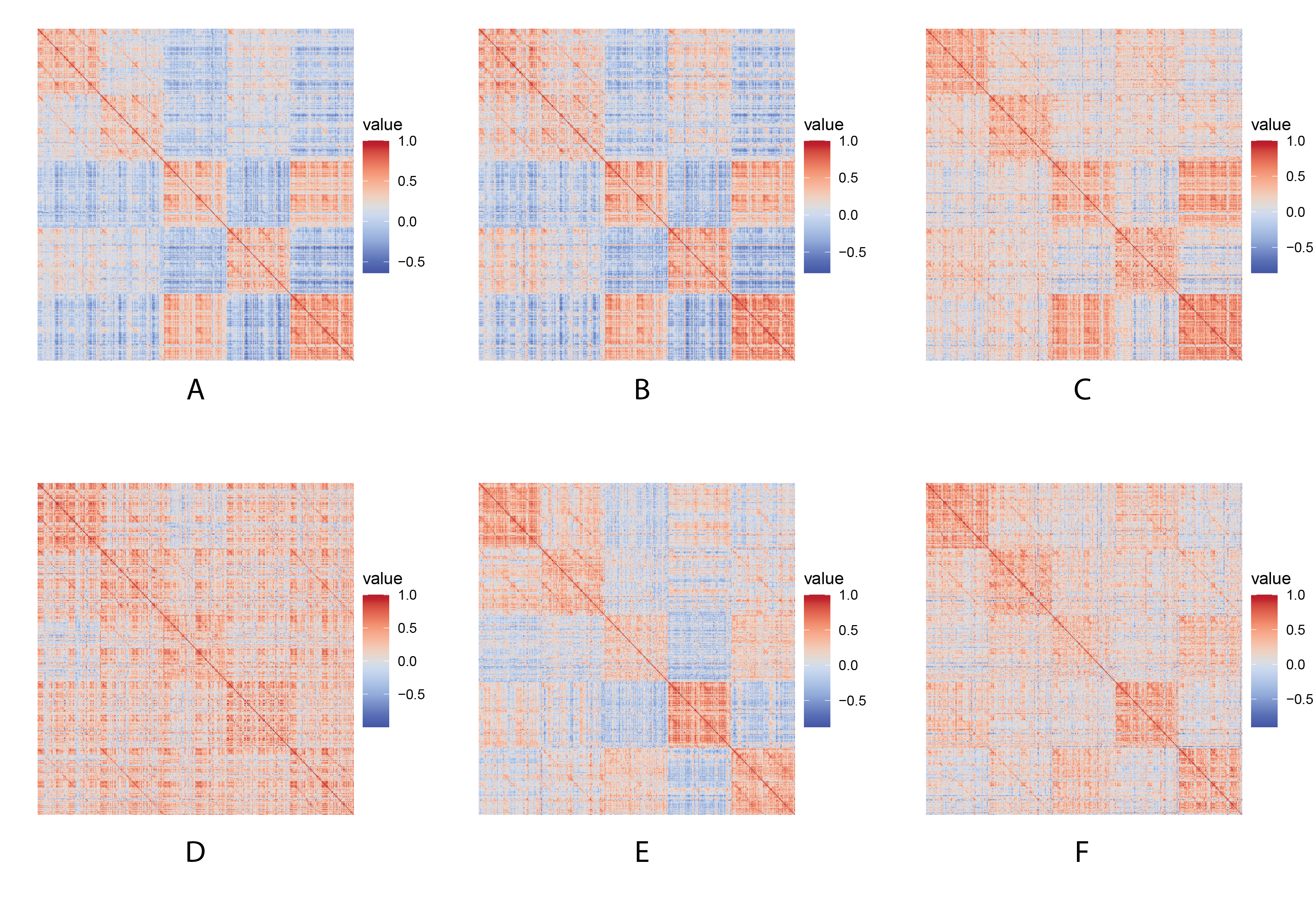} 
		\caption{
			Heatmaps ($445$ by $445$) exhibit correlation matrices for all participants (A), female participants (B), male participants (C), participants aged $12$--$30$ (D), 
			participants aged $31$--$50$ (E),
			participants aged $51$--$77$ (F) using ``pair'' method, where (diagonally from top to bottom) blocks represent choline-containing compounds (Cho), creatine-containing compounds (Cre), glutamate and glutamine (Glx), myo-inositol(Mi), and $N$-acetylaspartate (NAA), respectively.
		}
		\label{Fig:real}
	\end{figure}

	\textbf{Covariance structure.}
	We explored the relationships between measurements from $89$ brain regions (see the regions in the \href{Supplementary Material.pdf}{Supplementary Material}) and $5$ neurometabolites: choline-containing compounds (Cho), creatine-containing compounds (Cre), glutamate and glutamine (Glx), myo-inositol (Mi), and $N$-acetylaspartate (NAA).
	As illustrated in Figure~\ref{Fig:real}, 
	the correlation matrix of the regional neurometabolite measurements naturally exhibited a uniform-block structure, where each block represents the measurements of a single neurometabolite across $89$ brain regions.
	This motivated our use of the proposed approach to determine the dependence relationships between regional neurometabolites and to test whether their covariance structures and mean levels were stratified by gender and age, considering the uniform-block covariance structures.
	
	\textbf{Data analysis for all participants.}
	The EPSI dataset consisted of $n = 78$ participants, 
	with an average age of $42.1$ and a standard deviation of $18.8$.
	Each participant had $p = 445$ measurements, i.e., $\bm{X}_i ^ {(m)} \in \mathbb{R} ^ {p}$ with $m = M = 1$, 
	representing combinations of $K = 5$ neurometabolites across $89$ brain regions, i.e., $p_1 = \cdots = p_5 = 89$. 
	The sample correlation matrices of all participants, as shown in Figure~\ref{Fig:real}, displayed the structure of uniform blocks.
	
	Two primary challenges existed for classical hypothesis testing. 
	The high-dimensionality issue: $n = 78 < p = 445$; 
	and the presence of missing values: calculating the sample covariance matrix $\textbf{S} ^ {(1)}$ revealed that $76.65\%$ of its entries were missing.
	To address the high-dimensionality issue, we considered the proposed LRT statistics and Gessier's information statistics rather than the classical Hotelling's $T ^ 2$ statistics.
	To address the missing data, we employed three approaches.
	The ``full'' method: 
	calculate $\textbf{S} ^ {(1)}$ using all entries, 
	including those with missing values in $\bm{X}_i ^ {(m)}$.
	The ``pair(wise)'' method:
	calculate $\textbf{S} ^ {(1)}$ based on all complete pairs of variables, excluding any observations with missing values.
	The ``impu(ted)'' method:
	impute missing observations in all $\bm{X}_i ^ {(m)}$ using a numerical approach \citep{StekhovenBuhlmann2011} and then calculate $\textbf{S} ^ {(1)}$.
	Thus, the ``pair'' and ``impu'' methods resulted in $\textbf{S} ^ {(1)}$ matrices without missing values. 
	
	Given an empirical center $1.645$, i.e., the average of components in $\bar{\bm{X}}$, 
	we tested the hypothesis that the population mean vector has a homogeneous structure: $\bm{\mu} ^ {(1)} = 1.645 \times \textbf{1}_{p \times 1}$,  
	using the proposed $\Lambda_2$, $G_2$, and $\textbf{S} ^ {(1)}$ matrices obtained from ``full'', ``pair'', and ``impu'' methods, respectively. 	
	The estimates and testing results were detailed in the \href{Supplementary Material.pdf}{Supplementary Material}.
	Under the ``full'' framework, 
	only $G_2$ was available due to the extensive missing data,
	and it suggested statistically rejecting the above hypothesis due to the extremely small $p$-value (i.e., $< 10 ^ {-4}$).
 	Under the ``pair'' and ``impu'' frameworks, 
 	both $\Lambda_2$ and $G_2$ were available, 
	and they also indicated rejection of the homogeneous mean hypothesis due to the extremely small $p$-values (i.e., $< 10 ^ {-4}$).
	
	Additionally, we tested whether the marginal mean components were simultaneously equal to the empirical center: $\mu_j ^ {(1)} = 1.645$ for $j \in \{1, \ldots, 445\}$, 
	using the proposed FDP estimation procedure.
	The adjusted $p$-values for the marginal tests, 
	the estimated FDPs, 
	and the corresponding $t_{\alpha}$ values were reported in the \href{Supplementary Material.pdf}{Supplementary Material},
	for the ``full'', ``pair'', and ``impu'' methods, 
	with the respective estimated FDPs of $0.0492$, $0.0489$, and $0.0499$.
	Given $\alpha = 0.05$, 
	the regional neurometabolites whose marginal means were statistically different from the empirical center of $1.645$ were also reported for ``full'', ``pair'', and ``impu'' methods.
	
	\textbf{Data analysis for the gender-specific participants.}
	We also investigated the influence of gender.
	With $M = 2$, 
	the EPSI dataset comprised an equal number of male and female participants, e.g., $n_1 = n_2 = 39$.
	The sample correlation matrices for male and female participants,
	depicted in Figure~\ref{Fig:real}, 
	exhibited the structures of uniform blocks, 
	sharing a common $\bm{p} = (p_1, \ldots, p_5) ^ \top$.
	Missing values in $\bm{X}_i ^ {(m)}$ were handled separately for each gender group using the ``full'', ``pair'', and ``impu'' methods, respectively. 
	
	We tested the equality of population covariance structures using the proposed $\Lambda_3$.
	The estimated population covariance matrices were available in the \href{Supplementary Material.pdf}{Supplementary Material}.
	Testing results of $\Lambda_3$ suggested a statistical difference between gender-specific covariance structures at a significance level of $0.05$ for both ``pair'' and ``impu'' methods. 
	Additionally, 
	we tested the equality of population mean structures using the proposed $\Lambda_4$ and $G_4$.
	The testing results, 
	provided in the \href{Supplementary Material.pdf}{Supplementary Material}, 
	indicated significant heterogeneity in mean structures at the $0.05$ level. 
	
	\textbf{Data analysis for the age-specific participants.}
	With $M = 3$, we considered three age-specific groups \citep{LecoeurDomengeFayol2022}: $n_1 = 24$ participants aged $12$--$30$, $n_2 = 27$ participants aged $31$--$50$, 
	and $n_3 = 27$ participants aged $51$--$77$.
	Other age group definitions were possible. 
	The sample correlation matrices for these age groups,
	as depicted in Figure~\ref{Fig:real}, exhibited the structures of uniform blocks, 
	sharing a common $\bm{p} = (p_1, \ldots, p_5) ^ \top$.
	We conducted hypothesis tests for the equality of population covariance structures and mean structures across age-specified groups using the proposed $\Lambda_3$, $\Lambda_4$, $G_4$ with the ``full'', ``pair'', and ``impu'' methods.
	
	Following these imputation methods, we calculated the estimates of population matrices.
	The statistics and $p$-values were in the \href{Supplementary Material.pdf}{Supplementary Material}.
	The testing results demonstrated statistically significant heterogeneity in both population covariance and mean structures at the $0.05$ level. 

	\textbf{Results.}
	In contrast to conventional Hotelling-type statistics, 
	we have demonstrated the advantages of the proposed LRT and Geisser's information statistics in handling high-dimensionality and missing data.
	The proposed estimators produced reliable estimates of population parameters compared to imputation methods, but with reduced computational time. 
	Using all participants, the off-diagonal elements of estimated $\textbf{B}$ (presented in the \href{Supplementary Material.pdf}{Supplementary Material}) represented the estimated covariances between regional neurometabolite levels.
	We observed relatively strong positive dependence between Cre and Mi (glia-related metabolites), and between Glx and NAA (neuron-related metabolites), probably due to their metabolic interconnections \citep{ClarkDoepkeFilosa2006, LindBoraxbekkPetersen2020, BissonnetteFrancisMacNeil2022}.
	Using the gender- or age-specific participants, 
	the results consistently indicated that both covariance and mean structures were not identical, as evidenced by extremely small $p$-values at a significance level of $0.05$.
	These differences in gender- or age-specific participants highlighted the necessity of stratification in subsequent analyses. 
	These findings suggested that future research could achieve more precise statistical inferences by incorporating the interconnected relationship between neurometabolites and adjusting for gender and age.
	
\section{Discussion \label{Sec:discussion}}
	
	In this paper, we focus on high-dimensional hypothesis testing under uniform-block structures, motivated by their three features. 
	First, the uniform-block pattern has been popularly discovered in plenty of high-dimensional data across various fields, including biomedicine, engineering, and finance. 
	Second, it offers clear interpretability: variables grouped within the same community (or block) tend to exhibit stochastic equivalence or comparable patterns, while those from different communities maintain coherent connections at the community level.
	Third, this structure enables efficient computation in high-dimensional (with missing data) settings by leveraging its block pattern.
	Therefore, compared to conventional structures such as diagonal or block-diagonal models, the proposed uniform-block structure provides greater flexibility, making it particularly well-suited for large-scale or high-dimensional multivariate data analysis.
	Additionally, by incorporating information from the non-null off-diagonal blocks, this structure has the potential to reveal valuable insights into underlying scientific mechanisms. 
	
	To establish the algebraic properties of uniform-block matrices, we introduced an innovative block Hadamard product representation.
	This representation allows the decomposition of a large-scale uniform-block matrix into two lower-dimensional matrices and an integer-valued vector, thereby simplifying matrix computations.
	These properties enhance the applicability of uniform-block structures to various statistical problems, particularly in hypothesis testing. 
	
	We addressed hypothesis testing problems related to covariance structures, mean structures, and marginal mean components, respectively, under uniform-block structures. 
	In contrast to traditional methods, the proposed test statistics are applicable in high-dimensional scenarios, with their null distributions derived in closed form.  
	Furthermore, these test statistics demonstrated robustness against minor or moderate structural perturbations and missing data.
	
	In conclusion, hypothesis testing under uniform-block structures, along with its associated algebraic properties, has broad applicability across various disciplines, encompassing fields such as linear algebra, statistics, economics, and numerous others. 
	The online \href{Supplementary Material.pdf}{Supplementary Material} contains the proofs, additional simulation results, and the R code for numerical analysis, which is also available at https://github.com/yiorfun/UB.

\section*{Acknowledgments}

	The authors wish to express their gratitude to Dr. Joshua Chiappelli for providing the dataset and thank the anonymous referees, an Associate Editor, and the Editor for their constructive comments that improved the quality of this paper. 
	This work was partially supported by the National Institute on Drug Abuse of the National Institutes of Health under Award Number 1DP1DA048968 and by the National Heart, Lung, and Blood Institute of the National Institutes of Health under Grant Number 1R01HL175410.


\bibliographystyle{myjmva}

\bibliography{paper_ref_UB}

\end{document}


\maketitle
	
	\begin{abstract}
		The present supplementary material contains two sections. 
		Section~\ref{Sec:numerical} has the additional numerical results for both ``Simulation studies" and ``Real data analysis" sections. 
		Section~\ref{Sec:proof} provides the technical proofs.
	\end{abstract}
	
	\appendix
	
\section{ADDITIONAL NUMERICAL STUDIES RESULTS \label{Sec:numerical}}
	
	
	
	\subsection{Simulation Study 3}
	
		For testing results related to the proposed FDP estimation procedure, please refer to the other supplementary materials.
	
	\subsection{Simulation Study 4}

		To evaluate robustness against random noise,
		the empirical statistical powers (for $\ell \neq 1$) and sizes (for $\ell = 1$) of the proposed joint statistics are reported in Table~\ref{Tab:Study4_1} for $M = 1$ and Table~\ref{Tab:Study4_2} for $M = 3$.
		The results indicate that the proposed test statistics exhibit only minor reductions in empirical statistical powers and slight deviations in sizes from the nominal level when the uniform-block covariance structure is slightly or moderately disrupted.
		
		To assess robustness against random missingness,
		the empirical statistical powers (for $\ell \neq 1$) and sizes (for $\ell = 1$) of the proposed statistics $G_2$ and $G_4$ are summarized in Table~\ref{Tab:Study4_3} for $M = 1$ and $M = 3$.
		The results demonstrate that the proposed statistics $G_2$ and $G_4$ maintain reasonable and reliable empirical performance, even with moderate levels of random missingness in the sample covariance structure.
		
		\begin{table}[!htb]
			\scriptsize
			\centering
			\caption{
				Rejection proportions among $10 ^ 4$ Monte Carlo replicates for $M = 1$.
				``Pattern" refers to the equal, linearly increasing, and linearly decreasing configurations for the false null hypotheses. 
				``Proportion" denotes the proportion of true null hypotheses among all $p$ null hypotheses. 
				``$\epsilon$" represents the random noise level in $\textbf{E}_{\epsilon} \sim W_p(\epsilon \textbf{I}_p, p)$.
			}
			\begin{tabular}{ccccccccccc}
				\hline
				\multirow{3}{*}{Pattern} & \multirow{3}{*}{Proportion} & \multicolumn{9}{c}{$M = 1$} \\ \cline{3-11} 
				&  & \multicolumn{3}{c}{$\epsilon = 0\%$} & \multicolumn{3}{c}{$\epsilon = 1\%$} & \multicolumn{3}{c}{$\epsilon = 5\%$} \\ \cline{3-11} 
				&  & $\Lambda_2$-$\beta$ & $\Lambda_2$-$F$ & $G_2$ & $\Lambda_2$-$\beta$ & $\Lambda_2$-$F$ & $G_2$ & $\Lambda_2$-$\beta$ & $\Lambda_2$-$F$ & $G_2$ \\ \hline
				\multirow{7}{*}{equal} & 0.00 & 0.126 & 0.124 & 0.125 & 0.141 & 0.140 & 0.136 & 0.139 & 0.138 & 0.134 \\
				& 0.10 & 1.000 & 1.000 & 1.000 & 0.271 & 0.268 & 0.261 & 0.160 & 0.158 & 0.153 \\
				& 0.20 & 1.000 & 1.000 & 1.000 & 0.272 & 0.268 & 0.262 & 0.166 & 0.163 & 0.158 \\
				& 0.50 & 0.997 & 0.997 & 0.997 & 0.196 & 0.194 & 0.189 & 0.148 & 0.145 & 0.141 \\
				& 0.80 & 1.000 & 1.000 & 1.000 & 0.833 & 0.831 & 0.860 & 0.301 & 0.298 & 0.301 \\
				& 0.90 & 1.000 & 1.000 & 1.000 & 0.790 & 0.787 & 0.794 & 0.246 & 0.243 & 0.236 \\
				& 1.00 & 0.052 & 0.051 & 0.046 & 0.100 & 0.098 & 0.092 & 0.110 & 0.108 & 0.101 \\ \cline{1-1}
				\multirow{7}{*}{increasing} & 0.00 & 0.864 & 0.861 & 0.895 & 0.276 & 0.272 & 0.295 & 0.169 & 0.167 & 0.164 \\
				& 0.10 & 1.000 & 1.000 & 1.000 & 0.673 & 0.670 & 0.681 & 0.231 & 0.227 & 0.223 \\
				& 0.20 & 1.000 & 1.000 & 1.000 & 0.710 & 0.706 & 0.710 & 0.244 & 0.242 & 0.237 \\
				& 0.50 & 1.000 & 1.000 & 1.000 & 0.529 & 0.525 & 0.530 & 0.205 & 0.202 & 0.198 \\
				& 0.80 & 1.000 & 1.000 & 1.000 & 0.992 & 0.992 & 0.994 & 0.425 & 0.422 & 0.425 \\
				& 0.90 & 1.000 & 1.000 & 1.000 & 0.961 & 0.961 & 0.964 & 0.309 & 0.306 & 0.300 \\
				& 1.00 & 0.052 & 0.051 & 0.046 & 0.100 & 0.098 & 0.092 & 0.110 & 0.108 & 0.101 \\ \cline{1-1}
				\multirow{7}{*}{decreasing} & 0.00 & 0.861 & 0.858 & 0.895 & 0.274 & 0.272 & 0.292 & 0.164 & 0.163 & 0.159 \\
				& 0.10 & 1.000 & 1.000 & 1.000 & 0.675 & 0.671 & 0.679 & 0.232 & 0.229 & 0.224 \\
				& 0.20 & 1.000 & 1.000 & 1.000 & 0.703 & 0.701 & 0.703 & 0.241 & 0.238 & 0.233 \\
				& 0.50 & 1.000 & 1.000 & 1.000 & 0.530 & 0.527 & 0.532 & 0.206 & 0.204 & 0.198 \\
				& 0.80 & 1.000 & 1.000 & 1.000 & 0.992 & 0.992 & 0.994 & 0.426 & 0.421 & 0.423 \\
				& 0.90 & 1.000 & 1.000 & 1.000 & 0.964 & 0.963 & 0.967 & 0.307 & 0.304 & 0.296 \\
				& 1.00 & 0.052 & 0.051 & 0.046 & 0.100 & 0.098 & 0.092 & 0.110 & 0.108 & 0.101 \\ \hline
			\end{tabular}
			\label{Tab:Study4_1}
		\end{table}

		\begin{table}[!htb]
			\scriptsize
			\centering
			\caption{
				Rejection proportions among $10 ^ 4$ Monte Carlo replicates for $M = 3$.
				``Pattern" refers to the equal, linearly increasing, and linearly decreasing configurations for the false null hypotheses. 
				``Proportion" denotes the proportion of true null hypotheses among all $p$ null hypotheses. 
				``$\epsilon$" represents the random noise level in $\textbf{E}_{\epsilon} \sim W_p(\epsilon \textbf{I}_p, p)$.
			}
			\begin{tabular}{ccccccccccc}
				\hline
				\multirow{3}{*}{Pattern} & \multirow{3}{*}{Proportion} & \multicolumn{9}{c}{$M = 3$} \\ \cline{3-11} 
				&  & \multicolumn{3}{c}{$\epsilon = 0\%$} & \multicolumn{3}{c}{$\epsilon = 1\%$} & \multicolumn{3}{c}{$\epsilon = 5\%$} \\ \cline{3-11} 
				&  & $\Lambda_4$-$\beta$ & $\Lambda_4$-$F$ & $G_4$ & $\Lambda_4$-$\beta$ & $\Lambda_4$-$F$ & $G_4$ & $\Lambda_4$-$\beta$ & $\Lambda_4$-$F$ & $G_4$ \\ \hline
				\multirow{7}{*}{equal} & 0.00 & 0.162 & 0.178 & 0.176 & 0.159 & 0.173 & 0.166 & 0.150 & 0.163 & 0.156 \\
				& 0.10 & 1.000 & 1.000 & 1.000 & 0.383 & 0.404 & 0.395 & 0.191 & 0.204 & 0.198 \\
				& 0.20 & 1.000 & 1.000 & 1.000 & 0.375 & 0.398 & 0.388 & 0.193 & 0.209 & 0.202 \\
				& 0.50 & 1.000 & 1.000 & 1.000 & 0.252 & 0.268 & 0.262 & 0.166 & 0.178 & 0.173 \\
				& 0.80 & 1.000 & 1.000 & 1.000 & 0.984 & 0.987 & 0.991 & 0.418 & 0.439 & 0.439 \\
				& 0.90 & 1.000 & 1.000 & 1.000 & 0.975 & 0.978 & 0.979 & 0.332 & 0.353 & 0.346 \\
				& 1.00 & 0.049 & 0.057 & 0.053 & 0.095 & 0.103 & 0.099 & 0.106 & 0.115 & 0.110 \\ \cline{1-1}
				\multirow{7}{*}{increasing} & 0.00 & 0.992 & 0.993 & 0.996 & 0.381 & 0.400 & 0.417 & 0.193 & 0.204 & 0.202 \\
				& 0.10 & 1.000 & 1.000 & 1.000 & 0.909 & 0.918 & 0.920 & 0.304 & 0.325 & 0.318 \\
				& 0.20 & 1.000 & 1.000 & 1.000 & 0.934 & 0.940 & 0.940 & 0.322 & 0.340 & 0.334 \\
				& 0.50 & 1.000 & 1.000 & 1.000 & 0.777 & 0.793 & 0.795 & 0.260 & 0.277 & 0.272 \\
				& 0.80 & 1.000 & 1.000 & 1.000 & 1.000 & 1.000 & 1.000 & 0.629 & 0.648 & 0.649 \\
				& 0.90 & 1.000 & 1.000 & 1.000 & 1.000 & 1.000 & 1.000 & 0.446 & 0.467 & 0.460 \\
				& 1.00 & 0.049 & 0.057 & 0.053 & 0.095 & 0.103 & 0.099 & 0.106 & 0.115 & 0.110 \\ \cline{1-1}
				\multirow{7}{*}{decreasing} & 0.00 & 0.992 & 0.993 & 0.994 & 0.387 & 0.406 & 0.424 & 0.192 & 0.206 & 0.202 \\
				& 0.10 & 1.000 & 1.000 & 1.000 & 0.911 & 0.919 & 0.920 & 0.305 & 0.320 & 0.315 \\
				& 0.20 & 1.000 & 1.000 & 1.000 & 0.931 & 0.938 & 0.938 & 0.321 & 0.339 & 0.332 \\
				& 0.50 & 1.000 & 1.000 & 1.000 & 0.772 & 0.787 & 0.789 & 0.259 & 0.274 & 0.268 \\
				& 0.80 & 1.000 & 1.000 & 1.000 & 1.000 & 1.000 & 1.000 & 0.630 & 0.647 & 0.648 \\
				& 0.90 & 1.000 & 1.000 & 1.000 & 1.000 & 1.000 & 1.000 & 0.441 & 0.466 & 0.457 \\
				& 1.00 & 0.049 & 0.057 & 0.053 & 0.095 & 0.103 & 0.099 & 0.106 & 0.115 & 0.110 \\ \hline
			\end{tabular}
			\label{Tab:Study4_2}
		\end{table}

		\begin{table}[!htb]
			\scriptsize
			\centering
			\caption{
				Rejection proportions among $10 ^ 4$ Monte Carlo replicates for $M = 1$ and $M = 3$.
				``Pattern" refers to the equal, linearly increasing, and linearly decreasing configurations for the false null hypotheses.
				``Proportion" denotes the proportion of true null hypotheses among all $p$ null hypotheses. 
				``Miss" represents the proportion of missing values in $\textbf{S} ^ {(m)}$ for $m \in \{1, \ldots, M\}$.
			}
			\begin{tabular}{cccccccccc}
				\hline
				\multirow{3}{*}{Pattern} & \multirow{3}{*}{Proportion} & \multicolumn{4}{c}{$M = 1$} & \multicolumn{4}{c}{$M = 3$} \\ \cline{3-10} 
				&  & \multicolumn{4}{c}{$G_2$} & \multicolumn{4}{c}{$G_4$} \\ \cline{3-10} 
				&  & Miss $0\%$ & Miss $5\%$ & Miss $10\%$ & Miss $15\%$ & Miss $0\%$ & Miss $5\%$ & Miss $10\%$ & Miss $15\%$ \\ \hline
				\multirow{7}{*}{equal} & 0.00 & 0.124 & 0.192 & 0.266 & 0.320 & 0.174 & 0.212 & 0.250 & 0.286 \\
				& 0.10 & 1.000 & 1.000 & 1.000 & 1.000 & 1.000 & 1.000 & 1.000 & 1.000 \\
				& 0.20 & 1.000 & 1.000 & 1.000 & 0.999 & 1.000 & 1.000 & 1.000 & 1.000 \\
				& 0.50 & 0.996 & 0.995 & 0.991 & 0.988 & 1.000 & 1.000 & 1.000 & 1.000 \\
				& 0.80 & 1.000 & 1.000 & 1.000 & 1.000 & 1.000 & 1.000 & 1.000 & 1.000 \\
				& 0.90 & 1.000 & 1.000 & 1.000 & 1.000 & 1.000 & 1.000 & 1.000 & 1.000 \\
				& 1.00 & 0.050 & 0.117 & 0.186 & 0.240 & 0.054 & 0.089 & 0.125 & 0.161 \\ \cline{1-1}
				\multirow{7}{*}{increasing} & 0.00 & 0.897 & 0.900 & 0.904 & 0.900 & 0.995 & 0.994 & 0.993 & 0.992 \\
				& 0.10 & 1.000 & 1.000 & 1.000 & 1.000 & 1.000 & 1.000 & 1.000 & 1.000 \\
				& 0.20 & 1.000 & 1.000 & 1.000 & 1.000 & 1.000 & 1.000 & 1.000 & 1.000 \\
				& 0.50 & 1.000 & 1.000 & 1.000 & 1.000 & 1.000 & 1.000 & 1.000 & 1.000 \\
				& 0.80 & 1.000 & 1.000 & 1.000 & 1.000 & 1.000 & 1.000 & 1.000 & 1.000 \\
				& 0.90 & 1.000 & 1.000 & 1.000 & 1.000 & 1.000 & 1.000 & 1.000 & 1.000 \\
				& 1.00 & 0.050 & 0.117 & 0.186 & 0.240 & 0.054 & 0.089 & 0.125 & 0.161 \\ \cline{1-1}
				\multirow{7}{*}{decreasing} & 0.00 & 0.894 & 0.897 & 0.899 & 0.899 & 0.996 & 0.994 & 0.994 & 0.994 \\
				& 0.10 & 1.000 & 1.000 & 1.000 & 1.000 & 1.000 & 1.000 & 1.000 & 1.000 \\
				& 0.20 & 1.000 & 1.000 & 1.000 & 1.000 & 1.000 & 1.000 & 1.000 & 1.000 \\
				& 0.50 & 1.000 & 1.000 & 1.000 & 1.000 & 1.000 & 1.000 & 1.000 & 1.000 \\
				& 0.80 & 1.000 & 1.000 & 1.000 & 1.000 & 1.000 & 1.000 & 1.000 & 1.000 \\
				& 0.90 & 1.000 & 1.000 & 1.000 & 1.000 & 1.000 & 1.000 & 1.000 & 1.000 \\
				& 1.00 & 0.050 & 0.117 & 0.186 & 0.240 & 0.054 & 0.089 & 0.125 & 0.161 \\ \hline
			\end{tabular}
			\label{Tab:Study4_3}
		\end{table}

	\subsection{Real Data Analysis}
	
	Given $K = 5$, $p = 445$, and $\bm{p} = \left(p_1, \ldots, p_5\right) ^ \top = \left(89, \ldots, 89\right) ^ \top$, 
	suppose the estimated covariance matrices for all participants were denoted as $\widehat{\bm{\Sigma}} ^ {\text{(full)}} \left[\widehat{\textbf{A}} ^ {\text{(full)}}, \widehat{\textbf{B}} ^ {\text{(full)}}, \bm{p} \right]$, 
	$\widehat{\bm{\Sigma}} ^ {\text{(pair)}} \left[\widehat{\textbf{A}} ^ {\text{(pair)}}, \widehat{\textbf{B}} ^ {\text{(pair)}}, \bm{p} \right]$, 
	and 
	$\widehat{\bm{\Sigma}} ^ {\text{(impu)}} \left[\widehat{\textbf{A}} ^ {\text{(impu)}}, \widehat{\textbf{B}} ^ {\text{(impu)}}, \bm{p} \right]$, 
	using the methods of ``full", ``pair", and ``impu", 
	respectively.
	The estimated results ($\times 10^{-4}$) were
	\begin{equation*}
		\scriptsize
		\begin{split}
			\widehat{\textbf{A}} ^ {\text{(full)}} = 
			\begin{pmatrix*}[r]
				0.0015 & & & & \\
				& 0.0181 & & & \\
				& & 0.0334 & & \\
				& & & 0.0318 & \\
				& & & & 0.0277
			\end{pmatrix*}, \quad 
			\widehat{\textbf{B}} ^ {\text{(full)}} = 
			\begin{pmatrix*}[r]
			0.0015 & 0.0027 & -0.0005 &  0.0040 &  0.0000 \\
		  		   & 0.0145 &  0.0023 &  0.0110 &  0.0020 \\
		  		   &        &  0.0363 & -0.0007 &  0.0356 \\
				   & 		& 		  &  0.0317 & -0.0021 \\
				   & 	    & 		  & 	    &  0.0446
			\end{pmatrix*}, \\
			\widehat{\textbf{A}} ^ {\text{(pair)}} = 
			\begin{pmatrix*}[r]
				0.0026 & & & & \\
				& 0.0324 & & & \\
				& & 0.0573 & & \\
				& & & 0.0581 & \\
				& & & & 0.0395
			\end{pmatrix*}, \quad 
			\widehat{\textbf{B}} ^ {\text{(pair)}} = 
			\begin{pmatrix*}[r]
		 0.0014 & 0.0021 & -0.0003 &  0.0029 &  0.0003 \\
				& 0.0119 &  0.0017 &  0.0062 &  0.0021 \\
				&        &  0.0266 & -0.0046 &  0.0288 \\
				& 		 & 		   &  0.0251 & -0.0068 \\
				& 	     & 		   & 	     &  0.0417
			\end{pmatrix*}, \\
			\widehat{\textbf{A}} ^ {\text{(impu)}} = 
			\begin{pmatrix*}[r]
				0.0024 & & & & \\
				& 0.0289 & & & \\
				& & 0.0517 & & \\
				& & & 0.0511 & \\
				& & & & 0.0362
			\end{pmatrix*}, \quad 
			\widehat{\textbf{B}} ^ {\text{(impu)}} = 
			\begin{pmatrix*}[r]
		 0.0013 & 0.0021 & -0.0001 &  0.0029 &  0.0004 \\
				& 0.0118 &  0.0020 &  0.0065 &  0.0024 \\
				&        &  0.0263 & -0.0037 &  0.0286 \\
				& 		 & 		   &  0.0247 & -0.0056 \\
				& 	     & 		   & 	     &  0.0407
			\end{pmatrix*}.
		\end{split}
	\end{equation*}
	We observed relatively strong positive dependence between Cre and Mi (i.e., $0.0110$, $0.0062$, and $0.0065$ for three methods), and between Glx and NAA (i.e., $0.0356$, $0.0288$, and $0.0286$ for three methods), probably due to their metabolic interconnections.
	
	Tables~\ref{Tab:real_1},~\ref{Tab:real_2}, and~\ref{Tab:real_3} present the results for joint hypothesis tests related to all, gender-specific, and age-specific participants, respectively.
	For the remaining testing results, please refer to the other supplementary materials.
	
	\begin{table}[!htb]
		\scriptsize
		\centering
		\caption{
			Results for testing $\bm{\mu} ^ {(1)} = 1.645 \times \textbf{1}_{p \times 1}$ in all participants using ``full", ``pair", and ``impu" methods, 
			where ``cut-off point" denotes the critical value, the $(1 - \alpha)$th percentile of the theoretical null distribution,  and ``rejection" denotes the statistical decision regarding the null hypothesis. 
		}
		\begin{tabular}{ccccccccccccc}
			\hline
			& \multicolumn{4}{c}{full} & \multicolumn{4}{c}{pair} & \multicolumn{4}{c}{impu} \\ \cline{2-13} 
			& $\Lambda_1$ & \begin{tabular}[c]{@{}c@{}}$\Lambda_2$-$\beta$\end{tabular} & \begin{tabular}[c]{@{}c@{}}$\Lambda_2$-$F$\end{tabular} & $G_2$ & $\Lambda_1$ & \begin{tabular}[c]{@{}c@{}}$\Lambda_2$-$\beta$\end{tabular} & \begin{tabular}[c]{@{}c@{}}$\Lambda_2$-$F$\end{tabular} & $G_2$ & $\Lambda_1$ & \begin{tabular}[c]{@{}c@{}}$\Lambda_2$-$\beta$\end{tabular} & \begin{tabular}[c]{@{}c@{}}$\Lambda_2$-$F$\end{tabular} & $G_2$ \\ \hline
			statistic & NA & NA & NA & 310365.1 & NA & 21855.55 & 21855.55 & 239953.9 & NA & 23127.87 & 23127.87 & 259072.6 \\
			cut-off point & NA & 249.326 & 249.172 & 496.484 & NA & 249.326 & 249.172 & 496.484 & NA & 249.326 & 249.172 & 496.484 \\
			rejection & NA & NA & NA & Yes & NA & Yes & Yes & Yes & NA & Yes & Yes & Yes \\
			$p$-value & NA & NA & NA & 0 & NA & 0 & 0 & 0 & NA & 0 & 0 & 0 \\ \hline
		\end{tabular}
		\label{Tab:real_1}
	\end{table}

	\begin{table}[!htb]
		\scriptsize
		\centering
		\caption{
			Results for testing $\bm{\Sigma} ^ {\text{(male)}} = \bm{\Sigma} ^ {\text{(female)}}$ (obtained from $\Lambda_3$) and $\bm{\mu} ^ {\text{(male)}} = \bm{\mu} ^ {\text{(female)}}$ (obtained from $\Lambda_4$ and $G_4$) in gender-specific participants using ``full", ``pair", and ``impu" methods, 
			where ``cut-off point" denotes the critical value, the $(1 - \alpha)$th percentile of the theoretical null distribution,  and ``rejection" denotes the statistical decision regarding the null hypothesis. 
		}
		\begin{tabular}{ccccccccccccc}
			\hline
			& \multicolumn{4}{c}{full} & \multicolumn{4}{c}{pair} & \multicolumn{4}{c}{impu} \\ \cline{2-13} 
			& $\Lambda_3$ & \begin{tabular}[c]{@{}c@{}}$\Lambda_4$-$\beta$\end{tabular} & \begin{tabular}[c]{@{}c@{}}$\Lambda_4$-$F$\end{tabular} & $G_4$ & $\Lambda_3$ & \begin{tabular}[c]{@{}c@{}}$\Lambda_4$-$\beta$\end{tabular} & \begin{tabular}[c]{@{}c@{}}$\Lambda_4$-$F$\end{tabular} & $G_4$ & $\Lambda_3$ & \begin{tabular}[c]{@{}c@{}}$\Lambda_4$-$\beta$\end{tabular} & \begin{tabular}[c]{@{}c@{}}$\Lambda_4$-$F$\end{tabular} & $G_4$ \\ \hline
			statistic & NA & NA & NA & 1711.901 & 20.899 & 510.566 & 510.566 & 1011.329 & 27.915 & 567.522 & 567.522 & 1126.098 \\
			cut-off point & 9.456 & 252.446 & 252.613 & 496.080 & 9.456 & 252.446 & 252.613 & 496.080 & 9.456 & 252.446 & 252.613 & 496.080 \\
			rejection & NA & NA & NA & Yes & Yes & Yes & Yes & Yes & Yes & Yes & Yes & Yes \\
			$p$-value & NA & NA & NA & 0 & 0 & 0 & 0 & 0 & 0 & 0 & 0 & 0 \\ \hline
		\end{tabular}
		\label{Tab:real_2}
	\end{table}

	\begin{table}[!htb]
		\scriptsize
		\centering
		\caption{
			Results for testing $\bm{\Sigma} ^ {(12-30)} = \bm{\Sigma} ^ {(31-50)} = \bm{\Sigma} ^ {(51-77)}$ (obtained from $\Lambda_3$) and $\bm{\mu} ^ {(12-30)} = \bm{\mu} ^ {(31-50)} = \bm{\mu} ^ {(51-77)}$ (obtained from $\Lambda_4$ and $G_4$) in age-specific participants using ``full", ``pair", and ``impu" methods, 
			where ``cut-off point" denotes the critical value, the $(1 - \alpha)$th percentile of the theoretical null distribution,  and ``rejection" denotes the statistical decision regarding the null hypothesis. 
		}
		\begin{tabular}{ccccccccccccc}
			\hline
			& \multicolumn{4}{c}{full} & \multicolumn{4}{c}{pair} & \multicolumn{4}{c}{impu} \\ \cline{2-13} 
			& $\Lambda_3$ & \begin{tabular}[c]{@{}c@{}}$\Lambda_4$-$\beta$\end{tabular} & \begin{tabular}[c]{@{}c@{}}$\Lambda_4$-$F$\end{tabular} & $G_4$ & $\Lambda_3$ & \begin{tabular}[c]{@{}c@{}}$\Lambda_4$-$\beta$\end{tabular} & \begin{tabular}[c]{@{}c@{}}$\Lambda_4$-$F$\end{tabular} & $G_4$ & $\Lambda_3$ & \begin{tabular}[c]{@{}c@{}}$\Lambda_4$-$\beta$\end{tabular} & \begin{tabular}[c]{@{}c@{}}$\Lambda_4$-$F$\end{tabular} & $G_4$ \\ \hline
			statistic & NA & NA & NA & 5036.739 & 166.801 & 1496.888 & 1496.888 & 3053.145 & 171.366 & 1667.290 & 1667.290 & 3417.295 \\
			cut-off point & 16.528 & 492.714 & 493.155 & 962.752 & 16.528 & 492.714 & 493.155 & 962.752 & 16.528 & 492.714 & 493.155 & 962.752 \\
			rejection & NA & NA & NA & Yes & Yes & Yes & Yes & Yes & Yes & Yes & Yes & Yes \\
			$p$-value & NA & NA & NA & 0 & 0 & 0 & 0 & 0 & 0 & 0 & 0 & 0 \\ \hline
		\end{tabular}
		\label{Tab:real_3}
	\end{table}

\section{TECHNICAL PROOFS \label{Sec:proof}}

	\subsection{Proofs in Section 2 \label{Subsec:sec2}}

	\begin{proof}[Proof of Lemma 1 (block Hadamard product representation)]
		This proof is straightforward.
		To show the uniqueness, consider two equal uniform-block matrices $\textbf{N}_1 \left[\textbf{A}_1, \textbf{B}_1, \bm{p} \right] \triangleq \left(\textbf{N}_{1, kk'}\right)$ and $\textbf{N}_2\left[\textbf{A}_2, \textbf{B}_2, \bm{p} \right] \triangleq \left(\textbf{N}_{2, kk'} \right)$ with a common predetermined partition-size vector $\bm{p} \triangleq \left(p_1, \ldots, p_K\right) ^ \top$ satisfying that $p_k > 1$ for every $k$ and $p = p_1 + \cdots + p_K$, where $\textbf{A}_i \triangleq \text{diag}\left(a_{i, 11}, \ldots, a_{i, KK}\right)$ is a diagonal matrix and $\textbf{B}_i \triangleq \left(b_{i, kk'}\right)$ is a symmetric matrix with $b_{i,k'k} = b_{i, kk'}$ for $i = 1, 2$.
		By the equality, $\textbf{N}_{1, kk'} = \textbf{N}_{2, kk'}$ for every $k$ and $k'$.
		If $k \neq k'$, then $b_{1, kk'} \textbf{1}_{p_k \times p_{k'}} = b_{2, kk'} \textbf{1}_{p_k \times p_{k'}}$ and therefore $b_{1, kk'} = b_{2, kk'}$.
		If $k = k'$, then $a_{1, kk} \textbf{I}_{p_k} + b_{1, kk} \textbf{J}_{p_k} = a_{2, kk} \textbf{I}_{p_k} + b_{2, kk} \textbf{J}_{p_k}$, equivalently, $\left\{\left(a_{1,kk} - a_{2,kk}\right) + \left(b_{1,kk} - b_{2,kk}\right) \right\} \textbf{I}_{p_k} + \left(b_{1,kk} - b_{2,kk} \right) \textbf{J}_{p_k} = \textbf{0}_{p_k \times p_k}$.
		Due to $p_k > 1$, off-diagonally, $b_{1,kk} = b_{2,kk}$; diagonally, $a_{1,kk} = a_{2,kk}$.
		Eventually, $\textbf{A}_1 = \textbf{A}_2$ and $\textbf{B}_1 = \textbf{B}_2$.
	\end{proof}
	
	\begin{proof}[Proof of Lemma 2 (properties of uniform-block matriecs)]
		
		(1) holds straightforwardly.
		(2) and (3) hold because of the following equalities
		\begin{equation*}
			\begin{split}
				\left( \textbf{A}_1 \circ \textbf{I} \left[\bm{p} \right] \right)  \left( \textbf{A}_2 \circ \textbf{I} \left[\bm{p} \right] \right)
				& = \left( \textbf{A}_1 \textbf{A}_2 \right) \circ \textbf{I} \left[\bm{p} \right],
				\quad
				\left(\textbf{B}_1 \circ \textbf{J} \left[\bm{p} \right]\right)  \left(\textbf{B}_2 \circ \textbf{J} \left[\bm{p} \right]\right) 
				= \left(\textbf{B}_1  \textbf{P} \textbf{B}_2 \right) \circ \textbf{J} \left[\bm{p} \right], \\
				\left(\textbf{B}_1 \circ \textbf{J} \left[\bm{p} \right]\right)  \left( \textbf{A}_2 \circ \textbf{I} \left[\bm{p}\right] \right)
				& = \left( \textbf{B}_1  \textbf{A}_2 \right) \circ \textbf{J} \left[\bm{p} \right], 
				\quad
				\left( \textbf{A}_1 \circ \textbf{I}\left[\bm{p}\right] \right)  \left(\textbf{B}_2 \circ \textbf{J}\left[\bm{p} \right]\right)
				= \left( \textbf{A}_1  \textbf{B}_2 \right) \circ \textbf{J} \left[\bm{p}\right].
			\end{split}
		\end{equation*}
		They are easy to verify since $\textbf{A}_i$ and $\textbf{P}$ are diagonal for $i = 1, 2$ and $\textbf{B}_i$ is symmetric for $i = 1, 2$. 
		
		(4) and (5) are proved using the induction approach with respect to $K$.  
		Specifically, when $K = 1$, the equality $\left(a \textbf{I}_p + b \textbf{J}_p \right) ^ {- 1} = a ^ {- 1} \textbf{I}_p - b / \{a(a + bp)\} \textbf{J}_p$ establishes (4) and (5). 
		Now, we assume that both (4) and (5) hold for the case of $K$ and check whether they hold for the case of $K + 1$. 
		Let $\bm{\eta} \triangleq \left(\eta_1, \ldots, \eta_K \right) ^ \top \in \mathbb{R} ^ {K}$, $a, b \in \mathbb{R}$, $q \in \mathbb{N}$ and $q > 1$.
		Denote $\textbf{A} ^ * \triangleq \operatorname{Bdiag}\left(\textbf{A}, a\right) \in \mathbb{R} ^{(K+1) \times (K+1)}$, $\textbf{P} ^ * \triangleq \operatorname{Bdiag}\left(\textbf{P}, q\right) \in \mathbb{R} ^ {(K+1) \times (K+1)}$, and $\textbf{B} ^ * \triangleq \left(\textbf{B}, \bm{\eta}; \bm{\eta} ^ \top, b \right) \in \mathbb{R} ^ {(K+1) \times (K + 1)}$.
		We would like to obtain the eigenvalues of the $(p + q)$ by $(p + q)$ uniform-block matrix $\textbf{N} ^ * \left[\textbf{A} ^ *, \textbf{B} ^ *, \bm{p} ^ * \right]$ with $(K + 1)$ diagonal blocks, where $\bm{p} ^ * \triangleq \left(\bm{p} ^ \top, q \right) ^ \top$.
		By definition, the eigenvalues of $\textbf{N} ^ * \left[\textbf{A} ^ *, \textbf{B} ^ *, \bm{p} ^ * \right]$ are the solutions to the characteristic equation $\det\left(\textbf{N} ^ * \left[\textbf{A} ^ *, \textbf{B} ^ *, \bm{p} ^ * \right] - \lambda \textbf{I}_{p + q} \right) = 0$. 
		Equivalently, 
		\begin{equation*}
			\begin{split}
				0 
				= \det \left(\textbf{N} ^ * \left[\textbf{A} ^ *, \textbf{B} ^ *, \bm{p} ^ * \right] - \lambda \textbf{I}_{p + q} \right)
				= \det \begin{pmatrix}
					\left(\textbf{A} - \lambda \textbf{I}_K \right) \circ \textbf{I} \left[\bm{p} \right] + \textbf{B} \circ \textbf{J} \left[\bm{p} \right] & \left(\eta_k \textbf{1}_{p_k \times q} \right) \\
					\left(\eta_k \textbf{1}_{q \times p_k} \right) & \left(a - \lambda \right) \textbf{I}_q + b \textbf{J}_q
				\end{pmatrix},
			\end{split}
		\end{equation*}
		where $\left(\eta_k \textbf{1}_{p_k \times q} \right) \in \mathbb{R} ^ {p \times q}$ and $\left(b_k \textbf{1}_{q \times p_k} \right) \in \mathbb{R} ^ {q \times p}$.
		
		Without loss of generality, assume that $\left(a - \lambda\right) \textbf{I}_q + b \textbf{J}_q$ is invertible, or we can consider $\epsilon > 0$ satisfying that $\left\{\left(a - \lambda\right) \textbf{I}_q + b \textbf{J}_q\right\} + \epsilon \textbf{I}_q$ is positive definite and let $\epsilon \to 0 ^ +$.
		Then, the characteristic equation can be written as 
		\begin{align}
			0 
			\label{Eq:det1}
			& = \det \left\{\left(a - \lambda \right) \textbf{I}_q + b \textbf{J}_q \right\} \\
			\label{Eq:det2}
			& \times \det \left[
			\left\{ \left(\textbf{A} - \lambda \textbf{I}_K \right) \circ \textbf{I}\left[\bm{p} \right] + \textbf{B} \circ \textbf{J} \left[\bm{p} \right] \right\}
			- \left(\eta_k \textbf{1}_{p_k \times q} \right) \left\{\left(a - \lambda\right) \textbf{I}_q + b \textbf{J}_q \right\} ^ {-1}
			\left(\eta_k \textbf{1}_{q \times p_k} \right)
			\right].
		\end{align}
		Thus, the eigenvalues of $\textbf{N} ^ * \left[\textbf{A} ^ *, \textbf{B} ^ *, \bm{p} ^ * \right]$ consist of some eigenvalues of $\left(a \textbf{I}_q + b \textbf{J}_q \right)$ and some roots of the rational equation~\eqref{Eq:det2} $= 0$ (note the rational equation~\eqref{Eq:det2} $= 0$ is not a polynomial equation, so, it is not a characteristic equation).
		
		First, it is easy to observe that the eigenvalues of $\left(a \textbf{I}_q + b \textbf{J}_q \right)$ are exactly $a$ with multiplicity $(q - 1)$ and $(a + b q)$.
		We will discard $(a + b q)$ because it is not the eigenvalue of $\textbf{N} ^ * \left[\textbf{A} ^ *, \textbf{B} ^ *, \bm{p} ^ * \right]$.
		Second, using the fact $\left\{\left(a - \lambda\right) \textbf{I}_q + b \textbf{J}_q \right\} ^ {- 1} = \left(a - \lambda \right) ^ {-1} \textbf{I}_q - b \left(a - \lambda \right) ^ {-1} \left (a - \lambda + b q \right) ^ {-1} \textbf{J}_q$, 
		we simplify~\eqref{Eq:det2} to~\eqref{Eq:det3} as below, 
		\begin{equation}
			\label{Eq:det3}
			\begin{split}
				\det \left\{
				\left(\textbf{A} - \lambda \textbf{I}_K \right) \circ \textbf{I} \left[\bm{p}\right] + \left(\textbf{B} - \frac{q}{a - \lambda + b q} \bm{\eta} \bm{\eta} ^ \top \right) \circ \textbf{J}\left[\bm{p} \right]
				\right\},
			\end{split}
		\end{equation}
		which is the determinant of a uniform-block matrix with $K$ diagonal blocks. 
		Thus, use the induction assumption and $\textbf{A} \triangleq \operatorname{diag} \left(a_{11}, \ldots, a_{KK} \right)$, 
		this determinant equals to the product of $\prod_{k = 1}^{K} \left(a_{kk} - \lambda \right) ^ {p_k - 1}$ and ~\eqref{Eq:det4}, given by 
		\begin{equation}
			\label{Eq:det4}
			\begin{split}
				\det \left \{\left(\textbf{A} - \lambda \textbf{I}_K \right) + \left(\textbf{B} - \frac{q}{a - \lambda + b q} \bm{\eta} \bm{\eta} ^ \top \right) \textbf{P} \right \}
				= \det \left\{
				\left(\bm{\Delta} - \lambda \textbf{I}_K \right) - \frac{q}{a + b q - \lambda} \bm{\eta} \bm{\eta} ^ \top \textbf{P}
				\right\}.
			\end{split}
		\end{equation} 
		Thus, the rational equation~\eqref{Eq:det2} $= 0$ yields the solutions consisting of $a_{kk}$ with multiplicity $(p_k - 1)$ for $k \in \{1, \ldots, K\}$ and the roots of the rational equation~\eqref{Eq:det4} $= 0$, which are the eigenvalues of $\bm{\Delta} ^ * \triangleq \textbf{A} ^ * + \textbf{B} ^ * \textbf{P} ^ *$.
		It is because, 
		\begin{equation*}
			\begin{split}
				\bm{\Delta} ^ * = \begin{pmatrix}
					\textbf{A} & \textbf{0}_{K \times 1} \\ \textbf{0}_{1 \times K} & a
				\end{pmatrix} + 
				\begin{pmatrix}
					\textbf{B} & \bm{\eta} \\ \bm{\eta} ^ \top & b
				\end{pmatrix}  
				\begin{pmatrix}
					\textbf{P} & \textbf{0}_{K \times 1} \\ \textbf{0}_{1 \times K} & q
				\end{pmatrix} =
				\begin{pmatrix}
					\bm{\Delta} & q \bm{\eta} \\ \bm{\eta} ^ \top \textbf{P} & a + b q
				\end{pmatrix}.
			\end{split}
		\end{equation*}
		Assuming that $\left(a + b q - \lambda \right)$ is invertible, we obtain 
		\begin{equation*}
			\begin{split}
				0 = \det \left(\bm{\Delta} ^ * - \lambda \textbf{I}_{K + 1} \right) 
				= \det \left(a + b q - \lambda \right)
				\det \left\{\left(\bm{\Delta} - \lambda \textbf{I}_K \right) - \frac{q}{a + bq - \lambda} \bm{\eta} \bm{\eta} ^ \top \textbf{P}  \right\}.
			\end{split}
		\end{equation*} 
		Since $\left(a + b q \right)$ cannot be the eigenvalue of $\bm{\Delta} ^ *$, 
		all eigenvalues of $\bm{\Delta} ^ *$ are the roots of the rational equation~\eqref{Eq:det4} $= 0$.
		Therefore, the eigenvalues of $\textbf{N} ^ * \left[\textbf{A} ^ *, \textbf{B} ^ *, \bm{p} ^ * \right]$ consist of $a$ with multiplicity $(q - 1)$, 
		$a_{kk}$ with multiplicity $(p_k - 1)$ for $k \in \{1, \ldots, K\}$, and all eigenvalues of $\bm{\Delta} ^ * = \textbf{A} ^ * + \textbf{B} ^ * \textbf{P} ^ *$.
		
		(6) Let $\textbf{A} ^ * = \textbf{A} ^ {-1}$ and $\textbf{B} ^ * = - \bm{\Delta} ^ {- 1} \textbf{B} \textbf{A} ^ {-1}$, where both inverses exist by the assumption. 
		Thus, $\textbf{A} ^ *$ is a diagonal matrix and $\textbf{B} ^ *$ is a symmetric matrix.  
		It is because both $\textbf{A}$ and $\textbf{P}$ are diagonal matrices, so they commute, and the following equalities are equivalent
		\begin{equation*}
			\begin{split}
				\textbf{B} \textbf{P} \textbf{A} ^ {-1} \textbf{B} = \textbf{B} \textbf{A} ^ {-1} \textbf{P} \textbf{B} 
				& \Leftrightarrow
				\left(\textbf{A} + \textbf{B} \textbf{P} \right) \textbf{A} ^ {-1} \textbf{B} 
				= \textbf{B} \textbf{A} ^ {-1} \left( \textbf{A} + \textbf{P} \textbf{B} \right) \\
				& \Leftrightarrow
				- \textbf{A} ^ {-1} \textbf{B} \left( \textbf{A} + \textbf{P} \textbf{B} \right) ^ {- 1}
				= - \left(\textbf{A} + \textbf{B} \textbf{P} \right) ^ {- 1} \textbf{B} \textbf{A} ^ {-1} \\
				& \Leftrightarrow
				\textbf{B} ^ {*, \top} = \textbf{B} ^ *.
			\end{split}
		\end{equation*}
		Then, we follow analogous lines of argument for the square formula, 
		\begin{equation*}
			\begin{split}
				\textbf{A} \textbf{A} ^ * = \textbf{I}_{K}, \quad
				\textbf{A} \textbf{B} ^ * + \textbf{B} \textbf{A} ^ * + \textbf{B} \textbf{P} \textbf{B} ^ * = \textbf{0}_{K \times K},
			\end{split}
		\end{equation*}
		thus, $\left(\textbf{A} \circ \textbf{I} \left[\bm{p} \right] + \textbf{B} \circ \textbf{J} \left[\bm{p} \right]\right) \left(\textbf{A} ^ * \circ \textbf{I} \left[\bm{p} \right] + \textbf{B} ^ * \circ \textbf{J} \left[\bm{p} \right] \right) = \textbf{I}_{K} \circ \textbf{I} \left[\bm{p} \right] + \textbf{0}_{K \times K} \circ \textbf{J} \left[\bm{p} \right] = \textbf{I}_{p}$.  
		
		(7) Suppose $\lambda$ and $\bm{\xi} \triangleq \left(\xi_1, \ldots, \xi_K\right) ^ \top$ is the eigenvalue and eigenvector of $\bm{\Delta}$,
		i.e.,
		$\bm{\Delta} \bm{\xi} = \lambda \bm{\xi}$ and 
		\begin{equation*}
			\begin{pmatrix}
				a_{11} + b_{11} p_1 & b_{12} p_2 & \dots & b_{1K} p_K \\
				b_{21} p_2 & a_{22} + b_{22} p_2 & \dots & b_{2K} p_K \\
				\vdots & \vdots & \ddots & \vdots \\
				b_{K1} p_1 & b_{K2} p_2 & \dots & a_{KK} + b_{KK} p_K
			\end{pmatrix}
			\begin{pmatrix}
				\xi_1 \\ \xi_2 \\ \vdots \\ \xi_K
			\end{pmatrix}
			= \lambda 
			\begin{pmatrix}
				\xi_1 \\ \xi_2 \\ \vdots \\ \xi_K
			\end{pmatrix}.
		\end{equation*}
		Then, we claim that $\lambda$ and $\left(\xi_1 \textbf{1}_{1 \times p_1}, \ldots, \xi_K \textbf{1}_{1 \times p_K}\right)$ are a eigenvalue and the corresponding eigenvector of $\textbf{N}\left[\textbf{A}, \textbf{B}, \bm{p}\right]$, 
		because the following equation holds:
		\begin{equation*}
			\begin{split}
				& \textbf{N}\left[\textbf{A}, \textbf{B}, \bm{p}\right]
				\begin{pmatrix}
					\xi_1 \textbf{1}_{p_1 \times 1} \\
					\xi_2 \textbf{1}_{p_2 \times 1} \\
					\vdots \\
					\xi_K \textbf{1}_{p_K \times 1}
				\end{pmatrix}
				= 
				\begin{pmatrix}
					\sum_{k = 1}^{K} \textbf{N}_{1, k} \xi_k \textbf{1}_{p_k \times 1} \\
					\sum_{k = 1}^{K} \textbf{N}_{2, k} \xi_k \textbf{1}_{p_k \times 1} \\
					\vdots \\
					\sum_{k = 1}^{K} \textbf{N}_{K, k} \xi_k \textbf{1}_{p_k \times 1}
				\end{pmatrix} \\
				& = \begin{pmatrix}
					\left(a_{11} \textbf{I}_{p_1} + b_{11} \textbf{1}_{p_1 \times 1} \textbf{1}_{1 \times p_1}\right) \xi_1 \textbf{1}_{p_1 \times 1} + 
					b_{12} \textbf{1}_{p_1 \times 1} \textbf{1}_{1 \times p_2} \xi_2 \textbf{1}_{p_2 \times 1} + \cdots + 
					b_{1K} \textbf{1}_{p_1 \times 1} \textbf{1}_{1 \times p_K} \xi_K \textbf{1}_{p_K \times 1} \\ \vdots \\
					b_{K1} \textbf{1}_{p_K \times 1} \textbf{1}_{1 \times p_1} \xi_1 \textbf{1}_{p_1 \times 1} + 
					b_{K2} \textbf{1}_{p_K \times 1} \textbf{1}_{1 \times p_2} \xi_2 \textbf{1}_{p_2 \times 1} + \cdots + 
					\left(a_{KK} \textbf{I}_{p_K} + b_{KK} \textbf{1}_{p_K \times 1} \textbf{1}_{1 \times p_K}\right) \xi_K \textbf{1}_{p_K \times 1}
				\end{pmatrix} \\
				& = 
				\begin{pmatrix}
					a_{11} \xi_1 \textbf{1}_{p_1 \times 1} + 
					b_{11} \xi_1 p_1 \textbf{1}_{p_1 \times 1} + 
					b_{12} \xi_2 p_2 \textbf{1}_{p_1 \times 1} + 
					\dots + 
					b_{1K} \xi_K p_K \textbf{1}_{p_1 \times 1} \\ \vdots \\
					b_{K1} \xi_1 p_1 \textbf{1}_{p_K \times 1} + 
					b_{K2} \xi_2 p_2 \textbf{1}_{p_K \times 1} + 
					\dots + 
					a_{KK} \xi_K \textbf{1}_{p_K \times 1} +
					b_{KK} \xi_K p_K \textbf{1}_{p_K \times 1}
				\end{pmatrix} \\
				& = 
				\begin{pmatrix}
					\left(a_{11} \xi_{1} + b_{11} \xi_1 p_1 + \cdots + b_{1K} \xi_K p_K \right) \textbf{1}_{p_1 \times 1} \\ \vdots \\
					\left(a_{KK} \xi_{K} + b_{K1} \xi_1 p_1 + \cdots + b_{KK} \xi_K p_K \right) \textbf{1}_{p_K \times 1}
				\end{pmatrix} 
				= 
				\begin{pmatrix}
					\lambda \xi_{1} \textbf{1}_{p_1 \times 1} \\ \vdots \\
					\lambda \xi_{K} \textbf{1}_{p_K \times 1}
				\end{pmatrix} 
				= \lambda 
				\begin{pmatrix}
					\xi_1 \textbf{1}_{p_1 \times 1} \\
					\vdots \\
					\xi_K \textbf{1}_{p_K \times 1}
				\end{pmatrix}.
			\end{split}
		\end{equation*}
		Consider the normalizing $\sum_{k = 1}^{K} \xi_{k} ^ 2 p_k = 1$ and $a_{k k} \xi_{k} + \sum_{k' = 1}^{K} b_{k k'} p_{k'} \xi_{k'} = \lambda \xi_{k}$, we have  
		\begin{equation*}
			\begin{split}
				a_{k k} \xi_{k} ^ 2 p_k + \left(\sum_{k' = 1}^{K} b_{k k'} p_{k'} \xi_{k'}\right) \xi_{k} p_k = \lambda \xi_{k} ^ 2 p_k, \quad
				\sum_{k = 1}^{K}
				a_{k k} p_k \xi_{k} ^ 2  
				+ \sum_{k = 1}^{K} \sum_{k' = 1}^{K} b_{k k'} p_k p_{k'} \xi_{k} \xi_{k'} 
				= \lambda.
			\end{split}
		\end{equation*}
		In other words, for the $m$-th eigenvalue $\lambda_{\bar{p}_m}$ and eigenvector $\bm{\xi}_m = \left(\xi_{m, 1}, \ldots, \xi_{m, K}\right) ^ \top$, 
		$\sum_{k' = 1}^{K} a_{k' k'} p_{k'} \xi_{m, k'} ^ 2
		+ \sum_{k_1 = 1}^{K} \sum_{k_2 = 1}^{K}
		b_{k_1 k_2} p_{k_1} p_{k_2} \xi_{m, k_1} \xi_{m, k_2} = \lambda_{\bar{p}_m}$ holds, $m \in \{1, \ldots, K\}$.
		Using the relationship, 
		\begin{equation*}
			\bm{\Gamma} \textbf{N}\left[\textbf{A}, \textbf{B}, \bm{p}\right] \bm{\Gamma} ^ \top
			= \begin{pmatrix}
				\sum_{k_1 = 1}^{K} \sum_{k_2 = 1}^{K} \bm{\Gamma}_{m, k_1} \textbf{N}_{k_1, k_2} \bm{\Gamma}_{k_2, \ell} ^ \top
			\end{pmatrix}_{m, \ell = 1, \ldots, K}.
		\end{equation*}
		
		When $m = \ell$, 
		
		(1) $k_1 = m$:
		\begin{equation*}
			\begin{split}
				\bm{\Gamma}_{m, k_1} 
				\textbf{N}_{k_1, k_2} 
				\bm{\Gamma}_{k_2, \ell} ^ \top
				& = 
				\begin{cases}
					\begin{pmatrix}
						\tilde{\textbf{H}}_{m} \\ \xi_{m, m} \textbf{1}_{1 \times p_m}
					\end{pmatrix}
					\left(a_{m m} \textbf{I}_{p_m} + b_{m m}\textbf{J}_{m}\right)
					\begin{pmatrix}
						\tilde{\textbf{H}}_m ^ \top & \xi_{m, m} \textbf{1}_{p_m \times 1}
					\end{pmatrix},
					& k_2 = k_1 \\
					\begin{pmatrix}
						\tilde{\textbf{H}}_{m} \\ \xi_{m, m} \textbf{1}_{1 \times p_m}
					\end{pmatrix}
					b_{m, k_2} \textbf{1}_{p_m \times p_{k_2}}
					\begin{pmatrix}
						\tilde{\textbf{H}}_m ^ \top & \xi_{m, m} \textbf{1}_{p_m \times 1} & k_2 \neq k_1
					\end{pmatrix},
				\end{cases} \\
				& = 
				\begin{cases}
					\begin{pmatrix}
						a_{m m} \textbf{I}_{p_m - 1} & \textbf{0}_{(p_m - 1) \times 1} \\
						\textbf{0}_{1 \times (p_m - 1)} & a_{m m} p_m \xi_{m, m} ^ 2 + b_{m m} p_m ^ 2 \xi_{m, m} ^ 2
					\end{pmatrix}, & k_2 = k_1 \\
					\begin{pmatrix}
						\textbf{0}_{(p_m - 1) \times (p_m - 1)} & \textbf{0}_{(p_m - 1) \times 1} \\
						\textbf{0}_{1 \times (p_m - 1)} & b_{m, k_2} p_m ^ 2 \xi_{m, m} ^ 2 & k_2 \neq k_1
					\end{pmatrix}, & k_2 \neq k_1
				\end{cases}
			\end{split}
		\end{equation*}
		
		(2) $k_1 \neq m$:
		\begin{equation*}
			\begin{split}
				\bm{\Gamma}_{m, k_1} 
				\textbf{N}_{k_1, k_2} 
				\bm{\Gamma}_{k_2, \ell} ^ \top
				& = 
				\begin{cases}
					\begin{pmatrix}
						\textbf{0}_{(p_m - 1) \times p_{k_1}} \\ 
						\xi_{m, k_1} \textbf{1}_{1 \times p_{k_1}}
					\end{pmatrix}
					\left(a_{k_1 k_1} \textbf{I}_{p_{k_1}} + b_{k_1 k_1 }\textbf{J}_{k_1}\right)
					\begin{pmatrix}
						\textbf{0}_{p_{k_2} \times (p_m - 1)} & \xi_{m, k_2} \textbf{1}_{p_{k_2} \times 1}
					\end{pmatrix}, & k_2 = k_1 \\
					\begin{pmatrix}
						\textbf{0}_{(p_m - 1) \times p_{k_1}} \\ 
						\xi_{m, k_1} \textbf{1}_{1 \times p_{k_1}}
					\end{pmatrix}
					b_{k_1 k_2}\textbf{1}_{p_{k_1} \times p_{k_2}}
					\begin{pmatrix}
						\textbf{0}_{p_{k_2} \times (p_m - 1)} & \xi_{m, k_2} \textbf{1}_{p_{k_2} \times 1}
					\end{pmatrix}, & k_2 \neq k_1			
				\end{cases} \\
				& = 
				\begin{cases}
					\begin{pmatrix}
						\textbf{0}_{(p_m - 1) \times (p_m - 1)} & \textbf{0}_{(p_m - 1) \times 1} \\
						\textbf{0}_{1 \times (p_m - 1)} & 
						a_{k_1 k_1} \xi_{m, k_1} \xi_{m, k_2} p_{k_1} + 
						b_{k_1 k_2} \xi_{m, k_1} \xi_{m, k_2} p_{k_1} p_{k_2}
					\end{pmatrix}, & k_2 = k_1 \\
					\begin{pmatrix}
						\textbf{0}_{(p_m - 1) \times (p_m - 1)} & \textbf{0}_{(p_m - 1) \times 1} \\
						\textbf{0}_{1 \times (p_m - 1)} & 
						b_{k_1 k_2} \xi_{m, k_1} \xi_{m, k_2} p_{k_1} p_{k_2}
					\end{pmatrix}, & k_2 \neq k_1
				\end{cases}
			\end{split}
		\end{equation*}
		Thus, 
		$m = \ell = k$, the diagonal of $\bm{\Gamma} \textbf{N}\left[\textbf{A}, \textbf{B}, \bm{p}\right] \bm{\Gamma} ^ \top$ is 
		\begin{equation*}
			\begin{split}
				\sum_{k_1 = 1}^{K} \sum_{k_2 = 1}^{K} \bm{\Gamma}_{k, k_1} \textbf{N}_{k_1, k_2} \bm{\Gamma}_{k_2, k} ^ \top 
				& = 
				\begin{pmatrix}
					a_{k k} \textbf{I}_{p_k - 1} & \textbf{0}_{(p_k - 1) \times 1} \\ \textbf{0}_{1 \times (p_k - 1)} & 
					\sum_{k' = 1}^{K} a_{k' k'} p_{k'} \xi_{k, k'} ^ 2
					+ \sum_{k_1 = 1}^{K} \sum_{k_2 = 1}^{K}
					b_{k_1 k_2} p_{k_1} p_{k_2} \xi_{k, k_1} \xi_{k, k_2} 
				\end{pmatrix} \\
				& = \begin{pmatrix}
					a_{k k} \textbf{I}_{p_k - 1} & \textbf{0}_{(p_k - 1) \times 1} \\ \textbf{0}_{1 \times (p_k - 1)} & 
					\lambda_{\bar{p}_k}
				\end{pmatrix}.
			\end{split}
		\end{equation*}
		
		When $m \neq \ell$, use the property that the eigenvectors belonging to different eigenvalues are orthogonal.
		Thus, the off-diagonal of $\bm{\Gamma} \textbf{N}\left[\textbf{A}, \textbf{B}, \bm{p}\right] \bm{\Gamma} ^ \top$ is zero matrix.
		We complete the proof of (7).
	\end{proof}
	
	\begin{proof}[Proof of Lemma 3 (covariance matrix with a uniform-block structure)]
		Using the results in Lemma 2, the proof is straightforward. 
	\end{proof}	
	
	\subsection{Proofs in Section 3}
	
	Before proceeding to the proofs of the main results, 
	we emphasize the definition of Wishart distributions because they play important roles in the main results.
	
	\begin{dfn}[Definition of Wishart distribution]
		\label{Def:wishart}
		Let $\bm{X}_1, \ldots, \bm{X}_n \overset{\text{i.i.d.}}{\sim} N_p\left(\bm{\mu}, \bm{\Sigma}\right)$ be $p$-dimensional normal vectors. 
		Then, the $p$ by $p$ matrix $\textbf{W} \triangleq \sum_{i = 1} ^ {n} \bm{X}_i \bm{X}_i ^ \top$ is defined to follow a Wishart distribution, 
		denoted as $\textbf{W} \sim W_p \left(\bm{\Sigma}, n; \bm{\Omega}\right)$, 
		with the scale matrix $\bm{\Sigma}$, 
		the degrees of freedom parameter $n$, 
		and the noncentrality parameter $\bm{\Omega} \triangleq \bm{\mu} \bm{\mu} ^ \top \in \mathbb{R} ^ {p \times p}$.
		In particular, 
		
		(1) if $\bm{\mu} = \textbf{0}_{p \times 1}$, then $\bm{\Omega} = \textbf{0}_{p \times p}$ and $W_p\left(\bm{\Sigma}, n\right) \triangleq W_p\left(\bm{\Sigma}, n; \textbf{0}_{p \times p}\right)$ is called a central Wishart distribution; 
		otherwise, it is called a noncentral Wishart distribution.
		
		(2) if $n \geq p$, then $\textbf{W}$ is positive definite with probability $1$, and $W_p\left(\bm{\Sigma}, n; \bm{\Omega}\right)$ is called a nonsingular Wishart distribution;
		if $0 < n < p$, then $\textbf{W}$ is semi-positive definite and $\det \left(\textbf{W}\right) = 0$ with a non-zero probability, 
		so $W_p\left(\bm{\Sigma}, n; \bm{\Omega}\right)$ is called a singular Wishart distribution.
	\end{dfn}
	
	Definition~\ref{Def:wishart} can be found in \citet[Definition 2.4.1]{KolloRosen2005}; \citet[Definition 2b]{Seber2004}; \citet[Definition 8.1]{Eaton2007}; \citet[Definition 2.2.1 (2)]{FujikoshiUlyanovShimizu2011}; \citet[Definition 7.1]{BilodeauBrenner1999}.
	We remark that the condition $\bm{\Sigma} > 0$ is not necessary in the above definition; in other words, $\bm{\Sigma}$ can be semi-positive definite.  
	Additionally, \citet{DiazCarciaJaimezMardia1997} distinguishes the sources of singularities,   
	defining (singular or nonsingular) pseudo-Wishart distribution families if ($\bm{\Sigma}$ does not have or has full ranks) $n < p$.
	
	The following lemmas are helpful to prove our main theorems. 
	
	\begin{lemma}[Closure of Wishart distriubtions under linear transformations]
		\label{Lem:transform}
		Let $\textbf{W} \sim W_p\left(\bm{\Sigma}, n; \bm{\Omega}\right)$ and $\textbf{A} \in \mathbb{R} ^ {q \times p}$ with $\operatorname{rank}(\textbf{A}) = q$.
		Then, $\textbf{A} \textbf{W} \textbf{A} ^ \top \sim W_q \left(\textbf{A} \bm{\Sigma} \textbf{A} ^ \top, n; \textbf{A} \bm{\Omega} \textbf{A} ^ \top\right)$.
	\end{lemma}
	
	\begin{proof}[Proof of Lemma~\ref{Lem:transform}]
		Please refer to the proofs in \citet[Theorem 2.4.2]{KolloRosen2005}; \citet[Theorem 2.2]{Seber2004}; \citet[Proposition 8.1]{Eaton2007}; \citet[Theorem 2.2.1 (2)]{FujikoshiUlyanovShimizu2011}; \citet[Proposition 7.4]{BilodeauBrenner1999}.
	\end{proof}
	
	\begin{lemma}[Trace inequality of positive definite matrices]
		\label{Lem:trace}
		Suppose matrix $\textbf{A} \in \mathbb{R} ^ {p \times p}$ is positive definite.
		Then,
		\begin{equation*}
			\operatorname{tr}\left(\textbf{A}\right) - \log \det \left(\textbf{A}\right) \geq p
		\end{equation*}
		where the equality holds if and only if $\textbf{A} = \textbf{I}_p$.
	\end{lemma}
	
	\begin{proof}[Proof of Lemma~\ref{Lem:trace}]
		Consider the eigenvalues of $\textbf{A}$, denoted as $\lambda_1, \ldots, \lambda_p$, all are positive due to $\textbf{A} > 0$.
		Using the inequality $\log (1 + x) \leq x$ for $x + 1 > 0$, 
		we obtain that 
		\begin{equation*}
			\log \det \left(\textbf{A}\right)
			= \log \prod_{i = 1}^{p} \lambda_i
			= \sum_{i = 1}^{n} \log\left(\lambda_i\right)
			= \sum_{i = 1}^{n} \log\left(1 + \lambda_i - 1\right)
			\leq \sum_{i = 1}^{n} \left(\lambda_i - 1\right)
			= \operatorname{tr}\left(\textbf{A}\right) - p,
		\end{equation*}
		where the equality holds if and only if $\lambda_i - 1 = 0$ for $i = 1, \ldots, p$, i.e., $\textbf{A} = \textbf{I}_p$.
	\end{proof}
	
	\begin{lemma}[Distributions of quadratic form related to scatter matrices]
		\label{Lem:decompose}
		Suppose $p$-dimensional random vector $\bm{Y}_i \sim N_p\left(\omega_i \bm{\mu}_i, \bm{\Sigma}\right)$ are mutually independently distributed for $i = 1, 2, \dots, n$. 
		Let $\bar{\bm{Y}} \triangleq \left(\sum_{i = 1}^{n} \omega_i \bm{Y}_i \right) / \left(\sum_{i = 1}^{n} \omega_i ^ 2 \right)$ denote the overall mean. 
		
		(1) Suppose $\bm{\mu}_i \triangleq \bm{\mu}$ for $i = 1, \ldots, n$. The following decomposition holds: 
		\begin{equation*}
			\sum_{i = 1}^{n} \left(\bm{Y}_i - \omega_i \bm{\mu} \right)\left(\bm{Y}_i - \omega_i \bm{\mu} \right) ^ \top
			= 
			\sum_{i = 1}^{n}\left(\bm{Y}_i - \omega_i \bar{\bm{Y}} \right) \left(\bm{Y}_i - \omega_i \bar{\bm{Y}} \right) ^ \top 
			+ 
			\sum_{i = 1}^{n} \omega_i ^ 2 \left(\bar{\bm{Y}} - \bm{\mu} \right) \left(\bar{\bm{Y}} - \bm{\mu} \right) ^ \top
		\end{equation*}
		
		(2) The quadratic form $\sum_{i = 1}^{n} \left(\bm{Y}_i - \omega_i \bar{\bm{Y}} \right) \left(\bm{Y}_i - \omega_i \bar{\bm{Y}} \right) ^ \top$
		follows a $p$-dimensional Wishart distribution with a degree of freedom $\operatorname{rank}\left(\textbf{A}\right)$,
		a scale matrix $\bm{\Sigma}$, 
		and a noncentrality parameter $\bm{\Omega} \triangleq \textbf{M} \textbf{A} \textbf{M} ^ \top$, 
		i.e., $W_p\left(\bm{\Sigma}, \operatorname{rank}\left(\textbf{A}\right); \bm{\Omega} \right)$, 
		where $\textbf{A} \triangleq \textbf{I}_n - \left(\omega_1, \ldots, \omega_n \right) ^ \top \left(\omega_1, \ldots, \omega_n \right) / \left(\sum_{i = 1}^{n}\omega_i ^ 2\right) \in \mathbb{R} ^ {n \times n}$ and $\textbf{M} \triangleq \left(\operatorname{E}(\bm{Y}_1), \ldots, \operatorname{E}\left(\bm{Y}_n\right)\right) \in \mathbb{R} ^ {p \times n}$.
	\end{lemma}
	
	\begin{proof}[Proof of Lemma~\ref{Lem:decompose}]
		
		(1) We expand the sum of the squares, 
		\begin{equation*}
			\begin{split}
				& \sum_{i = 1}^{n} \left(\bm{Y}_i - \omega_i \bm{\mu} \right)\left(\bm{Y}_i - \omega_i \bm{\mu} \right) ^ \top \\
				& = \sum_{i = 1}^{n} \omega_i ^ 2
				\bigg\{
				\left(\frac{\bm{Y}_i}{\omega_i} - \bar{\bm{Y}}\right) \left(\frac{\bm{Y}_i}{\omega_i} - \bar{\bm{Y}}\right) ^ \top +
				\left(\frac{\bm{Y}_i}{\omega_i} - \bar{\bm{Y}}\right) \left(\bar{\bm{Y}} - \bm{\mu} \right) ^ \top + 
				\left(\bar{\bm{Y}} - \bm{\mu} \right) \left(\frac{\bm{Y}_i}{\omega_i} - \bar{\bm{Y}} \right) ^ \top \\
				& +
				\left(\bar{\bm{Y}} - \bm{\mu} \right) \left(\bar{\bm{Y}} - \bm{\mu} \right) ^ \top
				\bigg\},
			\end{split}
		\end{equation*}
		where the cross term equals $\textbf{0}_{p \times p}$ if and only if 
		\begin{equation*}
			\sum_{i = 1}^{n} \omega_i ^ 2 \left(\frac{\bm{Y}_i}{\omega_i} - \bar{\bm{Y}}\right) \left(\bar{\bm{Y}} - \bm{\mu} \right) ^ \top 
			= \left\{ \sum_{i = 1}^{n} \left( \omega_i \bm{Y}_i - \omega_i ^ 2 \bar{\bm{Y}} \right) \right\}
			\left(\bar{\bm{Y}} - \bm{\mu} \right) ^ \top
			= \textbf{0}_{p \times p}. 
		\end{equation*}
		This holds following the definition of $\bar{\bm{Y}}$.
		We rewrite the quadratic form as below,
		\begin{equation*}
			\sum_{i = 1}^{n} \left(\bm{Y}_i - \omega_i \bm{\mu} \right)\left(\bm{Y}_i - \omega_i \bm{\mu} \right) ^ \top 
			= \sum_{i = 1}^{n}
			\left\{
			\left(\bm{Y}_i - \omega_i \bar{\bm{Y}} \right) \left(\bm{Y}_i - \omega_i \bar{\bm{Y}} \right) ^ \top + 
			\omega_i ^ 2 \left(\bar{\bm{Y}} - \bm{\mu} \right) \left(\bar{\bm{Y}} - \bm{\mu} \right) ^ \top
			\right\}.
		\end{equation*}
		
		(2) Let $\alpha \triangleq \sum_{i = 1}^{n} \omega_i ^ 2$ and $\beta_i \triangleq \omega_i / \alpha$ for every $i$.
		Thus, $\bar{\bm{Y}}$, $\omega_i \bar{\bm{Y}}$ and $\bm{Y}_i - \omega_i \bar{\bm{Y}}$ can be written as 
		$\sum_{j = 1}^{n} \beta_j \bm{Y}_j$, $\sum_{j = 1}^{n} \omega_i \beta_j \bm{Y}_j$, and
		$\begin{pmatrix}
			\bm{Y}_1 & \bm{Y}_2 & \dots & \bm{Y}_n
		\end{pmatrix}_{p \times n} \begin{pmatrix}
			\omega_i \beta_1 & \dots & 1 - \omega_i \beta_i & \dots & \omega_i \beta_n
		\end{pmatrix}_{1 \times n} ^ \top$, respectively.
		%
		Let $\textbf{e}_i$ denote the unit vector containing $1$ only at $i$-th element, 
		let $\bm{\beta} \triangleq \left(\beta_1, \beta_2, \dots, \beta_n \right) ^ \top$ denote the $n$-dimensional column vector,
		and let $\textbf{Y} \triangleq \begin{pmatrix}
			\bm{Y}_1 & \bm{Y}_2 & \dots & \bm{Y}_n
		\end{pmatrix}$ denote the $p$ by $n$ data matrix.
		Thus, the objective quadratic form can be expressed by
		\begin{equation*}
			\sum_{i = 1}^{n} \left(\bm{Y}_i - \omega_i \bar{\bm{Y}} \right) \left(\bm{Y}_i - \omega_i \bar{\bm{Y}} \right) ^ \top
			= \sum_{i = 1}^{n} \left\{\textbf{Y} \left( \textbf{e}_i - \omega_i \bm{\beta} \right) \right\}
			\left[ \textbf{Y} \left( \textbf{e}_i - \omega_i \bm{\beta} \right) \right] ^ \top 
			\triangleq \textbf{Y} \textbf{A} \textbf{Y} ^ \top,
		\end{equation*}
		where $\textbf{A} \triangleq \sum_{i = 1}^{n} \left( \textbf{e}_i - \omega_i \bm{\beta} \right) \left( \textbf{e}_i - \omega_i \bm{\beta} \right) ^ \top \in \mathbb{R} ^ {n \times n}$ is a symmetric matrix. 
		Observing that $\sum_{i = 1}^{n} \omega_i \textbf{e}_i = \left(\omega_1, \dots, \omega_n \right) ^ \top = \alpha \left(\beta_1, \dots, \beta_n \right) ^ \top = \alpha \beta$, 
		we may simplify $\textbf{A}$ as 
		\begin{equation*}
			\begin{split}
				\textbf{A} 
				& = \sum_{i = 1}^{n} \left( \textbf{e}_i - \omega_i \bm{\beta} \right) \left( \textbf{e}_i - \omega_i \bm{\beta} \right) ^ \top \\
				& = \sum_{i = 1}^{n} \textbf{e}_i \textbf{e}_i ^ \top
				- \left(\sum_{i = 1}^{n} \textbf{e}_i \omega_i \right) \bm{\beta} ^ \top
				- \bm{\beta} \left(\sum_{i = 1}^{n} \omega_i \textbf{e}_i ^ \top \right)
				+ \sum_{i = 1}^{n} \omega_i ^ 2 \bm{\beta} \bm{\beta} ^ \top \\
				& = \textbf{I}_n
				- \left(\alpha \bm{\beta} \right) \bm{\beta} ^ \top
				- \bm{\beta} \left(\alpha \bm{\beta} ^ \top \right)
				+ \alpha \bm{\beta} \bm{\beta} ^ \top \\
				& = \textbf{I}_n
				- \alpha \bm{\beta} \bm{\beta} ^ \top.
			\end{split}
		\end{equation*}
		Furthermore, the above expression implies $\textbf{A}$ is an idempotent matrix because
		\begin{equation*}
			\begin{split}
				\textbf{A} ^ 2 & = \left( \textbf{I}_n - \alpha \bm{\beta} \bm{\beta} ^ \top\right) 
				\left( \textbf{I}_n - \alpha \bm{\beta} \bm{\beta} ^ \top\right) \\
				& =  \textbf{I}_n \textbf{I}_n - \textbf{I}_n \alpha \bm{\beta} \bm{\beta} ^ \top
				- \alpha \bm{\beta} \bm{\beta} ^ \top \textbf{I}_n + \alpha ^ 2
				\left( \bm{\beta} \bm{\beta} ^ \top \right)\left( \bm{\beta} \bm{\beta} ^ \top \right) \\
				& = \textbf{I}_n - 2 \alpha \bm{\beta} \bm{\beta} ^ \top
				+ \alpha ^ 2 \left( \bm{\beta} \bm{\beta} ^ \top \right)\left( \bm{\beta} \bm{\beta} ^ \top \right) \\
				& = \textbf{I}_n - 2 \alpha \bm{\beta} \bm{\beta} ^ \top
				+ \alpha ^ 2 \left( \bm{\beta} \bm{\beta} ^ \top  \bm{\beta} \bm{\beta} ^ \top \right) \\
				& = \textbf{I}_n - 2 \alpha \bm{\beta} \bm{\beta} ^ \top
				+ \alpha ^ 2 \left( \bm{\beta} \left( \sum_{i = 1}^{n} \frac{\omega_i ^ 2}{\alpha ^ 2} \right) \bm{\beta} ^ \top \right) \\
				& = \textbf{I}_n - 2 \alpha \bm{\beta} \bm{\beta} ^ \top
				+ \alpha \left( \bm{\beta} \bm{\beta} ^ \top \right) \\
				& = \textbf{A}.
			\end{split}
		\end{equation*}
		
		From the point of matrix quadratic forms, 
		$\textbf{Y} \in \mathbb{R} ^ {p \times n}$ follows a matrix normal distribution with mean matrix $\textbf{M} \in \mathbb{R} ^ {p \times n}$, a separable covariance matrix $\bm{\Sigma} \in \mathbb{R} ^ {p \times p}$ and $\bm{\Psi} \triangleq \textbf{I}_n$, i.e., 
		\begin{equation*}
			\textbf{Y} \triangleq \left(\bm{Y}_1, \ldots, \bm{Y}_n \right) \sim N_{p, n} \left(\textbf{M}, \bm{\Sigma}, \bm{\Psi} \right), \quad
			\textbf{M} \triangleq \operatorname{E}(\textbf{Y}), \quad
			\bm{\Sigma} \triangleq \operatorname{cov}\left(\bm{Y}_i\right), \quad
			\bm{\Psi} \triangleq \textbf{I}_n.
		\end{equation*}
		Furthermore, $\textbf{A} \in \mathbb{R} ^ {p \times p}$ is idempotent,  
		using Corollary 2.1 in \citet{SingullKoski2012}, 
		we conclude that 
		\begin{equation*}
			\textbf{Y} \textbf{A} \textbf{Y} ^ \top \sim W_p\left(\bm{\Sigma}, \operatorname{rank}\left(\textbf{A}\right); \bm{\Omega} \right),
		\end{equation*}
		where the noncentrality parameter $\bm{\Omega} \triangleq \textbf{M} \textbf{A} \textbf{M} ^ \top$. 
		We complete the proof of Lemma \ref{Lem:decompose}.
	\end{proof}
	
	\begin{lemma}[One-Sample LRTs]
		\label{Lem:1-sample}
		Suppose $\bm{Y}_1, \ldots, \bm{Y}_n \overset{\text{i.i.d.}}{\sim} N_p\left(\bm{\nu}, n \bm{\Xi}\right)$ with $\bm{\Xi} > 0$.
		Furthermore, define that $\bar{\bm{Y}} \triangleq \left(\bm{Y}_1 + \cdots + \bm{Y}_n\right) / n$ and $n \textbf{V} \triangleq \sum_{i = 1}^{n}\left(\bm{Y}_i - \bar{\bm{Y}}\right)\left(\bm{Y}_i - \bar{\bm{Y}}\right) ^ \top$.
		Let 
		\begin{equation*}
			\begin{split}
				\omega_1: & \text{ the region containing that } \bm{\nu} \text{ is unstructured and } \bm{\Xi} \text{ is unstructured, } \\
				\omega_2: & \text{ the region containing that } \bm{\nu} \text{ is unstructured and } \bm{\Xi} = \operatorname{diag}\left(a_{11}\textbf{1}_{1 \times p_1}, \lambda_{\bar{p}_1}, \ldots, a_{KK}\textbf{1}_{1 \times p_K}, \lambda_{\bar{p}_K}\right), \\
				\omega_3: & \text{ the region containing that } \bm{\nu} = \bm{\nu}_0 \text{ and } \bm{\Xi} = \operatorname{diag}\left(a_{11}\textbf{1}_{1 \times p_1}, \lambda_{\bar{p}_1}, \ldots, a_{KK}\textbf{1}_{1 \times p_K}, \lambda_{\bar{p}_K}\right).
			\end{split}
		\end{equation*}
		In particular, under $\omega_2$ or $\omega_3$, 
		$\textbf{A} \triangleq \operatorname{diag}\left(a_{11}, \ldots, a_{KK}\right) > 0$ and $\bm{\Delta} \triangleq \textbf{A} + \textbf{B} \textbf{P} > 0$,
		where $\textbf{B} \triangleq \left(b_{kk'}\right)$ with $b_{k'k} = b_{kk'}$ for $k \neq k'$, 
		$\textbf{P} \triangleq \operatorname{diag}\left(p_1, \ldots, p_K\right)$, satisfying that $p = p_1 + \cdots + p_K$ and $p_k > 1$ for $k = 1, \ldots, K$ are integers, 
		and $\bar{p}_k \triangleq \sum_{k' = 1}^{k} p_{k'}$ for $k = 1, \ldots, K$, 
		and $\lambda_{\bar{p}_1}, \ldots, \lambda_{\bar{p}_K}$ are $K$ eigenvalues of $\bm{\Delta}$ in the decreasing order.
		Thus, these conditions guarantee that $\bm{\Xi} > 0$.
		Under $\omega_3$, 
		$\bm{\nu}_0 \in \mathbb{R} ^ {p \times 1}$ is known.
		
		Additionally, if $n > p$,  
		then the likelihood ratio between $\omega_2$ and $\omega_1$ is 
		\begin{equation*}
			\begin{split}
				\frac{\sup \limits_{\omega_2} L\left(\bm{\nu}, \bm{\Xi}; \bm{Y}_1, \ldots, \bm{Y}_n\right)}{\sup \limits_{\omega_1} L\left(\bm{\nu}, \bm{\Xi}; \bm{Y}_1, \ldots, \bm{Y}_n\right)} 
				= \left[
				\left\{\prod_{k = 1}^{K} (p_k - 1) ^ {(p_k - 1)} \right\}
				\frac{\det \left(\textbf{V}\right)}
				{\prod_{k = 1}^{K} \textbf{V}_{\bar{p}_k, \bar{p}_k} \left(\sum_{j = \bar{p}_{k - 1} + 1}^{\bar{p}_k - 1} \textbf{V}_{j, j}\right) ^ {p_k - 1}}
				\right] ^ {n / 2};
			\end{split}
		\end{equation*}
		if $\widehat{\textbf{A}} > 0$ and $\widehat{\bm{\Delta}} = \widehat{\textbf{A}} + \widehat{\textbf{B}} \textbf{P} > 0$, 
		then the likelihood ratio between $\omega_3$ and $\omega_2$ is 
		\begin{equation*}
			\begin{split}
				\frac{\sup \limits_{\omega_3} L\left(\bm{\nu}, \bm{\Xi}; \bm{Y}_1, \ldots, \bm{Y}_n\right)}{\sup \limits_{\omega_2} L\left(\bm{\nu}, \bm{\Xi}; \bm{Y}_1, \ldots, \bm{Y}_n\right)}
				= \left[
				\prod_{k = 1}^{K}
				\frac{\textbf{V}_{\bar{p}_k, \bar{p}_k}}{\left(\textbf{H}_{\bar{p}_k, \bar{p}_k} + \textbf{V}_{\bar{p}_k, \bar{p}_k}\right)}
				\left(
				\frac{\sum_{j = \bar{p}_{k - 1} + 1}^{\bar{p}_k - 1} \textbf{V}_{j, j}}{\sum_{j = \bar{p}_{k - 1} + 1}^{\bar{p}_k - 1} \left(\textbf{H}_{j, j} + \textbf{V}_{j, j}\right)}
				\right) ^ {p_k - 1}
				\right] ^ {n / 2},
			\end{split}
		\end{equation*}		
		where $\textbf{H} \triangleq \left(\bar{\bm{Y}} - \bm{\nu}\right)\left(\bar{\bm{Y}} - \bm{\nu}\right) ^ \top$.
	\end{lemma}
	
	\begin{proof}[Proof of Lemma~\ref{Lem:1-sample}]
		
		Omitting the constant term $- n p \log(2\pi) / 2  - (n p \log n) / 2$, the (kernel of the) log-likelihood function is 
		\begin{equation*}
			\log L \left(\bm{\nu}, \bm{\Xi}; \bm{Y}_1, \ldots, \bm{Y}_n\right)
			= - \frac{n}{2} \log \det \left(\bm{\Xi}\right)
			- \frac{1}{2} \operatorname{tr}\left(\bm{\Xi} ^ {- 1}\textbf{V}\right)
			- \frac{n}{2}
			\left\{
			\left(\bar{\bm{Y}} - \bm{\nu}\right) ^ \top \left(n\bm{\Xi}\right) ^ {- 1}\left(\bar{\bm{Y}} - \bm{\nu}\right)
			\right\}.
		\end{equation*}
		
		Under $\omega_1$ or $\omega_2$, $\bm{\nu}$ is unspecified/unstructured,
		\begin{equation*}
			\log L \left(\bm{\nu}, \bm{\Xi}; \bm{Y}_1, \ldots, \bm{Y}_n\right)
			\leq 
			\log L \left(\bm{\nu} = \bar{\bm{Y}}, \bm{\Xi}; \bm{Y}_1, \ldots, \bm{Y}_n\right)
			= - \frac{n}{2} \log \det \left(\bm{\Xi}\right)
			- \frac{1}{2} \operatorname{tr}\left(\bm{\Xi} ^ {- 1}\textbf{V}\right), 
		\end{equation*}
		where the maximum likelihood estimator $\widehat{\bm{\nu}}(\omega_1) = \widehat{\bm{\nu}}(\omega_2) \triangleq \bar{\bm{Y}}$, 
		since $n \bm{\Xi} > 0$, 
		yielding that 
		$\left(\bar{\bm{Y}} - \bm{\nu}\right) ^ \top \left(n\bm{\Xi}\right) ^ {- 1}\left(\bar{\bm{Y}} - \bm{\nu}\right) > 0$ if $\bar{\bm{Y}} \neq \bm{\nu}$; 
		and $\left(\bar{\bm{Y}} - \bm{\nu}\right) ^ \top \left(n\bm{\Xi}\right) ^ {- 1}\left(\bar{\bm{Y}} - \bm{\nu}\right) = 0$ if $\bar{\bm{Y}} = \bm{\nu}$.
		In other words, for a fixed $\bm{\Xi} > 0$,
		\begin{equation*}
			\log L \left(\bm{\nu} = \bar{\bm{Y}}, \bm{\Xi}; \bm{Y}_1, \ldots, \bm{Y}_n\right) = \sup \limits_{\bm{\nu} \in \mathbb{R} ^ {p \times 1}} \log L \left(\bm{\nu}, \bm{\Xi}; \bm{Y}_1, \ldots, \bm{Y}_n\right).
		\end{equation*}
		
		\textbf{\emph{Derivation under $\omega_1$}}: $\bm{\Xi}$ is unstructured in the following 
		\begin{equation*}
			\begin{split}
				& \log L \left(\bm{\nu} = \bar{\bm{Y}}, \bm{\Xi}; \bm{Y}_1, \ldots, \bm{Y}_n\right) \\
				& = - \frac{n}{2} \log \det \left(\bm{\Xi}\right)
				- \frac{1}{2} \operatorname{tr}\left(\bm{\Xi} ^ {- 1}\textbf{V}\right) \\
				& = - \frac{n}{2}
				\left[
				\log \det \left(\bm{\Xi}\right)
				+ \operatorname{tr}\left\{
				\left(n \bm{\Xi}\right) ^ {- 1} \textbf{V}
				\right\}
				\right] \\
				& = - \frac{n}{2}
				\left[
				\log \det \left(\bm{\Xi}\right)
				+ \operatorname{tr}
				\left\{
				\left(n \bm{\Xi}\right) ^ {- 1} \textbf{V}
				\right\}
				- \log \det \left\{
				\left(n \bm{\Xi}\right) ^ {- 1} \textbf{V}
				\right\}
				+ \log \det \left\{
				\left(n \bm{\Xi}\right) ^ {- 1} \textbf{V}
				\right\}
				\right] \\
				& = - \frac{n}{2}
				\left[
				\operatorname{tr}\left\{
				\left(n \bm{\Xi}\right) ^ {- 1} \textbf{V}
				\right\}
				- \log \det \left\{
				\left(n \bm{\Xi}\right) ^ {- 1} \textbf{V}
				\right\}
				+ \log \det \left(\textbf{V} / n\right)
				\right],
			\end{split}
		\end{equation*}
		where we impose an additional condition $n > p$, 
		ensuring that $\textbf{V} > 0$ with probability $1$. 
		For $\bm{\Xi} > 0$, 
		there exists a unique square root matrix of $\bm{\Xi} ^ {- 1}$, 
		denoted as $\bm{\Xi} ^ {- 1/2}$, 
		so it is easy to check that 
		\begin{equation*}
			\operatorname{tr}\left\{
			\left(n \bm{\Xi}\right) ^ {- 1} \textbf{V}
			\right\}
			= \operatorname{tr}\left\{
			\bm{\Xi} ^ {-1/2} \left(\textbf{V} / n\right)
			\bm{\Xi} ^ {-1/2}
			\right\}, 
			\log \det \left\{
			\left(n \bm{\Xi}\right) ^ {- 1} \textbf{V}
			\right\}
			= \log \det \left\{
			\bm{\Xi} ^ {- 1/2} \left(\textbf{V} / n\right)
			\bm{\Xi} ^ {- 1/2}
			\right\},
		\end{equation*}
		and the matrix $\bm{\Xi} ^ {- 1/2} \left(\textbf{V} / n\right)
		\bm{\Xi} ^ {- 1/2}$ are positive definite, 
		because it is similar to $\bm{\Xi} ^ {-1}\left(\textbf{V}/n\right)$ by definition:  		
		$\bm{\Xi} ^ {- 1/2} \left(\textbf{V} / n\right)
		\bm{\Xi} ^ {- 1/2} = \bm{\Xi}^{1/2}\left\{\bm{\Xi} ^ {-1}\left(\textbf{V}/n\right)\right\}\bm{\Xi}^{- 1/2}$;
		thus $\bm{\Xi} ^ {- 1/2} \left(\textbf{V} / n\right)$ and $\bm{\Xi} ^ {-1}\left(\textbf{V}/n\right)$ share the same eigenvalues, i.e., all eigenvalues of $\bm{\Xi} ^ {-1}\left(\textbf{V}/n\right)$ are positive because $\bm{\Xi} ^ {-1}$ and $\textbf{V}$ are positive definite.
		
		Using the result in Lemma~\ref{Lem:trace}, 
		we conclude that 
		\begin{equation*}
			\begin{split}
				& \log L \left(\bm{\nu} = \bar{\bm{Y}}, \bm{\Xi}; \bm{Y}_1, \ldots, \bm{Y}_n\right) \\
				& =
				- \frac{n}{2}\left[
				\operatorname{tr}\left\{
				\bm{\Xi} ^ {-1/2} \left(\textbf{V} / n\right)
				\bm{\Xi} ^ {-1/2}
				\right\} - 
				\log \det \left\{
				\bm{\Xi} ^ {- 1/2} \left(\textbf{V} / n\right)
				\bm{\Xi} ^ {- 1/2}
				\right\}
				+ \log \det \left(\textbf{V} / n\right)
				\right] \\
				& \leq 
				- \frac{n}{2}\left[
				p + \log \det \left(\textbf{V} / n\right)
				\right] 
			\end{split}
		\end{equation*}
		where the equality holds if and only if $\bm{\Xi} ^ {- 1/2} \left(\textbf{V} / n\right)
		\bm{\Xi} ^ {- 1/2} = \textbf{I}_p$, i.e., $\bm{\Xi} = \textbf{V} / n$.
		
		Therefore, under $\omega_1$ and $n > p$, 
		we obtain that $\widehat{\bm{\nu}}(\omega_1) = \bar{\bm{Y}}$ and $\widehat{\bm{\Xi}}(\omega_1) = \textbf{V} / n$ are global maximizers of the log-likelihood function, and 
		\begin{equation*}
			\begin{split}
				\sup \limits_{\omega_1} \log L \left(\bm{\nu}, \bm{\Xi}; \bm{Y}_1, \ldots, \bm{Y}_n\right)
				& = \log L \left(\bm{\nu} = \bar{\bm{Y}}, \bm{\Xi} = \widehat{\bm{\Xi}}(\omega_1); \bm{Y}_1, \ldots, \bm{Y}_n\right) \\
				& = - n p \log(2\pi) / 2  - (n p \log n) / 2 - \frac{n}{2}\left[
				p + \log \det \left(\textbf{V} / n\right)
				\right].
			\end{split}
		\end{equation*}
		In particular, 
		$\widehat{\bm{\Xi}}(\omega_1) = \textbf{V} / n > 0$ with probability $1$ \citep{Dykstra1970}. 
		In other words, the value of the likelihood function is finite.
		Without the condition $n > p$, 
		the maximum likelihood estimators do not exist, 
		thus, the likelihood ratio test approach fails.
		
		\textbf{\emph{Derivation under $\omega_2$}}: 
		we start with 
		\begin{equation*}
			\begin{split}
				& \log L \left(\bm{\nu}, \bm{\Xi}; \bm{Y}_1, \ldots, \bm{Y}_n\right) \\
				& \leq 
				\log L \left(\bm{\nu} = \bar{\bm{Y}}, \bm{\Xi}; \bm{Y}_1, \ldots, \bm{Y}_n\right) \\
				& = - \frac{n}{2} \log \det \left(\bm{\Xi}\right)
				- \frac{1}{2} \operatorname{tr}\left(\bm{\Xi} ^ {- 1}\textbf{V}\right) \\
				& = - \frac{n}{2} \log \left\{
				\prod_{k = 1}^{K} \left(
				a_{kk} ^ {p_k - 1} \lambda_{\bar{p}_k}
				\right)
				\right\}
				- \frac{1}{2} \sum_{k = 1} ^ {K} \left\{ \left(\sum_{j = \bar{p}_{k - 1} + 1}^{\bar{p}_k - 1} \frac{\textbf{V}_{j, j}}{a_{kk}}\right) + \frac{\textbf{V}_{\bar{p}_k, \bar{p}_k}}{\lambda_{\bar{p}_k}} \right\} \\
				& = - \frac{n}{2} \sum_{k = 1}^{K}
				\left(p_k - 1\right) \log \left(a_{kk}\right)
				- \frac{n}{2} \sum_{k = 1}^{K} \log \left(\lambda_{\bar{p}_k}\right)
				- \frac{1}{2} \sum_{k = 1} ^ {K} \left\{ \left(\sum_{j = \bar{p}_{k - 1} + 1}^{\bar{p}_k - 1} \frac{\textbf{V}_{j, j}}{a_{kk}}\right) + \frac{\textbf{V}_{\bar{p}_k, \bar{p}_k}}{\lambda_{\bar{p}_k}} \right\},
			\end{split}
		\end{equation*}
		where $\bm{\Xi} = \operatorname{diag}\left(a_{11}\textbf{1}_{1 \times p_1}, \lambda_{\bar{p}_1}, \ldots, a_{KK}\textbf{1}_{1 \times p_K}, \lambda_{\bar{p}_K}\right)$.
		
		For a fixed $k$, we take the first-order partial derivatives of the log-likelihood function with respect to $a_{kk}$ and $\lambda_{\bar{p}_k}$, respectively, 
		\begin{equation*}
			\frac{\partial \log L}{\partial a_{kk}}
			= - \frac{n}{2} \left(p_k - 1\right) \frac{1}{a_{kk}}
			+ \frac{1}{2} \frac{\sum_{j = \bar{p}_{k - 1} + 1}^{\bar{p}_k - 1} \textbf{V}_{j, j}}{a_{kk} ^ 2}, \quad
			\frac{\partial \log L}{\partial \lambda_{\bar{p}_k}}
			= - \frac{n}{2} \frac{1}{\lambda_{\bar{p}_k}}
			+ \frac{1}{2} \frac{\textbf{V}_{\bar{p}_{k}, \bar{p}_k}}{\lambda_{\bar{p}_k} ^ 2}; 
		\end{equation*}
		we take the second-order partial derivatives of the log-likelihood function with respect to $a_{kk}$, $\lambda_{\bar{p}_k}$, and the cross term, respectively,
		\begin{equation*}
			\begin{split}
				\frac{\partial ^ 2 \log L}{\partial a_{kk} \partial x}
				& = 
				\begin{cases}
					\frac{n}{2}\left(p_k - 1\right) \frac{1}{a_{kk} ^ 2}
					- \frac{\sum_{j = \bar{p}_{k - 1} + 1}^{\bar{p}_k - 1} \textbf{V}_{j, j}}{a_{kk} ^ 3}, & x = a_{kk} \\
					0, & \text{ otherwise }
				\end{cases}, \\
				\frac{\partial ^ 2 \log L}{\partial \lambda_{\bar{p}_k} \partial y}
				& = 
				\begin{cases}
					\frac{n}{2} \frac{1}{\lambda_{\bar{p}_k} ^ 2} - \frac{\textbf{V}_{\bar{p}_{k}, \bar{p}_k}}{\lambda_{\bar{p}_k} ^ 3}, & y = \lambda_{\bar{p}_k} \\
					0, & \text{ otherwise }
				\end{cases}.
			\end{split}
		\end{equation*}
		Setting the first-order partial derivatives as $0$'s, 
		the candidates of the maximum likelihood estimators are
		\begin{equation*}
			\widehat{a}_{kk} = \sum_{j = \bar{p}_{k - 1} + 1}^{\bar{p}_k - 1} \textbf{V}_{j, j} / \left\{\left(p_k - 1\right) n\right\}, \quad
			\widehat{\lambda}_{\bar{p}_k} = \textbf{V}_{\bar{p}_{k}, \bar{p}_k} / n, \quad k = 1, \ldots, K.
		\end{equation*}
		Plugging them into the second-order partial derivatives, 
		we observe that 
		\begin{equation*}
			\begin{split}
				\frac{\partial ^ 2 \log L}{\partial a_{kk} \partial a_{kk}} \bigg |_{a_{kk} = \widehat{a}_{kk}}
				& = - \frac{\frac{1}{2} \sum_{j = \bar{p}_{k - 1} + 1}^{\bar{p}_k - 1} \textbf{V}_{j, j}}{\widehat{a}_{kk} ^ 3}
				= - \frac{\left(p_k - 1\right) n}{2 \widehat{a}_{kk} ^ 2} < 0, \quad k = 1, \ldots, K, \\
				\frac{\partial ^ 2 \log L}{\partial \lambda_{\bar{p}_k} \partial \lambda_{\bar{p}_k}} \bigg |_{\lambda_{\bar{p}_k} = \widehat{\lambda}_{\bar{p}_k}}
				& = - \frac{\frac{1}{2} \textbf{V}_{\bar{p}_k, \bar{p}_k}}{\widehat{\lambda}_{\bar{p}_k} ^ 3}
				= - \frac{n}{2 \widehat{\lambda}_{\bar{p}_k} ^ 2} < 0, \quad k = 1, \ldots, K.
			\end{split}
		\end{equation*}
		So, $\widehat{a}_{kk}$ and $\widehat{\lambda}_{\bar{p}_k}$ for $k = 1, \ldots, K$ are the only extreme points in the interior and they are maximum values, respectively.
		To verify they are global maximums, 
		we check the boundaries: 
		\begin{equation*}
			\begin{split}
				\lim\limits_{a_{kk} \to 0 ^ +} \log L = - \infty, 
				\lim\limits_{a_{kk} \to \infty} \log L = - \infty,
				\lim\limits_{\lambda_{\bar{p}_k} \to 0 ^ +} \log L = - \infty, 
				\lim\limits_{\lambda_{\bar{p}_k} \to \infty} \log L = - \infty,
			\end{split}
		\end{equation*} 
		for $k = 1, \ldots, K$, 
		yielding that the values of likelihood function are $0$.
		
		Therefore, under $\omega_2$, we obtain that $\widehat{\bm{\nu}}(\omega_2) = \bar{\bm{Y}}$,  $\widehat{a}_{kk}(\omega_2) = \sum_{j = \bar{p}_{k - 1} + 1}^{\bar{p}_k - 1} \textbf{V}_{j, j} / \left\{\left(p_k - 1\right) n\right\}$, 
		$\widehat{\lambda}_{\bar{p}_k}(\omega_2) = \textbf{V}_{\bar{p}_{k}, \bar{p}_k} / n$ for $k = 1, \ldots, K$,
		are global maximizers of the log-likelihood function, 
		and 
		\begin{equation*}
			\begin{split}
				& \sup \limits_{\omega_2} \log L \left(\bm{\nu}, \bm{\Xi}; \bm{Y}_1, \ldots, \bm{Y}_n\right) \\
				& = \log L \left(\bm{\nu} = \bar{\bm{Y}}, \bm{\Xi} = \widehat{\bm{\Xi}}(\omega_2); \bm{Y}_1, \ldots, \bm{Y}_n\right) \\
				& = - n p \log(2\pi) / 2  - (n p \log n) / 2 \\
				& - \frac{n}{2} \sum_{k = 1}^{K}
				\left(p_k - 1\right) \left\{\log \left(\sum_{j = \bar{p}_{k - 1} + 1}^{\bar{p}_k - 1} \textbf{V}_{j, j}\right) - \log(p_k - 1) - \log n \right\}
				- \frac{n}{2} \sum_{k = 1}^{K} \left(\log \textbf{V}_{\bar{p}_k, \bar{p}_k} - \log n\right)
				- \frac{p n}{2} \\
				& = - n p \log(2\pi) / 2  - (n p \log n) / 2 \\
				& - \frac{n}{2} \sum_{k = 1}^{K} \left(p_k - 1\right) 
				\log \left(\sum_{j = \bar{p}_{k - 1} + 1}^{\bar{p}_k - 1} \textbf{V}_{j, j}\right)
				+ \frac{n}{2} \sum_{k = 1}^{K} \left(p_k - 1\right) 
				\log(p_k - 1)
				- \frac{n}{2} \sum_{k = 1}^{K} 
				\log \textbf{V}_{\bar{p}_k, \bar{p}_k}
				+ \frac{n p}{2} \left(\log n - 1\right).
			\end{split}
		\end{equation*}
		In particular, 
		$\widehat{\bm{\Xi}}(\omega_2) = \operatorname{diag}\left(\widehat{a}_{11}(\omega_2)\textbf{1}_{1 \times p_1}, \widehat{\lambda}_{\bar{p}_1}(\omega_2), \ldots, \widehat{a}_{KK}(\omega_2)\textbf{1}_{1 \times p_K}, \widehat{\lambda}_{\bar{p}_K}(\omega_2)\right) > 0$ with probability $1$ since $\widehat{\textbf{A}} > 0$ and $\widehat{\bm{\Delta}} = \widehat{\textbf{A}} + \widehat{\textbf{B}}\textbf{P} > 0$ with probability $1$, 
		resulting in $\widehat{\bm{\Sigma}}(\omega_2) > 0$ with probability $1$.
		Thus, the terms inside the logarithm are positive with probability $1$. 
		
		\textbf{\emph{Derivation under $\omega_3$}}: 
		we rewrite the (kernel of the) log-likelihood function 
		\begin{equation*}
			\log L \left(\bm{\nu}, \bm{\Xi}; \bm{Y}_1, \ldots, \bm{Y}_n\right)
			= - \frac{n}{2} \log \det \left(\bm{\Xi}\right)
			- \frac{1}{2} \operatorname{tr}
			\left(
			\bm{\Xi} ^ {- 1} \textbf{D}
			\right),
		\end{equation*}
		where $\textbf{D} \triangleq \textbf{V} + \left(\bar{\bm{Y}} - \bm{\nu}_0\right)\left(\bar{\bm{Y}} - \bm{\nu}_0\right)^\top \in \mathbb{R} ^ {p \times p}$.
		
		Similar to the situation under $\omega_2$, 
		we obtain that $\widehat{\bm{\nu}}(\omega_3) = \bm{\nu}_0$, 
		$\widehat{a}_{kk}(\omega_3) = \sum_{j = \bar{p}_{k - 1} + 1}^{\bar{p}_k - 1} \textbf{D}_{j, j} / \left\{(p_k - 1) n\right\}$, 
		$\widehat{\lambda}_{\bar{p}_k}(\omega_3) = \textbf{D}_{\bar{p}_k, \bar{p}_k} / n$ for $k = 1, \ldots, K$, 
		are global maximizers of the log-likelihood function, 
		and 
		\begin{equation*}
			\begin{split}
				& \sup \limits_{\omega_3} \log L \left(\bm{\nu}, \bm{\Xi}; \bm{Y}_1, \ldots, \bm{Y}_n\right) \\
				& = \log L \left(\bm{\nu} = \bm{\nu}_0, \bm{\Xi} = \widehat{\bm{\Xi}}(\omega_3); \bm{Y}_1, \ldots, \bm{Y}_n\right) \\
				& = - n p \log(2\pi) / 2  - (n p \log n) / 2 \\
				& - \frac{n}{2} \sum_{k = 1}^{K} \left(p_k - 1\right) 
				\log \left(\sum_{j = \bar{p}_{k - 1} + 1}^{\bar{p}_k - 1} \textbf{D}_{j, j}\right)
				+ \frac{n}{2} \sum_{k = 1}^{K} \left(p_k - 1\right) 
				\log(p_k - 1)
				- \frac{n}{2} \sum_{k = 1}^{K} 
				\log \textbf{D}_{\bar{p}_k, \bar{p}_k}
				+ \frac{n p}{2} \left(\log n - 1\right).
			\end{split}
		\end{equation*}
		In particular, 
		$\widehat{\bm{\Xi}}(\omega_3) = \operatorname{diag}\left(\widehat{a}_{11}(\omega_3)\textbf{1}_{1 \times p_1}, \widehat{\lambda}_{\bar{p}_1}(\omega_3), \ldots, \widehat{a}_{KK}(\omega_3)\textbf{1}_{1 \times p_K}, \widehat{\lambda}_{\bar{p}_K}(\omega_3)\right) > 0$ with probability $1$ since $\widehat{\textbf{A}} > 0$ and $\widehat{\bm{\Delta}} = \widehat{\textbf{A}} + \widehat{\textbf{B}}\textbf{P} > 0$ with probability $1$, 
		and $\left(\bar{\bm{Y}} - \bm{\nu}_0\right)\left(\bar{\bm{Y}} - \bm{\nu}_0\right) ^ \top \geq 0$,
		resulting in $\textbf{D} > 0$ and $\widehat{\bm{\Sigma}}(\omega_3) > 0$ with probability $1$.
		Thus, the terms inside the logarithm are positive with probability $1$.
		
		Lastly, the likelihood ratio between $\omega_2$ and $\omega_1$ is 
		\begin{equation*}
			\begin{split}
				& \frac{\sup \limits_{\omega_2} L\left(\bm{\nu}, \bm{\Xi}; \bm{Y}_1, \ldots, \bm{Y}_n\right)}{\sup \limits_{\omega_1} L\left(\bm{\nu}, \bm{\Xi}; \bm{Y}_1, \ldots, \bm{Y}_n\right)} \\
				& = 
				\frac{\exp\left\{- \frac{n}{2} \log \prod_{k = 1}^{K} 
					\left(\sum_{j = \bar{p}_{k - 1} + 1}^{\bar{p}_k - 1} \textbf{V}_{j, j}\right) ^ {p_k - 1}
					+ \frac{n}{2} \log \prod_{k = 1}^{K} 
					(p_k - 1) ^ {p_k - 1}
					- \frac{n}{2} \log \prod_{k = 1}^{K} 
					\textbf{V}_{\bar{p}_k, \bar{p}_k}\right\}}
				{\exp\left(- \frac{n}{2} \log \det \textbf{V}
					\right)} \\
				& = 
				\frac{\exp\left[- \frac{n}{2} \log \prod_{k = 1}^{K}
					\left\{
					\left(\sum_{j = \bar{p}_{k - 1} + 1}^{\bar{p}_k - 1} \textbf{V}_{j, j}\right) ^ {p_k - 1}
					(p_k - 1) ^ {- (p_k - 1)}
					\textbf{V}_{\bar{p}_k, \bar{p}_k}
					\right\}
					\right]}
				{\exp\left(- \frac{n}{2} \log \det \textbf{V}
					\right)} \\
				& = \left[
				\frac{\prod_{k = 1}^{K}\left(\sum_{j = \bar{p}_{k - 1} + 1}^{\bar{p}_k - 1} \textbf{V}_{j, j}\right) ^ {p_k - 1}
					(p_k - 1) ^ {- (p_k - 1)}
					\textbf{V}_{\bar{p}_k, \bar{p}_k}}
				{\det \textbf{V}}
				\right] ^ {- n / 2} \\
				& = \left[
				\left\{\prod_{k = 1}^{K} (p_k - 1) ^ {(p_k - 1)} \right\}
				\frac{\det \left(\textbf{V}\right)}
				{\prod_{k = 1}^{K} \textbf{V}_{\bar{p}_k, \bar{p}_k} \left(\sum_{j = \bar{p}_{k - 1} + 1}^{\bar{p}_k - 1} \textbf{V}_{j, j}\right) ^ {p_k - 1}}
				\right] ^ {n / 2}.
			\end{split}
		\end{equation*}
		
		The likelihood ratio between $\omega_3$ and $\omega_2$ is 
		\begin{equation*}
			\begin{split}
				& \frac{\sup \limits_{\omega_3} L\left(\bm{\nu}, \bm{\Xi}; \bm{Y}_1, \ldots, \bm{Y}_n\right)}{\sup \limits_{\omega_2} L\left(\bm{\nu}, \bm{\Xi}; \bm{Y}_1, \ldots, \bm{Y}_n\right)} \\
				& = \frac{\exp\left\{- \frac{n}{2} \sum_{k = 1}^{K} \left(p_k - 1\right) 
					\log \left(\sum_{j = \bar{p}_{k - 1} + 1}^{\bar{p}_k - 1} \textbf{D}_{j, j}\right)
					- \frac{n}{2} \sum_{k = 1}^{K} 
					\log \textbf{D}_{\bar{p}_k, \bar{p}_k}\right\}}
				{\exp\left\{- \frac{n}{2} \sum_{k = 1}^{K} \left(p_k - 1\right) 
					\log \left(\sum_{j = \bar{p}_{k - 1} + 1}^{\bar{p}_k - 1} \textbf{V}_{j, j}\right)
					- \frac{n}{2} \sum_{k = 1}^{K} 
					\log \textbf{V}_{\bar{p}_k, \bar{p}_k}\right\}} \\
				& = \frac{\exp\left[- \frac{n}{2} 
					\log \prod_{k = 1}^{K} 
					\left\{\left(\sum_{j = \bar{p}_{k - 1} + 1}^{\bar{p}_k - 1} \textbf{D}_{j, j}\right) ^ {p_k - 1} 
					\textbf{D}_{\bar{p}_k, \bar{p}_k}
					\right\} \right]}
				{\exp\left[- \frac{n}{2} 
					\log \prod_{k = 1}^{K} 
					\left\{\left(\sum_{j = \bar{p}_{k - 1} + 1}^{\bar{p}_k - 1} \textbf{V}_{j, j}\right) ^ {p_k - 1} 
					\textbf{V}_{\bar{p}_k, \bar{p}_k}
					\right\} \right]} \\
				& = \left[
				\prod_{k = 1}^{K}
				\frac{\textbf{V}_{\bar{p}_k, \bar{p}_k}}{\textbf{D}_{\bar{p}_k, \bar{p}_k}}
				\left(
				\frac{\sum_{j = \bar{p}_{k - 1} + 1}^{\bar{p}_k - 1} \textbf{V}_{j, j}}{\sum_{j = \bar{p}_{k - 1} + 1}^{\bar{p}_k - 1} \textbf{D}_{j, j}}
				\right) ^ {p_k - 1}
				\right] ^ {n / 2}.
			\end{split}
		\end{equation*}
		We complete the proof.
	\end{proof}
	
	\begin{lemma}[Multiple-Sample LRTs]
		\label{Lem:M-sample}
		Suppose $\bm{Y}_1 ^ {(m)}, \ldots, \bm{Y}_{n ^ {(m)}} ^ {(m)} \overset{\text{i.i.d.}}{\sim} N_p \left(\bm{\nu} ^ {(m)}, n ^ {(m)} \bm{\Xi} ^ {(m)}\right)$ with $\bm{\Xi} ^ {(m)} > 0$ for $m = 1, \ldots, M$.
		Furthermore, define that $\bar{\bm{Y}} ^ {(m)} \triangleq \left(\bm{Y}_1 ^ {(m)} + \cdots + \bm{Y}_{n ^ {(m)}} ^ {(m)}\right) / n ^ {(m)}$, 
		$\bar{\bm{Y}} \triangleq \left(n ^ {(1)} \bar{\bm{Y}} ^ {(1)} + \cdots + n ^ {(M)} \bar{\bm{Y}} ^ {(M)} \right) / n$, 
		$n ^ {(m)} \textbf{V} ^ {(m)} \triangleq \sum_{i = 1}^{n ^ {(m)}} \left(\bm{Y}_i ^ {(m)} - \bar{\bm{Y}} ^ {(m)}\right)\left(\bm{Y}_i ^ {(m)} - \bar{\bm{Y}} ^ {(m)}\right) ^ \top$.
		Let 
		\begin{equation*}
			\begin{split}
				\omega_1: \text{ the region containing that } & \widetilde{\bm{\nu}} ^ {(m)} \text{ is unstructured and } \\ 
				& \bm{\Xi} ^ {(m)} = \operatorname{diag} \left(a_{11} ^ {(m)}\textbf{1}_{1 \times p_1}, \lambda_{\bar{p}_k} ^ {(m)}, \ldots, a_{KK} ^ {(m)}\textbf{1}_{1 \times p_K}, \lambda_{\bar{p}_K} ^ {(m)}\right), \quad m = 1, \ldots, M, \\
				\omega_2: \text{ the region containing that } & \widetilde{\bm{\nu}} ^ {(m)} \text{ is unstructured, } \quad m = 1, \ldots, M, \text{ and }  \\ 
				& \bm{\Xi} ^ {(1)} = \cdots \bm{\Xi} ^ {(M)} = \operatorname{diag} \left(a_{11} \textbf{1}_{1 \times p_1}, \lambda_{\bar{p}_k}, \ldots, a_{KK} \textbf{1}_{1 \times p_K}, \lambda_{\bar{p}_K} \right), \\
				\omega_3: \text{ the region containing that } & \widetilde{\bm{\nu}} ^ {(1)} = \cdots = \widetilde{\bm{\nu}} ^ {(M)} \text{ and } \\
				& \bm{\Xi} ^ {(1)} = \cdots \bm{\Xi} ^ {(M)} = \operatorname{diag} \left(a_{11} \textbf{1}_{1 \times p_1}, \lambda_{\bar{p}_k}, \ldots, a_{KK} \textbf{1}_{1 \times p_K}, \lambda_{\bar{p}_K} \right), 
			\end{split}
		\end{equation*}
		where $\widetilde{\bm{\nu}} ^ {(m)} \triangleq \bm{\nu} ^ {(m)} / \sqrt{n ^ {(m)}}$ for $m = 1, \ldots, M$.
		In particular, 
		under $\omega_1$, $\textbf{A} ^ {(m)} \triangleq \operatorname{diag}\left(a_{11} ^ {(m)}, \ldots, a_{KK} ^ {(m)}\right) > 0$ and $\bm{\Delta} ^ {(m)} \triangleq \textbf{A} ^ {(m)} + \textbf{B} ^ {(m)} \textbf{P} > 0$, where $\textbf{B} ^ {(m)} \triangleq \left(b_{kk'} ^ {(m)}\right)$ with $b_{k'k} ^ {(m)} = b_{kk'} ^ {(m)}$ for $k \neq k'$, 
		$\textbf{P} \triangleq \operatorname{diag}\left(p_1, \ldots, p_K\right)$, satisfying that $p = p_1 + \cdots + p_K$ and $p_k > 1$ for $k = 1, \ldots, K$ are integers, and $\bar{p}_{k} \triangleq \sum_{k' = 1}^{k} p_{k'}$ for $k = 1, \ldots, K$, and $\lambda_{\bar{p}_1} ^ {(m)}, \ldots, \lambda_{\bar{p}_K} ^ {(M)}$ are $K$ eigenvalues of $\bm{\Delta} ^ {(m)}$ in the decreasing order.
		Thus, these conditions guarantee that $\bm{\Xi} ^ {(m)} > 0$ for $m = 1, \ldots, M$. 
		Under $\omega_2$ or $\omega_3$, 
		$\textbf{A} \triangleq \operatorname{diag}\left(a_{11}, \ldots, a_{KK}\right) > 0$ and $\bm{\Delta} \triangleq \textbf{A} + \textbf{B} \textbf{P} > 0$,
		where $\textbf{B} \triangleq \left(b_{kk'}\right)$ with $b_{k'k} = b_{kk'}$ for $k \neq k'$, 
		and $\lambda_{\bar{p}_1}, \ldots, \lambda_{\bar{p}_K}$ are $K$ eigenvalues of $\bm{\Delta}$ in the decreasing order.
		Thus, these conditions guarantee that $\bm{\Xi} > 0$.
		Define $f ^ {(m)} \triangleq n ^ {(m)} / n$ for $m = 1, \ldots, M$.
		
		Additionally,
		if $\widehat{\textbf{A}} ^ {(m)} > 0$ and $\widehat{\bm{\Delta}} ^ {(m)} = \widehat{\textbf{B}} ^ {(m)} \textbf{P} > 0$ for $m = 1, \ldots, M$, and $\widehat{\textbf{A}} > 0$, $\widehat{\bm{\Delta}} \triangleq \widehat{\textbf{A}} + \widehat{\textbf{B}} \textbf{P} > 0$ with probability $1$, then the likelihood ratio between $\omega_2$ and $\omega_1$ is 
		\begin{equation*}
			\begin{split}
				& \frac{\sup\limits_{\omega_2}
					L \left(\bm{\nu} ^ {(m)}, \bm{\Xi} ^ {(m)}; \bm{Y}_1^{(m)}, \ldots, \bm{Y}_{n ^ {(m)}} ^ {(m)}, m = 1, \ldots, M\right)}
				{\sup\limits_{\omega_1}
					L \left(\bm{\nu} ^ {(m)}, \bm{\Xi} ^ {(m)}; \bm{Y}_1^{(m)}, \ldots, \bm{Y}_{n ^ {(m)}} ^ {(m)}, m = 1, \ldots, M\right)} \\
				& = \left\{\prod_{m = 1}^{M} f ^ {(m), f ^ {(m)}}\right\} ^ {- n p / 2}
				\prod_{m = 1}^{M}
				\left\{
				\prod_{k = 1}^{K}
				\frac
				{
					\textbf{V}_{\bar{p}_k, \bar{p}_k} ^ {(m)}
					\left(\sum_{j = \bar{p}_{k - 1} + 1}^{\bar{p}_k - 1} \textbf{V}_{j, j} ^ {(m)} \right) ^ {p_k - 1}}
				{
					\left(\sum_{m = 1}^{M} \textbf{V}_{\bar{p}_k, \bar{p}_k} ^ {(m)}\right) 
					\left(\sum_{j = \bar{p}_{k - 1} + 1}^{\bar{p}_k - 1} \sum_{m = 1}^{M} \textbf{V}_{j, j} ^ {(m)}\right) ^ {p_k - 1}}
				\right\} 
				^ {f ^ {(m)} n / 2};
			\end{split}
		\end{equation*}
		if $\widehat{\textbf{A}} > 0$ and $\widehat{\bm{\Delta}} = \widehat{\textbf{B}} \textbf{P} > 0$, then the likelihood ratio between $\omega_3$ and $\omega_2$ is
		\begin{equation*}
			\begin{split}
				& \frac{\sup\limits_{\omega_3}
					L \left(\bm{\nu} ^ {(m)}, \bm{\Xi} ^ {(m)}; \bm{Y}_1^{(m)}, \ldots, \bm{Y}_{n ^ {(m)}} ^ {(m)}, m = 1, \ldots, M\right)}
				{\sup\limits_{\omega_2}
					L \left(\bm{\nu} ^ {(m)}, \bm{\Xi} ^ {(m)}; \bm{Y}_1^{(m)}, \ldots, \bm{Y}_{n ^ {(m)}} ^ {(m)}, m = 1, \ldots, M\right)} \\
				& = 
				\prod_{k = 1}^{K}
				\left\{
				\frac{\sum_{m = 1}^{M} \textbf{V}_{\bar{p}_k, \bar{p}_k} ^ {(m)}}{\textbf{D}_{\bar{p}_k, \bar{p}_k}}
				\left(
				\frac{\sum_{j = \bar{p}_{k - 1} + 1}^{\bar{p}_k - 1} \sum_{m = 1}^{M} \textbf{V}_{j, j} ^ {(m)}}{\sum_{j = \bar{p}_{k - 1} + 1}^{\bar{p}_k - 1} \textbf{D}_{j, j}}
				\right) ^ {p_k - 1}
				\right\} ^ {n / 2}.
			\end{split}
		\end{equation*}
	\end{lemma}
	
	\begin{proof}[Proof of Lemma~\ref{Lem:M-sample}]
		
		Omitting the constant term $- n p \log(2\pi) / 2 - \sum_{m = 1}^{M} n ^ {(m)} p \log (n ^ {(m)}) / 2$, 
		the (kernel of the) log-likelihood function is 
		\begin{equation*}
			\begin{split}
				& \log L \left(\bm{\nu} ^ {(m)}, \bm{\Xi} ^ {(m)}; \bm{Y}_1^{(m)}, \ldots, \bm{Y}_{n ^ {(m)}} ^ {(m)}, m = 1, \ldots, M\right) \\
				& = - \sum_{m = 1}^{M} \frac{n ^ {(m)}}{2}
				\log \det \left(\bm{\Xi} ^ {(m)}\right)
				- \frac{1}{2} \sum_{m = 1}^{M} \sum_{i = 1}^{n ^ {(m)}}\left(\bm{Y}_i ^ {(m)} - \bm{\nu} ^ {(m)}\right) ^ \top \left(n ^ {(m)} \bm{\Xi} ^ {(m)}\right) ^ {- 1} \left(\bm{Y}_i ^ {(m)} - \bm{\nu} ^ {(m)}\right) \\
				& = - \sum_{m = 1}^{M} \frac{n ^ {(m)}}{2}
				\log \det \left(\bm{\Xi} ^ {(m)}\right)
				- \frac{1}{2} \sum_{m = 1}^{M} \sum_{i = 1}^{n ^ {(m)}} \operatorname{tr} \left\{
				\left(n ^ {(m)} \bm{\Xi} ^ {(m)}\right) ^ {- 1} \left(\bm{Y}_i ^ {(m)} - \bm{\nu} ^ {(m)}\right)
				\left(\bm{Y}_i ^ {(m)} - \bm{\nu} ^ {(m)}\right) ^ \top\right\} \\
				& = - \sum_{m = 1}^{M} \frac{n ^ {(m)}}{2}
				\log \det \left(\bm{\Xi} ^ {(m)}\right)
				- \frac{1}{2} \sum_{m = 1}^{M} \operatorname{tr} \left[
				\bm{\Xi} ^ {(m), - 1} \left\{
				\textbf{V} ^ {(m)}
				+ 
				\left(\bm{\nu} ^ {(m)} - \bar{\bm{Y}} ^ {(m)}\right)
				\left(\bm{\nu} ^ {(m)} - \bar{\bm{Y}} ^ {(m)}\right) ^ \top\right\}\right].
			\end{split}
		\end{equation*}
		
		Under $\omega_1$ or $\omega_2$, 
		$\widetilde{\bm{\nu}} ^ {(m)}$ or $\bm{\nu} ^ {(m)}$ is unsepcified/unstructured, 
		\begin{equation*}
			\begin{split}
				& \log L \left(\bm{\nu} ^ {(m)}, \bm{\Xi} ^ {(m)}; \bm{Y}_1^{(m)}, \ldots, \bm{Y}_{n ^ {(m)}} ^ {(m)}, m = 1, \ldots, M\right) \\
				& \leq 
				\log L \left(\bm{\nu} ^ {(m)} = \bar{\bm{Y}} ^ {(m)}, \bm{\Xi} ^ {(m)}; \bm{Y}_1^{(m)}, \ldots, \bm{Y}_{n ^ {(m)}} ^ {(m)}, m = 1, \ldots, M\right) \\
				& = 
				- \sum_{m = 1}^{M} \frac{n ^ {(m)}}{2}
				\log \det \left(\bm{\Xi} ^ {(m)}\right)
				- \frac{1}{2} \sum_{m = 1}^{M} \operatorname{tr}
				\left(\bm{\Xi} ^ {(m), - 1} \textbf{V} ^ {(m)} \right),
			\end{split}
		\end{equation*}
		where the maximum likelihood estimators $\widehat{\bm{\nu}} ^ {(m)}(\omega_1) = \widehat{\bm{\nu}} ^ {(m)}(\omega_2) \triangleq \bar{\bm{Y}} ^ {(m)}$, 
		since $n ^ {(m)} \bm{\Xi} ^ {(m)} > 0$, 
		yielding that 
		$\left(\bar{\bm{Y}} ^ {(m)} - \bm{\nu} ^ {(m)}\right) ^ \top \left(n ^ {(m)} \bm{\Xi} ^ {(m)} \right) ^ {- 1}\left(\bar{\bm{Y}} ^ {(m)} - \bm{\nu} ^ {(m)} \right) > 0$ if $\bar{\bm{Y}} ^ {(m)} \neq \bm{\nu} ^ {(m)}$; 
		and $\left(\bar{\bm{Y}} ^ {(m)} - \bm{\nu} ^ {(m)} \right) ^ \top \left(n ^ {(m)} \bm{\Xi}^ {(m)} \right) ^ {- 1}\left(\bar{\bm{Y}} ^ {(m)} - \bm{\nu} ^ {(m)} \right) = 0$ if $\bar{\bm{Y}} ^ {(m)} = \bm{\nu} ^ {(m)}$,
		for $m = 1, \ldots, M$.
		In other words, for fixed $\bm{\Xi} ^ {(1)}, \ldots, \bm{\Xi} ^ {(M)} > 0$,
		\begin{equation*}
			\begin{split}
				& \log L \left(\bm{\nu} ^ {(m)} = \bar{\bm{Y}} ^ {(m)}, \bm{\Xi} ^ {(m)}; \bm{Y}_1^{(m)}, \ldots, \bm{Y}_{n ^ {(m)}} ^ {(m)}, m = 1, \ldots, M\right) \\
				& = \sup \limits_{\bm{\nu} ^ {(1)}, \ldots, \bm{\nu} ^ {(M)} \in \mathbb{R} ^ {p \times 1}} \log L \left(\bm{\nu} ^ {(m)}, \bm{\Xi} ^ {(m)}; \bm{Y}_1^{(m)}, \ldots, \bm{Y}_{n ^ {(m)}} ^ {(m)}, m = 1, \ldots, M\right).
			\end{split}
		\end{equation*} 
		
		\textbf{\emph{Derivation under $\omega_1$}}: 
		following the lines of similar arguments for \textbf{\emph{Derivation under $\omega_2$}} in the proof of Lemma~\ref{Lem:1-sample}, 
		we obtain that $\widehat{\bm{\nu}} ^ {(m)}(\omega_1) = \bar{\bm{Y}} ^ {(m)}$, 
		$\widehat{a}_{kk} ^ {(m)}(\omega_1) = \sum_{j = \bar{p}_{k - 1} + 1}^{\bar{p}_k - 1} \textbf{V}_{j, j} ^ {(m)} / \left\{(p_k - 1) n ^ {(m)}\right\}$,
		$\widehat{\lambda}_{\bar{p}_k} ^ {(m)}(\omega_1) = \textbf{V}_{\bar{p}_k, \bar{p}_k} ^ {(m)} / n ^ {(m)}$
		for $k = 1, \ldots, K$ and $m = 1, \ldots, M$, 
		are global maximizers of the log-likelihood function, 
		and 
		\begin{equation*}
			\begin{split}
				& \sup\limits_{\omega_1}
				\log L \left(\bm{\nu} ^ {(m)}, \bm{\Xi} ^ {(m)}; \bm{Y}_1^{(m)}, \ldots, \bm{Y}_{n ^ {(m)}} ^ {(m)}, m = 1, \ldots, M\right) \\
				& = \log L \left(\bm{\nu} ^ {(m)} = \bar{\bm{Y}} ^ {(m)}, \bm{\Xi} ^ {(m)} = \widehat{\bm{\Xi}} ^ {(m)}(\omega_1); \bm{Y}_1^{(m)}, \ldots, \bm{Y}_{n ^ {(m)}} ^ {(m)}, m = 1, \ldots, M\right) \\
				& = - n p \log(2\pi) / 2 - \frac{n p}{2} + \frac{n}{2} \sum_{k = 1}^{K}
				\left(p_k - 1\right) \log (p_k - 1) \\
				& - \sum_{m = 1}^{M} \frac{n ^ {(m)}}{2} \sum_{k = 1}^{K}
				\left(p_k - 1\right) \log \left(\sum_{j = \bar{p}_{k - 1} + 1}^{\bar{p}_k - 1} \textbf{V}_{j, j} ^ {(m)} \right)
				- \sum_{m = 1}^{M} \frac{n ^ {(m)}}{2} \sum_{k = 1}^{K} \log \left(\textbf{V}_{\bar{p}_k, \bar{p}_k} ^ {(m)}\right). 
			\end{split}
		\end{equation*}
		In particular, 
		$\widehat{\bm{\Xi}} ^ {(m)}(\omega_1) = \operatorname{diag}\left(\widehat{a}_{11} ^ {(m)}(\omega_1)\textbf{1}_{1 \times p_1}, \widehat{\lambda}_{\bar{p}_1} ^ {(m)}(\omega_1), \ldots, \widehat{a}_{KK} ^ {(m)} (\omega_1)\textbf{1}_{1 \times p_K}, \widehat{\lambda}_{\bar{p}_K} ^ {(m)} (\omega_1)\right) > 0$ with probability $1$ since $\widehat{\textbf{A}} ^ {(m)} > 0$ and $\widehat{\bm{\Delta}} ^ {(m)} = \widehat{\textbf{A}} ^ {(m)} + \widehat{\textbf{B}} ^ {(m)} \textbf{P} > 0$ with probability $1$, 
		resulting in $\widehat{\bm{\Sigma}} ^ {(m)} (\omega_1) > 0$ with probability $1$, for $m = 1, \ldots, M$.
		Thus, the terms inside the logarithm are positive with probability $1$.
		
		\textbf{\emph{Derivation under $\omega_2$}}: 
		we rewrite 
		\begin{equation*}
			\log L \left(\bm{\nu} ^ {(m)} = \bar{\bm{Y}} ^ {(m)}, \bm{\Xi}; \bm{Y}_1^{(m)}, \ldots, \bm{Y}_{n ^ {(m)}} ^ {(m)}, m = 1, \ldots, M\right)
			= 
			- \frac{n}{2}
			\log \det \left(\bm{\Xi}\right)
			- \frac{1}{2} \sum_{m = 1}^{M} \operatorname{tr}
			\left(\bm{\Xi} ^ {- 1} \textbf{V} ^ {(m)} \right),
		\end{equation*}
		where $\bm{\Xi} ^ {(1)} = \cdots = \bm{\Xi} ^ {(M)} = \bm{\Xi}$.
		Again, 
		following the lines of similar arguments for \textbf{\emph{Derivation under $\omega_2$}} in the proof of Lemma~\ref{Lem:1-sample}, 
		we obtain that $\widehat{\bm{\nu}} ^ {(m)}(\omega_2) = \bar{\bm{Y}} ^ {(m)}$ for $m = 1, \ldots, M$, 
		$\widehat{a}_{kk}(\omega_2) = \sum_{j = \bar{p}_{k - 1} + 1}^{\bar{p}_k - 1} \sum_{m = 1}^{M} \textbf{V}_{j, j} ^ {(m)} / \left\{(p_k - 1) n \right\}$,
		$\widehat{\lambda}_{\bar{p}_k} (\omega_2) = \sum_{m = 1}^{M} \textbf{V}_{\bar{p}_k, \bar{p}_k} ^ {(m)} / n$
		for $k = 1, \ldots, K$, 
		are global maximizers of the log-likelihood function, 
		and 
		\begin{equation*}
			\begin{split}
				& \sup\limits_{\omega_2}
				\log L \left(\bm{\nu} ^ {(m)}, \bm{\Xi} ^ {(m)}; \bm{Y}_1^{(m)}, \ldots, \bm{Y}_{n ^ {(m)}} ^ {(m)}, m = 1, \ldots, M\right) \\
				& = \log L \left(\bm{\nu} ^ {(m)} = \bar{\bm{Y}} ^ {(m)}, \bm{\Xi} ^ {(m)} = \widehat{\bm{\Xi}} (\omega_2); \bm{Y}_1 ^ {(m)}, \ldots, \bm{Y}_{n ^ {(m)}} ^ {(m)}, m = 1, \ldots, M\right) \\
				& = - n p \log(2\pi) / 2 - \sum_{m = 1}^{M} n ^ {(m)} p \log (n ^ {(m)}) / 2 - \frac{np}{2} 
				+ \frac{n}{2} \sum_{k = 1}^{K}
				\left(p_k - 1\right) \log \left((p_k - 1) \right)
				+ n p \log \left(n\right) / 2 \\
				& - \frac{n}{2} \sum_{k = 1}^{K}
				\left(p_k - 1\right) \log \left(\sum_{j = \bar{p}_{k - 1} + 1}^{\bar{p}_k - 1} \sum_{m = 1}^{M} \textbf{V}_{j, j} ^ {(m)}\right)
				- \frac{n}{2} \sum_{k = 1}^{K} \log \left(\sum_{m = 1}^{M} \textbf{V}_{\bar{p}_k, \bar{p}_k} ^ {(m)}\right).
			\end{split}
		\end{equation*}
		In particular, 
		$\widehat{\bm{\Xi}}(\omega_2) = \operatorname{diag}\left(\widehat{a}_{11} (\omega_2)\textbf{1}_{1 \times p_1}, \widehat{\lambda}_{\bar{p}_1} (\omega_2), \ldots, \widehat{a}_{KK} (\omega_2) \textbf{1}_{1 \times p_K}, \widehat{\lambda}_{\bar{p}_K} (\omega_2)\right) > 0$ with probability $1$ since $\widehat{\textbf{A}} > 0$ and $\widehat{\bm{\Delta}} = \widehat{\textbf{A}} + \widehat{\textbf{B}} \textbf{P} > 0$ with probability $1$, 
		resulting in $\widehat{\bm{\Sigma}} (\omega_2) > 0$ with probability $1$, for $m = 1, \ldots, M$.
		Thus, the terms inside the logarithm are positive with probability $1$.
		
		\textbf{\emph{Derivation under $\omega_3$}}: 
		we rewrite $\bm{\nu} ^ {(1)} / \sqrt{n ^ {(1)}} = \cdots = \bm{\nu} ^ {(M)} / \sqrt{n ^ {(M)}} = \widetilde{\bm{\nu}}$ and $\bm{\Xi} ^ {(1)} = \cdots = \bm{\Xi} ^ {(M)} = \bm{\Xi}$ in the (kernel of the) log-likelihood function, 
		\begin{equation*}
			\begin{split}
				& \log L \left(\widetilde{\bm{\nu}}, \bm{\Xi}; \bm{Y}_1^{(m)}, \ldots, \bm{Y}_{n ^ {(m)}} ^ {(m)}, m = 1, \ldots, M\right) \\
				& = - \sum_{m = 1}^{M} \frac{n}{2}
				\log \det \left(\bm{\Xi} \right)
				- \frac{1}{2} \operatorname{tr} \left[
				\bm{\Xi} ^ {- 1} \left\{
				\sum_{m = 1}^{M} \textbf{V} ^ {(m)}
				+ 
				\sum_{m = 1}^{M} \left(\sqrt{n ^ {(m)}} \widetilde{\bm{\nu}} - \bar{\bm{Y}} ^ {(m)}\right)
				\left(\sqrt{n ^ {(m)}} \widetilde{\bm{\nu}} - \bar{\bm{Y}} ^ {(m)}\right) ^ \top\right\}\right] \\
				& = - \sum_{m = 1}^{M} \frac{n}{2}
				\log \det \left(\bm{\Xi} \right)
				- \frac{1}{2} \operatorname{tr}
				\left\{\bm{\Xi} ^ {- 1} \left(\sum_{m = 1}^{M} \textbf{V} ^ {(m)}\right)\right\} \\
				& - \frac{1}{2} \operatorname{tr}
				\left[
				\bm{\Xi} ^ {- 1} \left\{
				\sum_{m = 1}^{M} \left(\bar{\bm{Y}} ^ {(m)} - \sqrt{n ^ {(m)}} \bar{\bm{Y}}\right)
				\left(\bar{\bm{Y}} ^ {(m)} - \sqrt{n ^ {(m)}} \bar{\bm{Y}}\right) ^ \top
				+ 
				n \left(\bar{\bm{Y}} - \widetilde{\bm{\nu}}\right)
				\left(\bar{\bm{Y}} - \widetilde{\bm{\nu}}\right) ^ \top\right\}\right],
			\end{split}
		\end{equation*}
		using the result in Lemma~\ref{Lem:decompose},
		where $\bar{\bm{Y}} \triangleq \left(\sqrt{n ^ {(1)}} \bar{\bm{Y}} ^ {(1)} + \cdots + \sqrt{n ^ {(M)}} \bar{\bm{Y}} ^ {(M)} \right) / n \in \mathbb{R} ^ {p \times 1}$.
		Therefore, 
		\begin{equation*}
			\begin{split}
				& \log L \left(\widetilde{\bm{\nu}}, \bm{\Xi}; \bm{Y}_1^{(m)}, \ldots, \bm{Y}_{n ^ {(m)}} ^ {(m)}, m = 1, \ldots, M\right) \\
				& \leq 
				\log L \left(\widetilde{\bm{\nu}} = \bar{\bm{Y}}, \bm{\Xi}; \bm{Y}_1^{(m)}, \ldots, \bm{Y}_{n ^ {(m)}} ^ {(m)}, m = 1, \ldots, M\right) \\
				& = - \sum_{m = 1}^{M} \frac{n}{2}
				\log \det \left(\bm{\Xi} \right)
				- \frac{1}{2} \operatorname{tr}
				\left\{\bm{\Xi} ^ {- 1} \left(\sum_{m = 1}^{M} \textbf{V} ^ {(m)} + \textbf{H}\right)\right\},
			\end{split}
		\end{equation*}
		where $\textbf{H} \triangleq \sum_{m = 1}^{M} \left(\bar{\bm{Y}} ^ {(m)} - \sqrt{n ^ {(m)}} \bar{\bm{Y}}\right)
		\left(\bar{\bm{Y}} ^ {(m)} - \sqrt{n ^ {(m)}} \bar{\bm{Y}}\right) ^ \top \in \mathbb{R} ^ {p \times p}$, 
		and the maximum likelihood estimator $\widehat{\widetilde{\bm{\nu}}} (\omega_3) = \bar{\bm{Y}}$, 
		since $\bm{\Xi} > 0$, 
		yielding that 
		$\left(\bar{\bm{Y}} - \widetilde{\bm{\nu}} \right) ^ \top \left(\bm{\Xi} / n \right) ^ {- 1} \left(\bar{\bm{Y}} - \widetilde{\bm{\nu}} \right) > 0$ if $\bar{\bm{Y}} \neq \widetilde{\bm{\nu}}$; 
		and $\left(\bar{\bm{Y}} - \widetilde{\bm{\nu}} \right) ^ \top \left(\bm{\Xi} / n \right) ^ {- 1} \left(\bar{\bm{Y}} - \widetilde{\bm{\nu}} \right) = 0$ if $\bar{\bm{Y}} = \widetilde{\bm{\nu}}$,
		for $m = 1, \ldots, M$.
		In other words, for fixed $\bm{\Xi} > 0$,
		\begin{equation*}
			\begin{split}
				& \log L \left(\widetilde{\bm{\nu}} = \bar{\bm{Y}}, \bm{\Xi}; \bm{Y}_1^{(m)}, \ldots, \bm{Y}_{n ^ {(m)}} ^ {(m)}, m = 1, \ldots, M\right) \\
				& = \sup \limits_{\widetilde{\bm{\nu}} \in \mathbb{R} ^ {p \times 1}} \log L \left(\widetilde{\bm{\nu}}, \bm{\Xi}; \bm{Y}_1^{(m)}, \ldots, \bm{Y}_{n ^ {(m)}} ^ {(m)}, m = 1, \ldots, M\right).
			\end{split}
		\end{equation*} 
		
		Let $\textbf{D} \triangleq \sum_{m = 1}^{M} \textbf{V} ^ {(m)} + \textbf{H} \in \mathbb{R} ^ {p \times p}$.
		Again, 
		following the lines of similar arguments for \textbf{\emph{Derivation under $\omega_2$}} in the proof of Lemma~\ref{Lem:1-sample}, 
		we obtain that $\widehat{\widetilde{\bm{\nu}}}(\omega_3) = \bar{\bm{Y}}$, 
		$\widehat{a}_{kk}(\omega_3) = \sum_{j = \bar{p}_{k - 1} + 1}^{\bar{p}_k - 1} \textbf{D}_{j, j} / \left\{(p_k - 1) n \right\}$,
		$\widehat{\lambda}_{\bar{p}_k} (\omega_3) =  \textbf{D}_{\bar{p}_k, \bar{p}_k} / n$
		for $k = 1, \ldots, K$, 
		are global maximizers of the log-likelihood function, 
		and 
		\begin{equation*}
			\begin{split}
				& \sup\limits_{\omega_3}
				\log L \left(\bm{\nu} ^ {(m)}, \bm{\Xi} ^ {(m)}; \bm{Y}_1^{(m)}, \ldots, \bm{Y}_{n ^ {(m)}} ^ {(m)}, m = 1, \ldots, M\right) \\
				& = \log L \left(\bm{\nu} ^ {(m)} = \sqrt{n ^ {(m)}} \bar{\bm{Y}}, \bm{\Xi} ^ {(m)} = \widehat{\bm{\Xi}} (\omega_3); \bm{Y}_1 ^ {(m)}, \ldots, \bm{Y}_{n ^ {(m)}} ^ {(m)}, m = 1, \ldots, M\right) \\
				& = - n p \log(2\pi) / 2 - \sum_{m = 1}^{M} n ^ {(m)} p \log (n ^ {(m)}) / 2 - \frac{n p}{2} 
				+ \frac{n}{2} \sum_{k = 1}^{K}
				\left(p_k - 1\right) \log \left((p_k - 1) \right)
				+ n p \log \left( n\right) / 2 \\
				& - \frac{n}{2} \sum_{k = 1}^{K}
				\left(p_k - 1\right) \log \left(\sum_{j = \bar{p}_{k - 1} + 1}^{\bar{p}_k - 1} \textbf{D}_{j, j}\right)
				- \frac{n}{2} \sum_{k = 1}^{K} \log \left(\textbf{D}_{\bar{p}_k, \bar{p}_k}\right).
			\end{split}
		\end{equation*}
		In particular, 
		$\widehat{\bm{\Xi}}(\omega_3) = \operatorname{diag}\left(\widehat{a}_{11} (\omega_3)\textbf{1}_{1 \times p_1}, \widehat{\lambda}_{\bar{p}_1} (\omega_3), \ldots, \widehat{a}_{KK} (\omega_3) \textbf{1}_{1 \times p_K}, \widehat{\lambda}_{\bar{p}_K} (\omega_3)\right) > 0$ with probability $1$ since $\widehat{\textbf{A}} > 0$ and $\widehat{\bm{\Delta}} = \widehat{\textbf{A}} + \widehat{\textbf{B}} \textbf{P} > 0$ with probability $1$, 
		resulting in $\widehat{\bm{\Sigma}} (\omega_3) > 0$ with probability $1$.
		Thus, the terms inside the logarithm are positive with probability $1$.
		
		Lastly, the likelihood ratio between $\omega_2$ and $\omega_1$ is 
		\begin{equation*}
			\begin{split}
				& \frac{\sup\limits_{\omega_2}
					L \left(\bm{\nu} ^ {(m)}, \bm{\Xi} ^ {(m)}; \bm{Y}_1^{(m)}, \ldots, \bm{Y}_{n ^ {(m)}} ^ {(m)}, m = 1, \ldots, M\right)}
				{\sup\limits_{\omega_1}
					L \left(\bm{\nu} ^ {(m)}, \bm{\Xi} ^ {(m)}; \bm{Y}_1^{(m)}, \ldots, \bm{Y}_{n ^ {(m)}} ^ {(m)}, m = 1, \ldots, M\right)} \\
				& = \left\{\frac{\prod_{m = 1}^{M} n ^ {(m), f ^ {(m)}} }{n}\right\} ^ {- n p / 2} 
				\frac{\exp\left\{
					- \frac{n}{2} \log \prod_{k = 1}^{K}
					\left(\sum_{m = 1}^{M} \textbf{V}_{\bar{p}_k, \bar{p}_k} ^ {(m)}\right)
					\left(\sum_{j = \bar{p}_{k - 1} + 1}^{\bar{p}_k - 1} \sum_{m = 1}^{M} \textbf{V}_{j, j} ^ {(m)}\right) ^ {p_k - 1}
					\right\}}
				{\exp\left\{ 
					- \frac{n}{2} \log \prod_{m = 1}^{M} \prod_{k = 1}^{K}
					\left(\textbf{V}_{\bar{p}_k, \bar{p}_k} ^ {(m)}\right) ^ {f ^ {(m)}}
					\left(\sum_{j = \bar{p}_{k - 1} + 1}^{\bar{p}_k - 1} \textbf{V}_{j, j} ^ {(m)} \right) ^ {(p_k - 1)f^{(m)}}  
					\right\}} \\
				& = \left\{\prod_{m = 1}^{M} f ^ {(m), f ^ {(m)}}\right\} ^ {- n p / 2}
				\prod_{m = 1}^{M}
				\left\{
				\prod_{k = 1}^{K}
				\frac
				{
					\textbf{V}_{\bar{p}_k, \bar{p}_k} ^ {(m)}
					\left(\sum_{j = \bar{p}_{k - 1} + 1}^{\bar{p}_k - 1} \textbf{V}_{j, j} ^ {(m)} \right) ^ {p_k - 1}}
				{
					\left(\sum_{m = 1}^{M} \textbf{V}_{\bar{p}_k, \bar{p}_k} ^ {(m)}\right) 
					\left(\sum_{j = \bar{p}_{k - 1} + 1}^{\bar{p}_k - 1} \sum_{m = 1}^{M} \textbf{V}_{j, j} ^ {(m)}\right) ^ {p_k - 1}}
				\right\} 
				^ {f ^ {(m)} n / 2}.
			\end{split}
		\end{equation*}
		
		The likelihood ratio between $\omega_3$ and $\omega_2$ is
		\begin{equation*}
			\begin{split}
				& \frac{\sup\limits_{\omega_3}
					L \left(\bm{\nu} ^ {(m)}, \bm{\Xi} ^ {(m)}; \bm{Y}_1^{(m)}, \ldots, \bm{Y}_{n ^ {(m)}} ^ {(m)}, m = 1, \ldots, M\right)}
				{\sup\limits_{\omega_2}
					L \left(\bm{\nu} ^ {(m)}, \bm{\Xi} ^ {(m)}; \bm{Y}_1^{(m)}, \ldots, \bm{Y}_{n ^ {(m)}} ^ {(m)}, m = 1, \ldots, M\right)} \\
				& = 
				\frac{
					\exp\left\{
					- \frac{n}{2} \log \prod_{k = 1}^{K}
					\textbf{D}_{\bar{p}_k, \bar{p}_k}
					\left(\sum_{j = \bar{p}_{k - 1} + 1}^{\bar{p}_k - 1} \textbf{D}_{j, j}\right) ^ {p_k - 1}
					\right\}
				}
				{
					\exp\left\{
					- \frac{n}{2} \log \prod_{k = 1}^{K}
					\left(\sum_{m = 1}^{M} \textbf{V}_{\bar{p}_k, \bar{p}_k} ^ {(m)}\right)
					\left(\sum_{j = \bar{p}_{k - 1} + 1}^{\bar{p}_k - 1} \sum_{m = 1}^{M} \textbf{V}_{j, j} ^ {(m)}\right) ^ {p_k - 1} \right\}
				} \\
				& = 
				\prod_{k = 1}^{K}
				\frac
				{\left\{ 
					\left(\sum_{m = 1}^{M} \textbf{V}_{\bar{p}_k, \bar{p}_k} ^ {(m)}\right)
					\left(\sum_{j = \bar{p}_{k - 1} + 1}^{\bar{p}_k - 1} \sum_{m = 1}^{M} \textbf{V}_{j, j} ^ {(m)}\right) ^ {p_k - 1}\right\} ^ {n / 2}
				}
				{\left\{
					\textbf{D}_{\bar{p}_k, \bar{p}_k}
					\left(\sum_{j = \bar{p}_{k - 1} + 1}^{\bar{p}_k - 1} \textbf{D}_{j, j}\right) ^ {p_k - 1}\right\} ^ {n / 2}
				}.
			\end{split}
		\end{equation*}
		We complete the proof.
	\end{proof}
		
	\begin{lemma}[Craig's Theorem]
		\label{Lem:independent}
		Suppose $\bm{X}$ is a $p$-dimensional normal vector with mean $\bm{\mu} \in \mathbb{R} ^ p$ and positive definite covariance matrix $\bm{\Sigma} \in \mathbb{R} ^ {p \times p}$.
		Let $\textbf{A}_1, \textbf{A}_2 \in \mathbb{R} ^ {p \times p}$ be two real-valued symmetric matrices.
		Then, $\bm{X} ^ \top \textbf{A}_1 \bm{X}$ and $\bm{X} ^ \top \textbf{A}_2 \bm{X}$ are independently distributed if and only if $\textbf{A}_1 \bm{\Sigma} \textbf{A}_2 = \textbf{0}_{p \times p}$.
	\end{lemma}
	
	\begin{proof}[Proof of Lemma~\ref{Lem:independent}]
		This result was originally proposed by \citet{Craig1943}. 
		Please refer the corrected proof and discussion in \citet[page 175, proof of Result 5.4.4]{Ravishanker2002}, \citet[page 209, proof of Theorem 5.2.1]{MathaiProvost1992}, \citet{DriscollGundberg1986} and \citet{OgawaOlkin2008}.
	\end{proof}
	
	\begin{lemma}[Distributions of quadratic form of multivariate normal vectors]
		\label{Lem:chisquare}
		Suppose $\bm{X}$ is a $p$-dimensional normal vector with mean $\bm{\mu} \in \mathbb{R} ^ p$ and positive definite covariance matrix $\bm{\Sigma} \in \mathbb{R} ^ {p \times p}$.
		Let $\textbf{A} \in \mathbb{R} ^ {p \times p}$ be a real-valued symmetric matrix.
		Then, $\bm{X} ^ \top \textbf{A} \bm{X}$ follows a non-central chi-squared distribution with degree of freedom $m$ and noncentrality parameter $\Omega = \bm{\mu} ^ \top  \textbf{A}  \bm{\mu}$ if and only if $\textbf{A} \bm{\Sigma}$ is idempotent with rank $m$.
	\end{lemma}
	
	\begin{proof}[Proof of Lemma~\ref{Lem:chisquare}]
		Please refer the proof and discussion in \citet[page 31, proof of Theorem 1.4.5]{Muirhead2005}, \citet[page 199, proof of Theorem 5.1.3]{MathaiProvost1992} and \citet{Zhang2018}.
	\end{proof}
	
	Now, we start to prove the main results in Section 3.
	
	\begin{proof}[Proof of Theorem 1 (LRT statistics related to covariance structures)]
		
		For simplicity, we suppress the superscript $(1)$, i.e., 
		$\bar{\bm{X}} \triangleq \bar{\bm{X}} ^ {(1)}$,
		$\bm{\mu} \triangleq \bm{\mu} ^ {(1)}$, 
		$\textbf{W} \triangleq \textbf{W} ^ {(1)}$,
		$\bm{\Sigma} \triangleq \bm{\Sigma} ^ {(1)}$,  
		$\bm{\Sigma}\left[\textbf{A}, \textbf{B}, \bm{p}\right] \triangleq \bm{\Sigma}\left[\textbf{A} ^ {(1)}, \textbf{B} ^ {(1)}, \bm{p}\right]$, 
		where $\textbf{A} \triangleq \operatorname{diag}\left(a_{11}, \ldots, a_{KK} \right) \triangleq \textbf{A} ^ {(1)}$,
		$\textbf{B} \triangleq \left(b_{k k'} \right) \triangleq \textbf{B} ^ {(1)}$ with $b_{k k'} = b_{k' k}$ for $k \neq k'$,
		and $\bm{p} \triangleq \left(p_1, \ldots, p_K\right) ^ \top$ with $p \triangleq p_1 + \cdots + p_K$.
		
		Given $\bm{\Sigma}\left[\textbf{A}, \textbf{B}, \bm{p}\right]$ in $H_1$, 
		there exists an orthogonal matrix $\bm{\Gamma}_1 \in \mathbb{R} ^ {p \times p}$, 
		determined by $\bm{\Delta} \triangleq \textbf{A} + \textbf{B} \textbf{P}$ with $\textbf{P} \triangleq \operatorname{diag}\left(p_1, \ldots, p_K\right)$, and $\bm{p}$ only, 
		satisfying that the diagonal matrix $\bm{\Gamma}_1 \bm{\Sigma} \left[\textbf{A}, \textbf{B}, \bm{p} \right] \bm{\Gamma}_1 ^ \top = \operatorname{diag}\left(a_{11} \textbf{1}_{1 \times p_{1}}, \lambda_{\bar{p}_1}, \ldots, a_{KK} \textbf{1}_{1 \times p_{K}}, \lambda_{\bar{p}_K} \right)$ is the canonical form of $\bm{\Sigma} \left[\textbf{A}, \textbf{B}, \bm{p}\right]$.
		We define that 
		$\bm{Y} \triangleq \sqrt{n} \bm{\Gamma}_1 \bar{\bm{X}}$,
		$\bm{\nu} \triangleq \sqrt{n} \bm{\Gamma}_1 \bm{\mu}$, 
		$\textbf{V} \triangleq \bm{\Gamma}_1 \textbf{W} \bm{\Gamma}_1 ^ \top$, 
		and 
		$\bm{\Xi} \triangleq \bm{\Gamma}_1 \bm{\Sigma} \bm{\Gamma}_1 ^ \top$ (for a general form of $\bm{\Sigma}$). 
		
		Following the first part of Lemma~\ref{Lem:1-sample}, 
		we obtain the formula of the likelihood ratio statistic $\Lambda_1 ^ {2 / n}$, raised to the power of $2 / n$, is computed as below, 
		\begin{equation*}
			\Lambda_1 ^ {2 / n}
			= \prod_{k = 1}^{K} \left\{(p_k - 1) ^ {p_k - 1}\right\}
			\dfrac{\det\left( \textbf{V} \right)}
			{\prod_{k = 1}^{K} \left\{
				\textbf{V}_{\bar{p}_k, \bar{p}_k}
				\left(\sum_{j = \bar{p}_{k - 1} + 1}^{\bar{p}_{k} - 1} \textbf{V}_{j, j} \right) ^ {p_k - 1}
				\right\}}.
		\end{equation*}
		Furthermore, 
		we define  
		\begin{equation*}
			\begin{split}
				\Lambda_1 ^ {(0)} = \left\{\frac{\det\left(\textbf{V}\right)}{\prod_{j = 1}^{p} \textbf{V}_{j, j}}\right\} ^ {n / 2}, 
				\Lambda_1 ^ {(k)} = \left\{
				(p_k - 1) ^ {p_k - 1} 
				\frac{\prod_{j = \bar{p}_{k-1} + 1}^{\bar{p}_k - 1} \textbf{V}_{j, j}}
				{\left( \sum_{j' = \bar{p}_{k - 1} + 1}^{\bar{p}_k - 1} \textbf{V}_{j',j'} \right) ^ {p_k - 1}}
				\right\} ^ {n / 2}, \quad k = 1, 2, \dots, K,
			\end{split}
		\end{equation*}
		satisfying that $\Lambda_1 = \prod_{k = 0}^{K} \Lambda_1 ^ {(k)}$.
		Under $H_1$, 
		applying 
		the results of (2.4) and (2.5) in \citet{MarquesCoelhoArnold2011}, 
		we conclude that
		\begin{equation*}
			\Lambda_1 ^ {(0)} \sim 
			\prod_{j = 1}^{p - 1} \beta_j ^ {n / 2},  \quad 
			\beta_j \sim \operatorname{Beta}\left(\frac{n - 1}{2} - \frac{p - j}{2}, \frac{p - j}{2}\right), 
			\quad j = 1, \ldots, p - 1.
		\end{equation*}
		Applying the results of (2.13) and (2.14) in \citet{MarquesCoelhoArnold2011}, 
		we conclude that  
		\begin{equation*}
			\Lambda_1 ^ {(k)} \sim \prod_{j = 2}^{p_k - 1} \beta_{j, k} ^ {n / 2},  \quad 
			\beta_{j, k} \sim \operatorname{Beta}\left(\frac{n - 1}{2}, \frac{j - 1}{p_k - 1}\right), \quad j = 2, \ldots, p_k - 1; k = 1, \ldots, K.
		\end{equation*}
		The above beta variates are mutually independent across $j$ and $k$.
		
		We complete the proof of the first part of Theorem 1, 
		where the additional conditions, $n > p$ and $p_k > 2$ for $k = 1, \ldots, K$, guarantee that $\textbf{V} > 0$ holds with probability $1$.
		
		
		Given $\bm{\Sigma} \left[\textbf{A}, \textbf{B}, \bm{p}\right] \triangleq \bm{\Sigma} ^ {(1)}\left[\textbf{A} ^ {(1)}, \textbf{B} ^ {(1)}, \bm{p}\right] = \cdots = \bm{\Sigma} ^ {(M)}\left[\textbf{A} ^ {(M)}, \textbf{B} ^ {(M)}, \bm{p}\right]$, i.e., the common covariance matrix in $H_3$, 
		with $\textbf{A} \triangleq \operatorname{diag}\left(a_{11}, \ldots, a_{KK}\right) \triangleq \textbf{A} ^ {(1)} = \cdots = \textbf{A} ^ {(M)}$, 
		$\textbf{B} \triangleq \left(b_{k k'}\right) \triangleq \textbf{B} ^ {(1)} = \cdots = \textbf{B} ^ {(M)}$ with $b_{kk'} = b_{k' k}$ for $k \neq k'$, 
		and $\bm{p} = \left(p_1, \ldots, p_K\right) ^ \top$ with $p = p_1 + \cdots + p_K$, 
		there exists an orthogonal matrix $\bm{\Gamma}_3 \in \mathbb{R} ^ {p \times p}$, 
		determined by $\textbf{A}$, $\textbf{B}$, $\bm{p}$ only, 
		satisfying that the diagonal matrix $\bm{\Gamma}_3 \bm{\Sigma}\left[\textbf{A}, \textbf{B}, \bm{p}\right] \bm{\Gamma}_3 ^ \top = \operatorname{diag}\left(a_{11} \textbf{1}_{1 \times p_{1}}, \lambda_{\bar{p}_1}, \ldots, a_{KK} \textbf{1}_{1 \times p_{K}}, \lambda_{\bar{p}_K} \right)$ is the canonical form of $\bm{\Sigma} \left[\textbf{A}, \textbf{B}, \bm{p}\right]$.
		We define that 
		$\bm{Y} ^ {(m)} \triangleq \sqrt{n ^ {(m)}}\bm{\Gamma}_3 \bar{\bm{X}} ^ {(m)}$,
		$\bm{\nu} ^ {(m)} \triangleq \sqrt{n ^ {(m)}} \bm{\Gamma}_3 \bm{\mu} ^ {(m)}$,  
		$\textbf{V} ^ {(m)} \triangleq \bm{\Gamma}_3 \textbf{W} ^ {(m)} \bm{\Gamma}_3 ^ \top$, 
		$\bm{\Xi} \triangleq \bm{\Gamma}_3 \bm{\Sigma} \left[\textbf{A}, \textbf{B}, \bm{p}\right] \bm{\Gamma}_3 ^ \top$, 
		and $\bm{\Xi} ^ {(m)} \triangleq \bm{\Gamma}_3 \bm{\Sigma} ^ {(m)} \bm{\Gamma}_3 ^ \top$ (for a general form of $\bm{\Sigma} ^ {(m)}$).
		
		Following the first part of Lemma~\ref{Lem:M-sample}, we obtain the formula of the likelihood ratio statistic $\Lambda_3 ^ {2 / n}$, raised to the power of $2 / n$, is computed as below,
		\begin{equation*}
			\Lambda_3 ^ {2 / n}
			= 
			\left\{\prod_{m = 1}^{M} f ^ {(m), f ^ {(m)}}\right\} ^ {- p}
			\prod_{m = 1}^{M}
			\left\{
			\prod_{k = 1}^{K}
			\frac
			{
				\textbf{V}_{\bar{p}_k, \bar{p}_k} ^ {(m)}
				\left(\sum_{j = \bar{p}_{k - 1} + 1}^{\bar{p}_k - 1} \textbf{V}_{j, j} ^ {(m)} \right) ^ {p_k - 1}}
			{
				\left(\sum_{m = 1}^{M} \textbf{V}_{\bar{p}_k, \bar{p}_k} ^ {(m)}\right) 
				\left(\sum_{j = \bar{p}_{k - 1} + 1}^{\bar{p}_k - 1} \sum_{m = 1}^{M} \textbf{V}_{j, j} ^ {(m)}\right) ^ {p_k - 1}}
			\right\} 
			^ {f ^ {(m)}}.
		\end{equation*}
		
		Furthermore, we derive the distributions for the above components. 
		Note that a Wishart distribution $W_p\left(\bm{\Sigma}, n\right)$ will be singular when $n < p$, hence, the density of a singular Wishart distribution is not available with respect to Lebesgue measure. 
		However, we just employ Lemma~\ref{Lem:transform}, which is available even if $n < p$.
		
		Specifically, 
		under $H_3$, 
		on the one hand, 
		taking element vector $\textbf{e}_k$ as $\textbf{A}$ with rank $1$ in Lemma~\ref{Lem:decompose}, we obtain that 
		$\textbf{V}_{\bar{p}_k, \bar{p}_k} ^ {(m)} \sim W_1\left(\bm{\Xi}_{\bar{p}_k, \bar{p}_k}, n ^ {(m)} - 1 \right)$,
		which is non-singular if $\textbf{V} ^ {(m)} > 0$, 
		requiring that $n ^ {(m)} - 1 \geq 1$ and both $\widehat{\textbf{A}} ^ {(m)}$ and $\widehat{\bm{\Delta}} ^ {(m)}$ are positive definite with probability $1$. 
		Using the independence,  
		we obtain that 
		$\sum_{m' \neq m} \textbf{V}_{\bar{p}_k, \bar{p}_k} ^ {(m')} \sim W_1 \left(\bm{\Xi}_{\bar{p}_k, \bar{p}_k}, n - n ^ {(m)} - (M - 1) \right)$, 
		which is non-singular, if $\textbf{V} ^ {(m')} > 0$, 
		requiring that $n - n ^ {(m)} - (M - 1) \geq 1$ and all $\widehat{\textbf{A}} ^ {(m')}$ and $\widehat{\bm{\Delta}} ^ {(m')}$ are positive definite with probability $1$ for $m' \neq m$,
		and is independently distributed of $\textbf{V}_{\bar{p}_k, \bar{p}_k} ^ {(m)}$.
		Therefore,
		\begin{equation*}
			\begin{split}
				\frac{\textbf{V}_{\bar{p}_k, \bar{p}_k} ^ {(m)}}{\sum_{m' = 1} ^ {M} \textbf{V}_{\bar{p}_k, \bar{p}_k} ^ {(m')}}
				& = \frac{\textbf{V}_{\bar{p}_k, \bar{p}_k} ^ {(m)}}{\textbf{V}_{\bar{p}_k, \bar{p}_k} ^ {(m)} + \sum_{m' \neq m} \textbf{V}_{\bar{p}_k, \bar{p}_k} ^ {(m')}} \\
				& \sim \frac{\bm{\Xi}_{\bar{p}_k, \bar{p}_k} \chi ^ 2 \left(n ^ {(m)} - 1 \right)}
				{\bm{\Xi}_{\bar{p}_k, \bar{p}_k} \chi ^ 2 \left(n ^ {(m)} - 1 \right) + 
					\bm{\Xi}_{\bar{p}_k, \bar{p}_k} \chi ^ 2 \left(n - n ^ {(m)} - (M - 1) \right)} \\
				& \sim 
				\operatorname{Beta}
				\left(\frac{n ^ {(m)} - 1}{2}, \frac{n - n ^ {(m)} - (M - 1)}{2} \right).
			\end{split}
		\end{equation*}
		where $\chi ^ 2(\cdot)$ denotes the chi-squared variate with the degrees of freedom parameter.
		
		On the other hand, 
		let $\bm{\Xi}_{j(k), j(k)}$ denote the common element within $\bar{p}_{k - 1} - 1 \leq j \leq \bar{p}_{k} - 1$, i.e., $a_{k k}$.
		So, 
		using the independence between diagonal elements due to a diagonal scale matrix $\bm{\Xi}$ and applying Lemma~\ref{Lem:transform} again, 
		we obtain that 
		$\sum_{j = \bar{p}_{k - 1} + 1} ^ {\bar{p}_{k} - 1} \textbf{V}_{j, j} ^ {(m)} \sim W_1\left(\bm{\Xi}_{j(k), j(k)}, (p_k - 1)(n ^ {(m)} - 1) \right)$ and $\sum_{j = \bar{p}_{k - 1} + 1} ^ {\bar{p}_{k} - 1} \sum_{m = 1}^{M} \textbf{V}_{j, j} ^ {(m)} \sim W_1 \left(\bm{\Xi}_{j(k), j(k)}, (p_k - 1)(n - M) \right)$,
		which are non-singular, 
		if $\textbf{V} ^ {(m)} > 0$, 
		requiring that $(p_k - 1)(n ^ {(m)} - 1) \geq 1$, $(p_k - 1)(n - M) \geq 1$ and all $\widehat{\textbf{A}} ^ {(m)}$ and $\widehat{\bm{\Delta}} ^ {(m)}$ are positive definite with probability $1$,
		and independently distributed.
		Therefore,
		\begin{equation*}
			\begin{split}
				& \frac{\sum_{j = \bar{p}_{k - 1} + 1} ^ {\bar{p}_{k} - 1} \textbf{V}_{j, j} ^ {(m)}}
				{\sum_{j = \bar{p}_{k - 1} + 1} ^ {\bar{p}_{k} - 1} \sum_{m' = 1}^{M} \textbf{V}_{j, j} ^ {(m')}}
				=
				\frac{\sum_{j = \bar{p}_{k - 1} + 1}^{\bar{p}_{k} - 1} \textbf{V}_{j,j} ^ {(m)}}
				{\sum_{j = \bar{p}_{k - 1} + 1}^{\bar{p}_{k} - 1} \textbf{V}_{j,j} ^ {(m)} + 
					\sum_{j = \bar{p}_{k - 1} + 1}^{\bar{p}_{k} - 1}
					\sum_{m' \neq m}
					\textbf{V}_{j,j} ^ {(m')}} \\
				& \sim 
				\frac{\bm{\Xi}_{j(k), j(k)} \chi ^ 2 \left((p_k - 1)(n ^ {(m)} - 1) \right)}
				{\bm{\Xi}_{j(k), j(k)} \chi ^ 2 \left((p_k - 1)(n ^ {(m)} - 1) \right) + \bm{\Xi}_{j(k), j(k)} \chi ^ 2 \left((p_k - 1) \left(n - n ^ {(m)} - (M - 1) \right) \right)} \\
				& \sim
				\operatorname{Beta}
				\left(
				\frac{(p_k - 1)\left(n ^ {(m)} - 1 \right)}{2}, 
				\frac{(p_k - 1)\left(n - n ^ {(m)} - (M - 1) \right)}{2}
				\right).
			\end{split}
		\end{equation*}
		Combining the above results and the relationship between the Dirichlet and beta variates, we complete the proof of the second part of Theorem 1.
	\end{proof}
	
	\begin{proof}[Proof of Theorem 2 (LRT statistics related to mean vectors)]
		
		For simplicity, we suppress the superscript $(1)$, i.e.,
		$\bar{\bm{X}} \triangleq \bar{\bm{X}} ^ {(1)}$, 
		$\bm{\mu} \triangleq \bm{\mu} ^ {(1)}$,
		$\textbf{W} \triangleq \textbf{W} ^ {(1)}$, 
		$\bm{\Sigma}\left[\textbf{A}, \textbf{B}, \bm{p}\right] \triangleq \bm{\Sigma}\left[\textbf{A} ^ {(1)}, \textbf{B} ^ {(1)}, \bm{p}\right]$ with $\textbf{A} \triangleq \operatorname{diag}\left(a_{11}, \ldots, a_{KK} \right) \triangleq \textbf{A} ^ {(1)}$,
		$\textbf{B} \triangleq \left(b_{k k'} \right) \triangleq\textbf{B} ^ {(1)}$ with $b_{k k'} = b_{k' k}$ for $k \neq k'$,
		and $\bm{p} = \left(p_1, \ldots, p_K\right) ^ \top$ with $p = p_1 + \cdots + p_K$.
		
		Given $\bm{\Sigma}\left[\textbf{A}, \textbf{B}, \bm{p}\right]$ in $H_2$, 
		there exists an orthogonal matrix $\bm{\Gamma}_2 \in \mathbb{R} ^ {p \times p}$, 
		determined by $\textbf{A}$, $\textbf{B}$, and $\bm{p}$ only, 
		satisfying that the diagonal matrix $\bm{\Gamma}_2 \bm{\Sigma}\left[\textbf{A}, \textbf{B}, \bm{p}\right] \bm{\Gamma}_2 ^ \top = \operatorname{diag}\left(a_{11} \textbf{1}_{1 \times p_{1}}, \lambda_{\bar{p}_1}, \ldots, a_{KK} \textbf{1}_{1 \times p_{K}}, \lambda_{\bar{p}_K} \right)$ is the canonical form of $\bm{\Sigma} \left[\textbf{A}, \textbf{B}, \bm{p}\right]$.
		We define that 
		$\bm{Y} \triangleq \sqrt{n} \bm{\Gamma}_2 \bar{\bm{X}}$, 
		$\bm{\nu} \triangleq \sqrt{n} \bm{\Gamma}_2 \bm{\mu}$,
		$\bm{\nu}_0 \triangleq \sqrt{n} \bm{\Gamma}_2 \bm{\mu}_0$,  
		$\bm{V} \triangleq \bm{\Gamma}_2 \textbf{W} \bm{\Gamma}_2 ^ \top$, 
		and 
		$\bm{\Xi} \triangleq \bm{\Gamma}_2 \bm{\Sigma}\bm{\Gamma}_2 ^ \top$ (for a general form of $\bm{\Sigma}$). 
		
		Following the second part of Lemma~\ref{Lem:1-sample}, 
		we obtain the formula of the likelihood ratio statistic $\Lambda_2 ^ {2 / n}$, raised to the power of $2 / n$, is computed as below, 
		\begin{equation*}
			\Lambda_2 ^ {2 / n} 
			= \prod_{k = 1}^{K}
			\dfrac{\textbf{V}_{\bar{p}_k, \bar{p}_k}
				\left(\sum_{j = \bar{p}_{k - 1} + 1}^{\bar{p}_{k} - 1} \textbf{V}_{j, j} \right) ^ {p_k - 1}}
			{\left(\textbf{H}_{\bar{p}_k, \bar{p}_k} + \textbf{V}_{\bar{p}_k, \bar{p}_k}\right)
				\left(\sum_{j = \bar{p}_{k - 1} + 1}^{\bar{p}_{k} - 1} \left(\textbf{H}_{j,j} + \textbf{V}_{j,j}\right) \right) ^ {p_k - 1}},	
		\end{equation*}
		where $\textbf{H} \triangleq \left(\bm{Y} - \bm{\nu}_0\right) \left(\bm{Y} - \bm{\nu}_0\right) ^ \top > 0$ if $\bm{Y} \neq \bm{\nu}_0$.
		
		Under $H_2$ with a general $\bm{\nu}_{0}$,
		let the noncentrality parameter $\bm{\Omega} \triangleq \left\{\operatorname{E}(\bm{Y}) - \bm{\nu}_0\right\} \left\{\operatorname{E}(\bm{Y}) - \bm{\nu}_0\right\} ^ \top = \left(\bm{\nu} - \bm{\nu}_0\right)\left(\bm{\nu} - \bm{\nu}_0\right) ^ \top$.
		Then, applying Lemma~\ref{Lem:transform}, 
		we obtain that  
		$\textbf{V}_{\bar{p}_k, \bar{p}_k} \sim W_1 \left(\bm{\Xi}_{\bar{p}_k, \bar{p}_k}, n - 1 \right)$ for every $k$, 
		$\sum_{j = \bar{p}_{k - 1} + 1}^{\bar{p}_{k} - 1} \textbf{V}_{j, j} \sim W_1 \left(\bm{\Xi}_{j(k), j(k)}, (n - 1)(p_k - 1) \right)$, here $\bm{\Xi}_{j(k), j(k)}$ is the same $a_{k k}$ across $j$ for every $k$,
		which are non-singular if $\textbf{V} > 0$, 
		requiring that $n - 1 \geq 1$, $(n - 1)(p_k - 1) > 1$, and both $\widehat{\textbf{A}}$ and $\widehat{\bm{\Delta}}$ are positive definite with probability $1$.
		Additionally,
		\begin{equation*}
			\begin{split}
				\textbf{H}_{\bar{p}_k, \bar{p}_k} & \sim W_1 \left(\bm{\Xi}_{\bar{p}_k, \bar{p}_k}, 1; \left(\bm{\Xi} ^ {- 1/2} \bm{\Omega} \bm{\Xi} ^ {- 1/2}\right)_{\bar{p}_k, \bar{p}_k} \right), \\
				\sum_{j = \bar{p}_{k - 1} + 1}^{\bar{p}_{k} - 1} \textbf{H}_{j, j} & \sim W_1 \left(\bm{\Xi}_{j(k), j(k)}, p_k - 1; \sum_{j = \bar{p}_{k - 1} + 1}^{\bar{p}_{k} - 1} \left(\bm{\Xi} ^ {- 1/2} \bm{\Omega} \bm{\Xi} ^ {- 1/2}\right)_{j, j} \right),
			\end{split}
		\end{equation*} 
		for $k \in \{1, \ldots, K\}$,
		here $\bm{\Xi}_{j(k), j(k)}$ is the same $a_{k k}$ across $j$ for every $k$, 
		which are non-singular if $\textbf{V} > 0$, 
		requiring that $p_k - 1 > 1$, 
		and both $\widehat{\textbf{A}}$ and $\widehat{\bm{\Delta}}$ are positive definite with probability $1$,
		and they are mutually independently distributed.
		Thus, 
		\begin{equation*}
			\begin{split}
				\frac{\textbf{V}_{\bar{p}_k, \bar{p}_k}}{\textbf{H}_{\bar{p}_k, \bar{p}_k} + \textbf{V}_{\bar{p}_k, \bar{p}_k}} 
				& \sim 
				\frac{\bm{\Xi}_{\bar{p}_k, \bar{p}_k} \chi ^ 2 (n - 1)}{\bm{\Xi}_{\bar{p}_k, \bar{p}_k} \chi ^ 2 \left(1; \left(\bm{\Xi} ^ {- 1/2} \bm{\Omega} \bm{\Xi} ^ {- 1/2}\right)_{\bar{p}_k, \bar{p}_k}\right) + \bm{\Xi}_{\bar{p}_k, \bar{p}_k} \chi ^ 2 (n - 1)} \\
				& \sim \operatorname{Beta}_{\text{II}} \left(\frac{n - 1}{2}, \frac{1}{2}; \left(\bm{\Xi} ^ {- 1/2} \bm{\Omega} \bm{\Xi} ^ {- 1/2}\right)_{\bar{p}_k, \bar{p}_k} \right), \\
				\frac{\sum_{j = \bar{p}_{k - 1} + 1}^{\bar{p}_{k} - 1} \textbf{V}_{j, j}}{\sum_{j = \bar{p}_{k - 1} + 1}^{\bar{p}_{k} - 1} \left(\textbf{H}_{j, j} + \textbf{V}_{j,j}\right)} 
				& \sim
				\frac{\bm{\Xi}_{j(k), j(k)} \chi ^ 2 ((n - 1)(p_k - 1))}{\bm{\Xi}_{j(k), j(k)} \chi ^ 2 \left(p_k - 1; \sum_{j = \bar{p}_{k - 1} + 1}^{\bar{p}_{k} - 1} \left(\bm{\Xi} ^ {- 1/2} \bm{\Omega} \bm{\Xi} ^ {- 1/2}\right)_{j, j}\right) + \bm{\Xi}_{j(k), j(k)} \chi ^ 2 ((n - 1)(p_k - 1))} \\
				& \sim \operatorname{Beta}_{\text{II}}\left( \frac{(n - 1)(p_k - 1)}{2}, \frac{p_k - 1}{2}; \sum_{j = \bar{p}_{k - 1} + 1}^{\bar{p}_{k} - 1} \left(\bm{\Xi} ^ {- 1/2} \bm{\Omega} \bm{\Xi} ^ {- 1/2}\right)_{j, j} \right), \quad k = 1, \ldots, K,
			\end{split}
		\end{equation*}
		where $\chi ^ 2 \left(\cdot; \cdot\right)$ denotes the noncentral chi-squared variate and $\operatorname{Beta}_{\text{II}}(\cdot, \cdot; \cdot)$ denotes the type II noncentral beta variate.
		Alternatively, it is easy to show that $\Lambda_2 ^ {- 2 / n}$ follows a product of mutually independent $F$ variates.
		
		Under $H_2$, $\bm{\nu} = \bm{\nu}_0$, the above noncentrality parameters are $0$.
		We complete the proof of the first part of Theorem 2.

		Given $\bm{\mu} \triangleq \bm{\mu} ^ {(1)} = \cdots = \bm{\mu} ^ {(M)}$, i.e., the common mean vector in $H_4$, 
		and $\bm{\Sigma} \left[\textbf{A}, \textbf{B}, \bm{p}\right] \triangleq \bm{\Sigma} ^ {(1)}\left[\textbf{A} ^ {(1)}, \textbf{B} ^ {(1)}, \bm{p}\right] = \cdots = \bm{\Sigma} ^ {(M)}\left[\textbf{A} ^ {(M)}, \textbf{B} ^ {(M)}, \bm{p}\right]$, i.e., the common covariance matrix in $H_4$, 
		with $\textbf{A} \triangleq \operatorname{diag}\left(a_{11}, \ldots, a_{KK}\right) \triangleq \textbf{A} ^ {(1)} = \cdots = \textbf{A} ^ {(M)}$,
		$\textbf{B} \triangleq \left(b_{k k'}\right) \triangleq \textbf{B} ^ {(1)} = \cdots = \textbf{B} ^ {(M)}$ with $b_{kk'} = b_{k' k}$ for $k \neq k'$, 
		and $\bm{p} = \left(p_1, \ldots, p_K\right) ^ \top$ with $p = p_1 + \cdots + p_K$, 
		there exists an orthogonal matrix $\bm{\Gamma}_4 \in \mathbb{R} ^ {p \times p}$, 
		determined by $\textbf{A}$, $\textbf{B}$, $\bm{p}$ only, 
		satisfying that $\bm{\Xi} = \operatorname{diag}\left(a_{11} \textbf{1}_{1 \times p_{1}}, \lambda_{\bar{p}_1}, \ldots, a_{KK} \textbf{1}_{1 \times p_{K}}, \lambda_{\bar{p}_K} \right)$ is the canonical form of $\bm{\Sigma} \left[\textbf{A}, \textbf{B}, \bm{p}\right]$.
		We define that 
		$\bm{Y} ^ {(m)} \triangleq \sqrt{n ^ {(m)}}\bm{\Gamma}_4 \bar{\bm{X}} ^ {(m)}$,
		$\widetilde{\bm{\nu}} ^ {(m)} \triangleq \bm{\Gamma}_4 \bm{\mu} ^ {(m)}$, 
		$\bm{\nu} ^ {(m)} \triangleq \sqrt{n ^ {(m)}} \widetilde{\bm{\nu}} ^ {(m)}$, 
		$\textbf{V} ^ {(m)} \triangleq \bm{\Gamma}_4 \textbf{W} ^ {(m)} \bm{\Gamma}_4 ^ \top$, 
		and $\bm{\Xi} ^ {(m)} \triangleq \bm{\Gamma}_4 \bm{\Sigma} ^ {(m)} \bm{\Gamma}_4 ^ \top$ (for a general form of $\bm{\Sigma} ^ {(m)}$).
		
		Following the second part of Lemma~\ref{Lem:M-sample}, 
		we obtain the formula of the likelihood ratio statistic $\Lambda_4 ^ {2 / n}$, raised to the power of $2 / n$, is computed as below,
		\begin{equation*}
			\Lambda_4 ^ {2 / n}
			= \prod_{k = 1}^{K}
			\left\{
			\frac{\sum_{m = 1}^{M} \textbf{V}_{\bar{p}_k, \bar{p}_k} ^ {(m)}}{\textbf{D}_{\bar{p}_k, \bar{p}_k}}
			\left(
			\frac{\sum_{j = \bar{p}_{k - 1} + 1}^{\bar{p}_k - 1} \sum_{m = 1}^{M} \textbf{V}_{j, j} ^ {(m)}}{\sum_{j = \bar{p}_{k - 1} + 1}^{\bar{p}_k - 1} \textbf{D}_{j, j}}
			\right) ^ {p_k - 1}
			\right\},
		\end{equation*}
		where $\textbf{D} \triangleq \sum_{m = 1}^{M} \textbf{V} ^ {(m)} + \textbf{H} \in \mathbb{R} ^ {p \times p}$,		
		$\textbf{H} \triangleq \sum_{m = 1}^{M} \left(\bm{Y} ^ {(m)} - \sqrt{n ^ {(m)}} \bar{\bm{Y}} \right)\left(\bm{Y} ^ {(m)} - \sqrt{n ^ {(m)}} \bar{\bm{Y}} \right) ^ \top$, 
		and $\bar{\bm{Y}} = \left(\sum_{m = 1}^{M} \sqrt{n ^ {(m)}} \bm{Y} ^ {(m)} \right) / n$.
		
		Given the above explicit expression of $\Lambda_4$,
		under $H_4$ with unstructured $\widetilde{\bm{\nu}} ^ {(m)}$,  
		we observe that $\bm{Y} ^ {(m)} \sim N_p \left(\sqrt{n ^ {(m)}} \widetilde{\bm{\nu}} ^ {(m)}, \bm{\Xi} \right)$.
		By the second part of Lemma~\ref{Lem:decompose}, 
		$\operatorname{rank}\left(\textbf{A}_Y\right) = M - 1$ with $\textbf{A}_{Y} \triangleq \textbf{I}_M - \left(\sqrt{n ^ {(1)}}, \ldots, \sqrt{n ^ {(M)}}\right) ^ \top \left(\sqrt{n ^ {(1)}}, \ldots, \sqrt{n ^ {(M)}}\right) / n = \textbf{I}_M - \left(\sqrt{f ^ {(1)}}, \ldots, \sqrt{f ^ {(M)}}\right) ^ \top \left(\sqrt{f ^ {(1)}}, \ldots, \sqrt{f ^ {(M)}}\right)$, 
		where $\left(\sqrt{f ^ {(1)}}, \ldots, \sqrt{f ^ {(M)}}\right) ^ \top \left(\sqrt{f ^ {(1)}}, \ldots, \sqrt{f ^ {(M)}}\right)$ is a rank-one matrix; 
		and the noncentrality parameter $\bm{\Omega} \triangleq \textbf{M}_Y \textbf{A}_Y \textbf{M}_Y ^ \top$ with $\textbf{M}_Y \triangleq \left(\operatorname{E} \left(\bm{Y} ^ {(1)}\right), \ldots, \operatorname{E}\left(\bm{Y} ^ {(M)}\right) \right) = \left(\bm{\nu} ^ {(1)}, \ldots, \bm{\nu} ^ {(M)}\right)$.
		So, $\textbf{H} \sim W_p \left(\bm{\Xi}, M - 1; \bm{\Omega} \right)$, 
		Furthermore, $\sum_{m = 1}^{M} \textbf{V} ^ {(m)} \sim W_p\left(\bm{\Xi}, n - M\right)$.
		Fixing $k$ and applying Lemma~\ref{Lem:transform}, we obtain that 
		\begin{equation*}
			\begin{split}
				\frac{\sum_{m = 1}^{M} \textbf{V}_{\bar{p}_k, \bar{p}_k} ^ {(m)}}{\textbf{D}_{\bar{p}_k, \bar{p}_k}}
				& \sim \frac{\bm{\Xi}_{\bar{p}_k, \bar{p}_k} \chi ^ 2 \left(n - M \right)}
				{\bm{\Xi}_{\bar{p}_k, \bar{p}_k} \chi ^ 2 \left(n - M \right) + 
					\bm{\Xi}_{\bar{p}_k, \bar{p}_k} \chi ^ 2 \left(M - 1; \left(\bm{\Xi} ^ {- 1/2} \bm{\Omega} \bm{\Xi} ^ {- 1/2} \right)_{\bar{p}_k, \bar{p}_k} \right)} \\
				& \sim 
				\operatorname{Beta}_{\text{II}} \left( \frac{n - M}{2}, \frac{M - 1}{2}; \left(\bm{\Xi} ^ {- 1/2} \bm{\Omega} \bm{\Xi} ^ {- 1/2} \right)_{\bar{p}_k, \bar{p}_k} \right), 
			\end{split}
		\end{equation*}
		where 
		the two Wishart variates in the first term are non-singular if $\textbf{V} > 0$, 
		requiring that $M - 1 > 1$, $n - M > 1$, 
		and both $\widehat{\textbf{A}}$ and $\widehat{\bm{\Delta}}$ are positive definite with probability $1$,
		$\chi ^ 2 (\cdot; \cdot)$ denotes the noncentral chi-squared variate and $\operatorname{Beta}_{\text{II}}(\cdot, \cdot; \cdot)$ denotes the type II noncentral beta variates, 
		and 
		\begin{equation*}
			\begin{split}
				& \frac{\sum_{j = \bar{p}_{k - 1} + 1}^{\bar{p}_k - 1} \sum_{m = 1}^{M} \textbf{V}_{j, j} ^ {(m)}}{\sum_{j = \bar{p}_{k - 1} + 1}^{\bar{p}_k - 1} \textbf{D}_{j, j}} \\
				& \sim \frac{\bm{\Xi}_{j(k), j(k)} \chi ^ 2 \left( \left(n - M \right)(p_k - 1) \right)}
				{\bm{\Xi}_{j(k), j(k)} \chi ^ 2 \left(\left(n - M \right)(p_k - 1) \right) + 
					\bm{\Xi}_{j(k), j(k)} \chi ^ 2 \left(\left(M - 1 \right)(p_k - 1); \sum_{j = \bar{p}_{k - 1} + 1}^{\bar{p}_k - 1} \left(\bm{\Xi} ^ {- 1/2} \bm{\Omega} \bm{\Xi} ^ {- 1/2} \right)_{j, j} \right)} \\
				& \sim 
				\operatorname{Beta}_{\text{II}}
				\left(
				\frac{\left(n - M \right)(p_k - 1)}{2}, 
				\frac{\left(M - 1 \right)(p_k - 1)}{2};
				\sum_{j = \bar{p}_{k - 1} + 1}^{\bar{p}_k - 1} \left(\bm{\Xi} ^ {- 1/2} \bm{\Omega} \bm{\Xi} ^ {- 1/2} \right)_{j, j}
				\right),
			\end{split}
		\end{equation*}
		where the two Wishart variates in the first term are non-singular if $\textbf{V} > 0$, 
		requiring that $(n - M)(p_k - 1) > 1$, $(M - 1)(p_k - 1) > 1$,  
		and both $\widehat{\textbf{A}}$ and $\widehat{\bm{\Delta}}$ are positive definite with probability $1$.
		Alternatively, it is easy to show that $\Lambda_4 ^ {- 2 / n}$ follows a product of mutually independent $F$ variates.
		
		Under $H_4$ with $\widetilde{\bm{\nu}} ^ {(1)} = \cdots = \widetilde{\bm{\nu}} ^ {(M)}$, 
		the noncentrality parameters are $0$.
		We complete the proof of the second part of Theorem 2.
	\end{proof}
	
	\begin{proof}[Proof of Theorem 3 (Geisser's information statistics related to mean vectors)]
		
		The proof of the first part of Theorem 3 can be accomplished through three steps.
		First, we need to decompose $G_2$ into a sum of several components.
		Second, we need to examine the independence of these components.
		Third, we need to determine the distribution of each component.
		
		\textbf{\emph{Step 1: decomposition of $G_2$}}. 
		The following equality demonstrates an elegant way to decompose $G_2$:
		\begin{equation*}
			\left(\widehat{\textbf{A}} \textbf{P} \right) ^ {- 1} - \widehat{\bm{\Delta}} ^ {- 1} \widehat{\textbf{B}} \widehat{\textbf{A}} ^ {- 1} = \left(\textbf{P} \widehat{\bm{\Delta}} \right) ^ {- 1}.
		\end{equation*}
		The above equality holds if and only if $\textbf{P} ^ {-1} \widehat{\textbf{A}} ^ {-1} - \widehat{\bm{\Delta}} ^ {-1} \widehat{\textbf{B}} \widehat{\textbf{A}} ^ {- 1} = \widehat{\bm{\Delta}} ^ {-1} \textbf{P} ^ {- 1}$.
		First, right-multiple $\widehat{\textbf{A}}$ on both sides and switch $\textbf{P} ^ {- 1}$ and $\widehat{\textbf{A}}$ on the right hand side since both $\widehat{\textbf{A}}$ and $\textbf{P}$ are diagonal matrices, yielding that The above equality holds if and only if $\textbf{P} ^ {- 1} - \widehat{\bm{\Delta}} ^ {- 1} \widehat{\textbf{B}} = \widehat{\bm{\Delta}} ^ {- 1} \widehat{\textbf{A}} \textbf{P} ^ {- 1}$.
		Then, right-multiply $\textbf{P}$ on both sides. 
		The above equality holds if and only if $\textbf{I}_K - \widehat{\bm{\Delta}} ^ {- 1} \widehat{\textbf{B}} \textbf{P} = \widehat{\bm{\Delta}} ^ {-1} \widehat{\textbf{A}}$.
		Next, left-multiply $\widehat{\bm{\Delta}}$ on both sides, yielding that the above equality holds if and only if $\widehat{\bm{\Delta}} - \widehat{\textbf{B}} \textbf{P} = \widehat{\textbf{A}}$,
		which holds by the definition of $\widehat{\bm{\Delta}}$.
		
		Using The above equality and substituting $\widehat{\textbf{A}}_{\bm{\Theta}}$ and $\widehat{\textbf{B}}_{\bm{\Theta}}$, we have 
		\begin{equation*}
			\widehat{\bm{\Theta}}\left[\widehat{\textbf{A}}_{\bm{\Theta}}, \widehat{\textbf{B}}_{\bm{\Theta}}, \bm{p}\right]
			= \widehat{\textbf{A}} ^ {- 1} \circ \textbf{I}[\bm{p}] - \left(\widehat{\textbf{A}} \textbf{P} \right) ^ {-1} \circ \textbf{J}[\bm{p}]
			+ \left(\textbf{P} \widehat{\bm{\Delta}} \right) ^ {-1} \circ \textbf{J}[\bm{p}].
		\end{equation*}
		where $\widehat{\textbf{A}} ^ {- 1} \circ \textbf{I}[\bm{p}] - \left(\widehat{\textbf{A}} \textbf{P} \right) ^ {- 1} \circ \textbf{J}[\bm{p}] = \operatorname{Bdiag} \left(\widehat{a}_{11} ^ {- 1} \textbf{I}_{p_1} - \widehat{a}_{11}^{-1}p_{1}^{-1} \textbf{J}_{p_1}, \ldots, \widehat{a}_{KK} ^ {- 1} \textbf{I}_{p_K} - \widehat{a}_{KK}^{-1}p_{K}^{-1} \textbf{J}_{p_K}\right)$.
		
		Let $\textbf{W}_k \triangleq \operatorname{Bdiag}\left(\textbf{0}_{p_1 \times p_1}, \ldots, \widehat{a}_{kk} ^ {-1} \textbf{I}_{p_k} - \widehat{a}_{kk} ^ {-1} p_k ^ {-1} \textbf{J}_{p_k}, \ldots, \textbf{0}_{p_K \times p_K}\right)$ for every $k$, 
		which is a uniform-block matrix, expressed by $\textbf{W}_k = \textbf{W}_k \left[\textbf{A}_k, \textbf{B}_k, \bm{p} \right]$ for every $k$,
		where both $\textbf{A}_k \triangleq \operatorname{diag}\left(0, \ldots, \widehat{a}_{kk} ^ {-1}, \ldots, 0 \right)$ and $\textbf{B}_{k} \triangleq \operatorname{diag}\left(0, \ldots, - \widehat{a}_{kk} ^ {- 1} p_k ^ {-1}, \ldots, 0 \right)$ have the non-zero values on the $(k, k)$th elements.
		Herein, $G_2$ can be written as a sum of $(K+1)$ components:
		\begin{equation*}
			\begin{split}
				G_2 
				& = n  \left(\bar{\bm{X}} - \bm{\mu}_0 \right) ^ \top  \left(
				\sum_{k = 1}^{K} \textbf{W}_k
				\right)  \left(\bar{\bm{X}} - \bm{\mu}_0 \right) 
				+ n  \left(\bar{\bm{X}} - \bm{\mu}_0 \right) ^ \top  \left\{
				\left(\textbf{P} \widehat{\bm{\Delta}} \right) ^ {- 1} \circ \textbf{J}[\bm{p}]
				\right\}  \left(\bar{\bm{X}} - \bm{\mu}_0 \right) \\
				& = \sum_{k = 1}^{K} n \left(\bar{\bm{X}} - \bm{\mu}_0 \right) ^ \top \textbf{W}_k \left(\bar{\bm{X}} - \bm{\mu}_0 \right)
				+ n \left(\bar{\bm{X}} - \bm{\mu}_0 \right) ^ \top \left\{
				\left(\textbf{P} \widehat{\bm{\Delta}} \right) ^ {- 1} \circ \textbf{J}[\bm{p}]
				\right\} \left(\bar{\bm{X}} - \bm{\mu}_0 \right) \\
				& \triangleq F_1 + F_2 + \cdots + F_K + F_{K + 1},
			\end{split}
		\end{equation*}
		where $F_k$ denotes the last term $n \left(\bar{\bm{X}} - \bm{\mu}_0 \right) ^ \top \textbf{W}_k \left(\bar{\bm{X}} - \bm{\mu}_0 \right)$, for $k \in \{1, \ldots, K\}$, and $F_{K + 1}$ denotes $n \left(\bar{\bm{X}} - \bm{\mu}_0 \right) ^ \top \left\{
		\left(\textbf{P} \widehat{\bm{\Delta}} \right) ^ {- 1} \circ \textbf{J}[\bm{p}] \right\} \left(\bar{\bm{X}} - \bm{\mu}_0 \right)$.
		Next, we will show that these $F_k$ are mutually independent in Step 2 and each follows an $F$-distribution in Step 3.
		
		\textbf{\emph{Step 2: independence}}.
		Recall 
		$\left(\bar{\bm{X}} - \bm{\mu}_0\right)$ follows $N_p \left(\bm{\mu} - \bm{\mu}_0, n ^ {-1} \bm{\Sigma} \left[\textbf{A}, \textbf{B}, \bm{p}\right] \right)$.
		To show $F_1, \ldots, F_{K + 1}$ are mutually independent, 
		by Lemma \ref{Lem:independent}, 
		we need to check $\textbf{W}_k \left(n ^ {-1} \bm{\Sigma} \left[\textbf{A}, \textbf{B}, \bm{p} \right] \right) \textbf{W}_{k'} = \textbf{0}_{p \times p}$ for every $k \neq k'$ and $\textbf{W}_k \left(n ^ {-1} \bm{\Sigma} \left[\textbf{A}, \textbf{B}, \bm{p} \right] \right) \left\{\left(\textbf{P} \widehat{\bm{\Delta}}\right) ^ {- 1} \circ \textbf{J}[\bm{p}]\right\} = \textbf{0}_{p \times p}$ for every $k$.
		It is easy to verify whether the former holds using the representation of $\textbf{W}_k = \textbf{W}_k \left[\textbf{A}_k, \textbf{B}_k, \bm{p}\right]$:
		\begin{equation*}
			\begin{split}
				& \textbf{W}_k \bm{\Sigma} \left[\textbf{A}, \textbf{B}, \bm{p} \right]  \textbf{W}_{k'} \\
				& = \left\{
				\left(\textbf{A}_k \textbf{A} \right) \circ \textbf{I}[\bm{p}] + 
				\left(\textbf{A}_k \textbf{B} + \textbf{B}_k \textbf{A} + \textbf{B}_k \textbf{P} \textbf{B} \right) \circ \textbf{J}[\bm{p}]
				\right\} 
				\left(\textbf{A}_{k'} \circ \textbf{I}[\bm{p}] + \textbf{B}_{k'} \circ \textbf{J}[\bm{p}] \right) \\
				& = \left(\textbf{A}_k \textbf{A} \textbf{A}_{k'}\right) \circ \textbf{I}[\bm{p}] 
				+ \left\{ \left(\textbf{A}_k \textbf{B} + \textbf{B}_k \textbf{A} + \textbf{B}_k \textbf{P} \textbf{B} \right) \textbf{A}_{k'} + \textbf{A}_k \textbf{A} \textbf{B}_{k'} 
				+ \left(\textbf{A}_k \textbf{B} + \textbf{B}_k \textbf{A} + \textbf{B}_k \textbf{P} \textbf{B} \right) \textbf{P} \textbf{B}_{k'} \right\} \circ \textbf{J}[\bm{p}] \\
				& = \textbf{0}_{K \times K} \circ \textbf{I}[\bm{p}] + \textbf{0}_{K \times K} \circ \textbf{J}[\bm{p}] \\
				& = \textbf{0}_{p \times p}.
			\end{split}
		\end{equation*}
		To check whether the latter one holds, 
		we use $\textbf{A}_k + \textbf{B}_{k} \textbf{P} = \textbf{0}_{K \times K}$, then
		\begin{equation*}
			\begin{split}
				\textbf{W}_k \bm{\Sigma} \left[\textbf{A}, \textbf{B}, \bm{p}\right]  \left\{\left(\textbf{P} \widehat{\bm{\Delta}}\right) ^ {-1} \circ \textbf{J}[\bm{p}]\right\} 
				& = \left\{
				\textbf{A}_k \textbf{A} \left(\textbf{P} \widehat{\bm{\Delta}}\right) ^ {-1} 
				+ \left(\textbf{A}_k \textbf{B} + \textbf{B}_k \textbf{A} + \textbf{B}_k \textbf{P} \textbf{B} \right) \textbf{P} \left(\textbf{P} \widehat{\bm{\Delta}}\right) ^ {-1}
				\right\} \circ \textbf{J}[\bm{p}] \\
				& = \left[
				\left\{ \left(\textbf{A}_k + \textbf{B}_k \textbf{P}\right) \textbf{A} 
				+ \left(\textbf{A}_k + \textbf{B}_k \textbf{P}\right) \textbf{B} \textbf{P} \right\}
				\left(\textbf{P} \widehat{\bm{\Delta}}\right) ^ {-1}
				\right] \circ \textbf{J}[\bm{p}] \\
				& = \textbf{0}_{K \times K} \circ \textbf{J}[\bm{p}] \\
				& = \textbf{0}_{p \times p}.
			\end{split}
		\end{equation*}
		
		\textbf{\emph{Step 3: distribution}}.
		Now, we specify the distributions for $F_1, \ldots, F_{K + 1}$,  respectively.		
		Furthermore, let $\bar{\bm{X}} \triangleq \left(\bar{\bm{X}} ^ {(1), \top}, \ldots, \bar{\bm{X}} ^ {(K), \top}\right) ^ \top$ and $\bm{\mu}_0 \triangleq \left(\bm{\mu}_0 ^ {(1), \top}, \ldots, \bm{\mu}_0 ^ {(K), \top}\right) ^ \top$, 
		where $\bar{\bm{X}} ^ {(k)}, \bm{\mu}_0 ^ {(k)} \in \mathbb{R} ^ {p_k}$ for every $k$.
		Focus on the first $K$ component, i.e., $F_k$ for $k \in \{1, \ldots, K\}$.
		Substituting the block diagonal matrix $\textbf{W}_k$, we have
		\begin{equation*}
			F_k 
			= n \left(\bar{\bm{X}} - \bm{\mu}_0\right) ^ \top \textbf{W}_k \left(\bar{\bm{X}} - \bm{\mu}_0 \right) 
			= \left(\bar{\bm{X}} ^ {(k)} - \bm{\mu}_0 ^ {(k)} \right) ^ \top 
			\left(
			n \widehat{a}_{kk} ^ {- 1} \textbf{I}_{p_k} - 
			n \widehat{a}_{kk} ^ {- 1} p_k ^ {- 1} \textbf{J}_{p_k} \right)
			\left(\bar{\bm{X}} ^ {(k)} - \bm{\mu}_0 ^ {(k)} \right).
		\end{equation*}
		Let $\textbf{U}_k \triangleq n \left(\bar{\bm{X}} ^ {(k)} - \bm{\mu}_0 ^ {(k)} \right) ^ \top \left\{ (1 / a_{kk}) \textbf{I}_{p_k} - 1 / (a_{kk} p_k) \textbf{J}_{p_k}\right\} \left(\bar{\bm{X}} ^ {(k)} - \bm{\mu}_0 ^ {(k)} \right)$ and $\textbf{V}_k \triangleq (p_k - 1)(n - 1) (\widehat{a}_{kk} / a_{kk})$ for every $k$.
		Thus, $\textbf{U}_k / \textbf{V}_k = F_k / \{(p_k - 1)(n - 1)\}$ for every $k$.
		Let $\textbf{M}_k = \textbf{I}_{p_k} - \textbf{J}_{p_k} / p_k$ for every $k$.
		It is clear that $\textbf{M}_k ^ 2 = \left(\textbf{I}_{p_k} - \textbf{J}_{p_k} / p_k \right) ^ 2 
		= \textbf{I}_{p_k} + \textbf{J}_{p_k} / p_k - 2 \textbf{J}_{p_k} / p_k = \textbf{M}_k$ therefore $\textbf{M}_k$ is idempotent.
		We can observe that 
		\begin{equation*}
			\begin{split}
				\operatorname{tr} \left(\textbf{M}_k \textbf{S}_{kk} \textbf{M}_k\right)
				& = \operatorname{tr} \left( \textbf{M}_k \textbf{S}_{kk} \right) 
				= \operatorname{tr} \left( \textbf{S}_{kk} - \textbf{1}_{p_k \times 1} \textbf{1}_{p_k \times 1} ^ \top \textbf{S}_{kk} p_{k} ^ {- 1} \right) 
				= \operatorname{tr} \left( \textbf{S}_{kk} \right) - \operatorname{tr}\left(\textbf{1}_{p_k \times 1} ^ \top \textbf{S}_{kk} \textbf{1}_{p_k \times 1} \right) p_{k} ^ {- 1} \\
				& = \operatorname{tr} \left(\textbf{S}_{kk}\right) - \operatorname{sum}\left(\textbf{S}_{kk}\right) p_{k} ^ {- 1} 
				= (p_k - 1) \widehat{a}_{kk}.
			\end{split}
		\end{equation*}
		Using Lemma~\ref{Lem:transform}, we obtain that $(n - 1) \textbf{S}_{kk} \sim W_{p_k} \left(\bm{\Sigma}_{kk}, n - 1\right)$,
		where $\bm{\Sigma}_{kk}$ is positive definite because $\left(\textbf{0}_{1 \times p_1}, \ldots, \bm{\alpha} ^ \top, \ldots, \textbf{0}_{1 \times p_K} \right) \bm{\Sigma} \left(\textbf{0}_{1 \times p_1}, \ldots, \bm{\alpha} ^ \top, \ldots, \textbf{0}_{1 \times p_K} \right) ^ \top = \bm{\alpha} ^ \top \bm{\Sigma}_{kk} \bm{\alpha} > 0$ for any $\bm{\alpha} \in \mathbb{R} ^ {p_k}$ and every $k$, 
		$(n - 1) \textbf{S}_{kk}$ can be expressed as $\sum_{j = 1}^{n - 1} \bm{Z}_j ^ {(k)} \bm{Z}_j ^ {(k), \top}$ by Definition~\ref{Def:wishart}, where $\bm{Z}_1 ^ {(k)}, \ldots, \bm{Z}_{n - 1} ^ {(k)}$ are mutually independently distributed as $N_{p_k} \left(\textbf{0}_{p_k \times 1}, \bm{\Sigma}_{kk} \right)$ for every $k$.
		We then observe that 
		\begin{equation*}
			\begin{split}
				a_{kk} \textbf{V}_{k} 
				& = (n - 1) (p_k - 1) \widehat{a}_{kk} 
				= \operatorname{tr} \left(\textbf{M}_k \left\{(n - 1)\textbf{S}_{kk}\right\} \textbf{M}_k \right) 
				= \operatorname{tr} \left(\textbf{M}_k \left(\sum_{j = 1}^{n - 1} \bm{Z}_j ^ {(k)} \bm{Z}_j ^ {(k), \top}\right) \textbf{M}_k \right) \\
				& = \sum_{j = 1}^{n - 1} \bm{Z}_j ^ {(k), \top} \textbf{M}_k \bm{Z}_j ^ {(k)}.
			\end{split}
		\end{equation*}
		
		On the one hand, $\bar{\bm{X}}$ and $\textbf{S}$ are independent, 
		so $\textbf{U}_k$ and $\textbf{V}_k$ for every $k$ are mutually independent.
		On the other hand, $\sqrt{n}\left(\bar{\bm{X}} ^ {(k)} - \bm{\mu}_0 ^ {(k)} \right) \sim N_{p_k} \left(\bm{\mu} ^ {(k)} - \bm{\mu}_0 ^ {(k)}, \bm{\Sigma}_{kk} \right)$.
		By Lemma~\ref{Lem:chisquare} and 
		\begin{equation*}
			\left(a_{kk} ^ {-1} \textbf{I}_{p_k} - a_{kk} ^ {-1} p_k ^{-1} \textbf{J}_{p_k} \right) \bm{\Sigma}_{kk}
			= \left(a_{kk} ^ {-1} \textbf{I}_{p_k} - a_{kk} ^ {-1} p_k ^{-1} \textbf{J}_{p_k} \right) 
			\left(a_{kk} \textbf{I}_{p_k} + b_{kk} \textbf{J}_{p_k} \right) 
			= \textbf{I}_{p_K} - p_k ^ {- 1} \textbf{J}_{p_k},
		\end{equation*}
		which is idempotent with rank $(p_k - 1)$.
		Therefore, $\textbf{U}_k$ follows a $\chi ^ 2\left(p_k - 1; \Omega_k\right)$ distribution with the noncentrality parameter
		\begin{equation*}
			\Omega_k = \left(\bm{\mu} ^ {(k)} - \bm{\mu}_0 ^ {(k)}\right) ^ \top \left(a_{kk} ^ {-1} \textbf{I}_{p_k} - a_{kk} ^ {-1} p_k ^ {-1} \textbf{J}_{p_k} \right) \left(\bm{\mu} ^ {(k)} - \bm{\mu}_0 ^ {(k)}\right), \quad k \in \{1, \ldots, K\}.
		\end{equation*}
		Since $\textbf{M}_k = \textbf{I}_{p_k} - \textbf{J}_{p_k} / p_k$ is also idempotent with rank $(p_k - 1)$.
		Therefore, $a_{kk} \textbf{V}_k$ follows a central $\chi ^ 2$-distribution with degree of freedom $(n - 1)(p_k - 1)$.
		Furthermore,
		\begin{equation*}
			\begin{split}
				F_k = (p_k - 1) \frac{\textbf{U}_k / (p_k - 1)}{\textbf{V}_k / \{(p_k - 1) (n - 1)\}}
				\sim (p_k - 1) F \left((p_k - 1), (n - 1)(p_k - 1); \Omega_k\right), \quad k \in \{1, \ldots, K\}.
			\end{split}
		\end{equation*} 
		
		For $F_{K + 1} = n \left(\bar{\bm{X}} - \bm{\mu}_0 \right) ^ \top \left\{\left(\textbf{P} \widehat{\bm{\Delta}}\right) ^ {- 1} \circ \textbf{J}[\bm{p}]\right\} \left(\bar{\bm{X}} - \bm{\mu}_0 \right)$,
		consider a transformation $\bm{Y} = \textbf{C}_{\bm{p}} \bm{X}$ and $\bm{Y}_i = \textbf{C}_{\bm{p}} \bm{X}_i$ for $i = 1, \ldots, n$, 
		where $\textbf{C}_{\bm{p}} \triangleq \operatorname{Bdiag}\left(\textbf{1}_{1 \times p_1} / p_1, \ldots, \textbf{1}_{1 \times p_K} / p_K \right) \in \mathbb{R} ^ {K \times p}$.
		As $\bm{X} \sim N_p \left(\bm{\mu}, \bm{\Sigma} [\textbf{A}, \textbf{B}, \bm{p}]\right)$ and
		$\bm{Y} \sim N_K \left(\bm{\mu}_y, \bm{\Sigma}_y\right)$, where 
		\begin{equation*}
			\bm{\mu}_y = \textbf{C}_{\bm{p}} \bm{\mu} = \begin{pmatrix}
				\textbf{1}_{1 \times p_1} \bm{\mu} ^ {(1)} p_1 ^ {-1} \\ \vdots \\ \textbf{1}_{1 \times p_K} \bm{\mu} ^ {(K)} p_K ^ {-1}
			\end{pmatrix}, \quad
			\bm{\Sigma}_y = \textbf{C}_{\bm{p}} \bm{\Sigma} [\textbf{A}, \textbf{B}, \bm{p}]  \textbf{C}_{\bm{p}} ^ \top 
			= \textbf{A} \textbf{P} ^ {-1} + \textbf{B} 
			= \bm{\Delta} \textbf{P} ^ {- 1}.
		\end{equation*}
		Let $\bm{\nu}_{0} \triangleq \textbf{C}_{\bm{p}} \bm{\mu}_0 = \left(\textbf{1}_{1 \times p_1} \bm{\mu}_0 ^ {(1)} / p_1 , \ldots, \textbf{1}_{1 \times p_K} \bm{\mu}_0 ^ {(K)} / p_K \right) ^ \top$.
		Noting that $(\textbf{P} \textbf{C}_{\bm{p}}) ^ \top \bm{\Gamma} (\textbf{P} \textbf{C}_{\bm{p}}) = \bm{\Gamma} \circ \textbf{J}[\bm{p}]$ for any $\bm{\Gamma} \in \mathbb{R} ^ {K \times K}$, 
		then, 
		\begin{equation*}
			\begin{split}
				F_{K + 1} 
				& = n \left(\bar{\bm{X}} - \bm{\mu}_0 \right) ^ \top \left\{\left(\textbf{P} \widehat{\bm{\Delta}}\right) ^ {- 1} \circ \textbf{J}[\bm{p}]\right\} \left(\bar{\bm{X}} - \bm{\mu}_0 \right) \\
				& = n \left(\bar{\bm{X}} - \bm{\mu}_0 \right) ^ \top 
				\left\{ \left(\textbf{P} \textbf{C}\right) ^ \top \left(\textbf{P} \widehat{\bm{\Delta}}\right) ^ {- 1} \left(\textbf{P} \textbf{C} \right) \right\} 
				\left(\bar{\bm{X}} - \bm{\mu}_0 \right) \\
				& = n \left(\bar{\bm{Y}} - \bm{\nu}_0 \right) ^ \top \left(\textbf{P} \widehat{\bm{\Delta}} ^ {- 1}\right)
				\left(\bar{\bm{Y}} - \bm{\nu}_0 \right) \\
				& = n \left(\bar{\bm{Y}} - \bm{\nu}_0 \right) ^ \top \left(\widehat{\bm{\Sigma}}_y ^ {- 1}\right)
				\left(\bar{\bm{Y}} - \bm{\nu}_0 \right),
			\end{split}
		\end{equation*}
		where $\textbf{P} \widehat{\bm{\Delta}} ^ {- 1} = \textbf{P} \left( \widehat{\textbf{A}} \textbf{P} ^ {-1} \textbf{P} + \widehat{\textbf{B}} \textbf{P} \right) ^ {- 1} = \textbf{P} \left\{\left(\widehat{\textbf{A}} \textbf{P} ^ {-1}  + \widehat{\textbf{B}} \right) \textbf{P} \right\} ^ {- 1} = \textbf{P} \textbf{P} ^ {- 1} \left( \widehat{\textbf{A}} \textbf{P} ^ {-1}  + \widehat{\textbf{B}} \right) ^ {- 1} = \widehat{\bm{\Sigma}}_y ^ {- 1}$.
		By the fact that $\sum_{i = 1}^{n}\left(\bm{Y}_i - \bar{\bm{Y}}\right) ^ \top \left(\bm{Y}_i - \bar{\bm{Y}}\right) / (n - 1) = \widehat{\textbf{A}} \textbf{P} ^ {- 1} + \widehat{\textbf{B}}$ and the definition of the Hotelling's $T ^ 2$ statistic, $F_{K + 1} \sim T ^ 2 = K(n - 1) / (n - K) F\left(K, n - K; \Omega_{K + 1}\right)$, 
		where $\Omega_{K + 1} \triangleq n \left(\bm{\mu}_y - \bm{\nu}_0\right) ^ \top \bm{\Sigma}_y ^ {-1} \left(\bm{\mu}_y - \bm{\nu}_0\right)$.
		
		In conclusion, $G_2$ is decomposed as a linear combination of mutually independent $F$-variates, distributed as $\sum_{k = 1}^{K} (p_k - 1) F_{k} \left((p_k - 1), (n - 1)(p_k - 1); \Omega_k\right) + K (n - 1) / (n - K) F_{K + 1} \left(K, n - K; \delta_{K + 1}\right)$.
		Under $H_0$, $\Omega_k = 0$ for $k \in \{1, \ldots, K + 1\}$.

		We also complete the proof of the second part of Theorem 3 through three steps. 
		
		\textbf{\emph{Step 1: decomposition of $G_4$}}. 
		Let $\bm{Z} \triangleq \left(\bar{\bm{X}} ^ {(1), \top}, \ldots, \bar{\bm{X}} ^ {(M), \top} \right) ^ \top$ denote a $(pM)$-dimensional normal vector with mean $\bm{\mu}_{\bm{Z}} \triangleq \left(\bm{\mu} ^ {(1), \top}, \ldots, \bm{\mu} ^ {(M), \top} \right) ^ \top$ and covariance matrix $\bm{\Sigma}_{\bm{Z}} \triangleq \textbf{N} ^ {-1} \otimes \bm{\Sigma} \left[\textbf{A}, \textbf{B}, \bm{p}\right]$, where $\textbf{N} \triangleq \operatorname{diag}\left(n ^ {(1)}, \ldots, n ^ {(M)}\right) \in \mathbb{R} ^ {M \times M}$ and $\otimes$ denotes the Kronecker product.
		Thus, there exits $\textbf{C}_{\bm{Z}} = \left\{\textbf{N} ^ {1/2} \left( \textbf{I}_M - n ^ {- 1} \textbf{J}_M \textbf{N} \right)\right\} \otimes \textbf{I}_p \in \mathbb{R} ^ {(pM) \times (pM)}$ such that
		\begin{equation*}
			\left(\sqrt{n ^ {(1)}} \left(\bar{\bm{X}} ^ {(1)} - \bar{\bm{X}} \right) ^ \top, \ldots, \sqrt{n ^ {(M)}} \left(\bar{\bm{X}} ^ {(M)} - \bar{\bm{X}} \right) ^ \top \right) ^ \top = \textbf{C}_Z \bm{Z}.
		\end{equation*} 
		Therefore, $G_4$ can be rewritten as 
		\begin{equation*}
			G_4 = \left(\textbf{C}_{\bm{Z}} \bm{Z} \right) ^ \top 
			\left(\textbf{I}_M \otimes \widehat{\bm{\Theta}}\left[\widehat{\textbf{A}}_{\bm{\Theta}}, \widehat{\textbf{B}}_{\bm{\Theta}}, \bm{p} \right]\right)
			\left(\textbf{C}_{\bm{Z}} \bm{Z} \right).
		\end{equation*}
		Using the same decomposition $\widehat{\bm{\Theta}}\left[\widehat{\textbf{A}}_{\bm{\Theta}}, \widehat{\textbf{B}}_{\bm{\Theta}}, \bm{p}\right] = \widehat{\textbf{A}} ^ {-1} \circ \textbf{I}[\bm{p}] - \left(\widehat{\textbf{A}} \textbf{P} \right) ^ {-1} \circ \textbf{J}[\bm{p}] + \left(\textbf{P} \widehat{\bm{\Delta}}\right) ^ {-1} \circ \textbf{J}[\bm{p}]$,  
		$G_4$ can be expressed by 
		\begin{equation*}
			\begin{split}
				G_4
				& = \left(\textbf{C}_{\bm{Z}} \bm{Z} \right) ^ \top 
				\left\{\textbf{I}_M \otimes \left( \sum_{k = 1}^{K} \textbf{W}_{k} \right) \right\}
				\left(\textbf{C}_{\bm{Z}} \bm{Z} \right) 
				+ \left(\textbf{C}_{\bm{Z}} \bm{Z} \right) ^ \top 
				\left(\textbf{I}_M \otimes \textbf{W}_{K + 1} \right)
				\left(\textbf{C}_{\bm{Z}} \bm{Z} \right) \\
				& = \sum_{k = 1}^{K} \left(\textbf{C}_{\bm{Z}} \bm{Z} \right) ^ \top 
				\left(\textbf{I}_M \otimes \textbf{W}_{k} \right)
				\left(\textbf{C}_{\bm{Z}} \bm{Z} \right) 
				+ \left(\textbf{C}_{\bm{Z}} \bm{Z} \right) ^ \top 
				\left(\textbf{I}_M \otimes \textbf{W}_{K + 1} \right)
				\left(\textbf{C}_{\bm{Z}} \bm{Z} \right) \\
				& \triangleq F_1 + \cdots + F_{K} + F_{K + 1},
			\end{split}
		\end{equation*}
		where $\textbf{W}_k \triangleq \operatorname{Bdiag}\left(\textbf{0}_{p_1 \times p_1}, \ldots, \widehat{a}_{kk} ^ {-1} - \widehat{a}_{kk} ^ {-1} p_k ^ {-1} \textbf{J}_{p_k}, \ldots, \textbf{0}_{p_K \times p_K}\right) = \textbf{W}_k\left[\textbf{A}_k, \textbf{B}_k, \bm{p}\right]$ with $\textbf{A}_k \triangleq \text{diag}\left(0, \ldots, \widehat{a}_{kk} ^ {-1}, \ldots, 0 \right) \in \mathbb{R} ^ {K \times K}$ and $\textbf{B}_k \triangleq \text{diag}\left(0, \ldots, - \widehat{a}_{kk} ^ {-1} p_k ^ {-1}, \ldots, 0 \right) \in \mathbb{R} ^ {K \times K}$ having the non-zero values on the $(k,k)$th elements, 
		and $\textbf{W}_{K + 1} \triangleq \left(\textbf{P} \widehat{\bm{\Delta}}\right) ^ {-1} \circ \textbf{J}[\bm{p}]$.
		
		\textbf{\emph{Step 2: independence}}. 
		By Lemma \ref{Lem:independent}, we need to check
		\begin{equation*}
			\begin{split}
				& \left\{\textbf{C}_{\bm{Z}} ^ \top (\textbf{I}_M \otimes \textbf{W}_k) \textbf{C}_{\bm{Z}} \right\} \left(\textbf{N} ^ {-1} \otimes \bm{ }[\textbf{A}, \textbf{B}, \bm{p}]\right) \left\{\textbf{C}_{\bm{Z}} ^ \top (\textbf{I}_M \otimes \textbf{W}_{k'}) \textbf{C}_{\bm{Z}} \right\} 
				= \textbf{0}_{(pM) \times (pM)}, \quad k \neq k', k, k' \in \{1, \ldots, K\} \\
				& \left\{\textbf{C}_{\bm{Z}} ^ \top (\textbf{I}_M \otimes \textbf{W}_k) \textbf{C}_{\bm{Z}} \right\} \left(\textbf{N} ^ {-1} \otimes \bm{\Sigma}[\textbf{A}, \textbf{B}, \bm{p}]\right) \left\{\textbf{C}_{\bm{Z}} ^ \top (\textbf{I}_M \otimes \textbf{W}_{K + 1}) \textbf{C}_{\bm{Z}} \right\} 
				= \textbf{0}_{(pM) \times (pM)}, \quad k \in \{1, \ldots, K\}.
			\end{split}
		\end{equation*}
		Let $\textbf{M} = \textbf{N} ^ {1/2} \left(\textbf{I}_M - n ^ {- 1} \textbf{J}_M  \textbf{N}\right)$.
		Given $k \neq k'$ and $\textbf{C}_{\bm{Z}} = \textbf{M} \otimes \textbf{I}_p$ and the fact $\left(A \otimes B\right) ^ \top = A ^ \top \otimes B ^ \top$, we can observe that 
		\begin{equation*}
			\textbf{C}_{\bm{Z}} ^ \top (\textbf{I}_M \otimes \textbf{W}_k) \textbf{C}_{\bm{Z}}
			= \left(\textbf{M} ^ \top \textbf{I}_M \textbf{M} \right) \otimes \left( \textbf{I}_p \textbf{W}_k \textbf{I}_p \right) 
			= \left(\textbf{M} ^ \top \textbf{M} \right) \otimes \textbf{W}_k, \quad
			\textbf{C}_{\bm{Z}} ^ \top (\textbf{I}_M \otimes \textbf{W}_{k'}) \textbf{C}_{\bm{Z}}
			= \left(\textbf{M} ^ \top \textbf{M} \right) \otimes \textbf{W}_{k'}.
		\end{equation*}
		On the one hand, we can calculate the following result: 
		\begin{equation*}
			\begin{split}
				& \left\{\textbf{C}_{\bm{Z}} ^ \top (\textbf{I}_M \otimes \textbf{W}_k) \textbf{C}_{\bm{Z}} \right\} \left(\textbf{N} ^ {-1} \otimes \bm{\Sigma} [\textbf{A}, \textbf{B}, \bm{p}]\right) \left\{\textbf{C}_{\bm{Z}} ^ \top (\textbf{I}_M \otimes \textbf{W}_{k'}) \textbf{C}_{\bm{Z}} \right\} \\
				& = \left\{ \left(\textbf{M} ^ \top \textbf{M}\right) \textbf{N} ^ {-1} \left(\textbf{M} ^ \top \textbf{M}\right) \right\} 
				\otimes \left\{ \textbf{W}_k \bm{\Sigma} \left[\textbf{A}, \textbf{B}, \bm{p}\right] \textbf{W}_{k'} \right\}.
			\end{split}
		\end{equation*}
		Noting that $\textbf{W}_k = \textbf{W}_k\left[\textbf{A}_k, \textbf{B}_k, \bm{p}\right]$ and $\textbf{W}_{k'} = \textbf{W}_{k'}\left[\textbf{A}_{k'}, \textbf{B}_{k'}, \bm{p}\right]$, we have the result $\textbf{W}_k \bm{\Sigma} \left[\textbf{A}, \textbf{B}, \bm{p}\right] \textbf{W}_{k'} = \textbf{0}_{p \times p}$. 
		Therefore, $\left\{\textbf{C}_{\bm{Z}} ^ \top (\textbf{I}_M \otimes \textbf{W}_k) \textbf{C}_{\bm{Z}} \right\} \left(\textbf{N} ^ {-1} \otimes \bm{\Sigma}[\textbf{A}, \textbf{B}, \bm{p}]\right) \left\{\textbf{C}_{\bm{Z}} ^ \top (\textbf{I}_M \otimes \textbf{W}_{k'}) \textbf{C}_{\bm{Z}} \right\} = \textbf{0}_{(pM) \times (pM)}$.
		On the other hand, we can calculate that 
		\begin{equation*}
			\begin{split}
				& \left\{\textbf{C}_{\bm{Z}} ^ \top (\textbf{I}_M \otimes \textbf{W}_k) \textbf{C}_{\bm{Z}} \right\} \left(\textbf{N} ^ {-1} \otimes \bm{\Sigma}[\textbf{A}, \textbf{B}, \bm{p}]\right) \left\{\textbf{C}_{\bm{Z}} ^ \top (\textbf{I}_M \otimes \textbf{W}_{K + 1}) \textbf{C}_{\bm{Z}} \right\} \\
				& = \left\{\left(\textbf{M} ^ \top \textbf{M}\right) \textbf{N} ^ {-1} \left(\textbf{M} ^ \top \textbf{M}\right) \right\} 
				\otimes \left( \textbf{W}_k \bm{\Sigma}\left[\textbf{A}, \textbf{B}, \bm{p}\right] \textbf{W}_{K + 1} \right) \\
				& = \textbf{0}_{(pM) \times (pM)}
			\end{split}
		\end{equation*}
		using the result $\textbf{W}_k \bm{\Sigma}\left[\textbf{A}, \textbf{B}, \bm{p}\right] \textbf{W}_{K + 1} = \textbf{0}_{p \times p}$.
		
		\textbf{\emph{Step 3: distribution}}. 
		By definition, for $k \in \{1, \ldots, K\}$,
		\begin{equation*}
			F_k 
			= \bm{Z} ^ \top 
			\left\{\textbf{C}_{\bm{Z}} ^ \top (\textbf{I}_M \otimes \textbf{W}_k) \textbf{C}_{\bm{Z}} \right\}
			\bm{Z} 
			= \bm{Z} ^ \top
			\left\{
			\left(\textbf{M} ^ \top \textbf{M} \right) \otimes \textbf{W}_k
			\right\} \bm{Z},
		\end{equation*}
		Let $\widetilde{\textbf{W}}_k \triangleq \operatorname{Bdiag}\left(\textbf{0}_{p_1 \times p_1}, a_{kk} ^ {-1} \textbf{I}_{p_k} - a_{kk}^{-1}p_k^{-1}\textbf{J}_{p_k}, \ldots, \textbf{0}_{p_K \times p_k}\right)$ have non-zero values on the $(k, k)$th element for every $k$.
		Let $\textbf{U}_k \triangleq \bm{Z} ^ \top \left\{\left(\textbf{M} ^ \top \textbf{M} \right) \otimes \widetilde{\textbf{W}}_k \right\} \bm{Z}$ and $\textbf{V}_k \triangleq (n - M)(p_k - 1) \widehat{a}_{kk} / a_{kk}$, 
		and therefore, $\textbf{U}_k / \textbf{V}_k = F_k / \left\{(n - M) (p_k - 1)\right\}$ for $k \in \{1, \ldots, K\}$. 
		
		Since $\textbf{S}$ is independent from $\bar{\bm{X}} ^ {(1)}, \ldots, \bar{\bm{X}} ^ {(M)}$, then $\textbf{S}$ is independent from $\bm{Z}$, and therefore $\textbf{U}_k$ and $\textbf{V}_k$ are independent for every $k$. 
		By Lemma~\ref{Lem:chisquare} and $\bm{Z} \sim N_{pM} \left(\bm{\mu}_{\bm{Z}}, \textbf{N} ^ {-1} \otimes \bm{\Sigma}[\textbf{A}, \textbf{B}, \bm{p}]\right)$, we obtain that 
		\begin{equation*}
			\left\{\left(\textbf{M} ^ \top \textbf{M} \right) \otimes \widetilde{\textbf{W}}_k\right\} \left(\textbf{N}^{-1} \otimes \bm{\Sigma} \left[\textbf{A}, \textbf{B}, \bm{p}\right]\right) 
			= \left(\textbf{M} ^ \top \textbf{M} \textbf{N} ^ {-1}\right) \otimes \left(\widetilde{\textbf{W}}_k \bm{\Sigma}[\textbf{A}, \textbf{B}, \bm{p}]\right),
		\end{equation*}
		whose square equals $\left(\textbf{M} ^ \top \textbf{M} \textbf{N} ^ {-1}\right) ^ 2 \otimes \left(\widetilde{\textbf{W}}_k \bm{\Sigma}[\textbf{A}, \textbf{B}, \bm{p}]\right) ^ 2 = \left(\textbf{M} ^ \top \textbf{M} \textbf{N} ^ {-1}\right) \otimes \left(\widetilde{\textbf{W}}_k \bm{\Sigma}[\textbf{A}, \textbf{B}, \bm{p}]\right)$ due to 
		\begin{equation*}
			\begin{split}
				\textbf{M} ^ \top \textbf{M} \textbf{N} ^ {-1} 
				& = \left\{ \textbf{N} ^ {1/2} \left(\textbf{I}_M - n ^ {-1} \textbf{J}_M \textbf{N} \right)\right\} ^ \top 
				\left\{ \textbf{N} ^ {1/2} \left(\textbf{I}_M - n ^ {-1} \textbf{J}_M \textbf{N} \right)\right\} \textbf{N} ^ {-1} \\
				& = \left(\textbf{N} - n ^ {-1} \textbf{N} \textbf{J}_M \textbf{N}\right)
				\left(\textbf{N} ^ {-1} - n ^ {-1} \textbf{J}_M\right) \\
				& = \textbf{I}_M - n ^ {-1} \textbf{N} \textbf{J}_M
				- n ^ {-1} \textbf{N} \textbf{J}_M
				+ n ^ {-2} \textbf{N} \textbf{J}_M \textbf{N} \textbf{J}_M \\
				& = \textbf{I}_M - n ^ {- 1} \textbf{N} \textbf{J}_M,
			\end{split}
		\end{equation*}
		and $\textbf{J}_M \textbf{N} \textbf{J}_M = n \textbf{J}_M$.
		It is clear that $\textbf{I}_M - \textbf{N} \textbf{J}_M / n$ is an idempotent matrix since its square equals $\textbf{I}_M + \textbf{N} \textbf{J}_M \textbf{N} \textbf{J}_M / n ^ 2 - 2 \textbf{N} \textbf{J}_M / n = \textbf{I}_M -  \textbf{N}\textbf{J}_M / n$.
		Also, the result shows that $\widetilde{\textbf{W}}_k \bm{\Sigma}[\textbf{A}, \textbf{B}, \bm{p}]$ is an idempotent matrix.  
		
		Therefore, $\textbf{U}_k$ follows a noncentral $\chi^ 2$-distribution with the noncentrality parameter $\Omega_k = \bm{\mu}_{\bm{Z}} ^ \top \left\{\left(\textbf{M}^\top \textbf{M}\right)\otimes \widetilde{\textbf{W}}_k\right\} \bm{\mu}_{\bm{Z}}$ and a degree of freedom $\operatorname{rank}\left(\left(\textbf{M}^\top \textbf{M}\right)\otimes \widetilde{\textbf{W}}_k \right) = \operatorname{rank} \left(\textbf{M}^\top\textbf{M}\right)  \operatorname{rank}\left(\widetilde{\textbf{W}}_k\right) = (M - 1)(p_k - 1)$
		by using the fact that $\operatorname{rank}\left(A \otimes B\right) = \operatorname{rank}(A) \operatorname{rank}(B)$.
		
		Given $a_{kk} \textbf{V}_k = (n - M)(p_k - 1) \widehat{a}_{kk}$,
		the result that $(p_k - 1) \widehat{a}_{kk} = \operatorname{tr} \left(\textbf{M}_k \textbf{S}_{kk} \textbf{M}_{kk}\right)$, 
		and the fact that $(n - M) \textbf{S}_{kk} \sim W_{p_k} (\bm{\Sigma}_{kk}, n - M)$, 
		we obtain that $a_{kk} \textbf{V}_{k}$ follows a central $\chi^2$-distribution with a degree of freedom $(n - M)(p_k - 1)$.
		Therefore, for $k \in \{1, \ldots, K\}$, 
		\begin{equation*}
			\begin{split}
				F_k 
				& = (n - M)(p_k - 1) \frac{\textbf{U}_k}{\textbf{V}_k} 
				\sim (n - M)(p_k - 1) \frac{\chi ^ 2\left((M - 1)(p_k - 1); \Omega_k\right)}{\chi ^ 2\left((n - M)(p_k - 1)\right)} \\
				& \sim (M - 1)(p_k - 1) F \left((M - 1)(p_k - 1), (n - M)(p_k - 1); \Omega_k \right).
			\end{split}
		\end{equation*}
		Let $\textbf{C}_{\bm{p}} \triangleq \operatorname{Bdiag} \left(\textbf{1}_{1 \times p_1} / p_1, \ldots, \textbf{1}_{1 \times p_K} / p_K \right)$,
		let $\bm{\nu} ^ {(m)} \triangleq \textbf{C}_{\bm{p}} \bm{\mu} ^ {(m)}$, 
		and let $\bm{Y}_j ^ {(m)} \triangleq \textbf{C}_{\bm{p}} \bm{X}_j ^ {(m)}$ for every $j$ and $m$ with mean $\bm{\nu} ^ {(m)} \triangleq \textbf{C}_{\bm{p}} \bm{\mu} ^ {(m)}$ and $\bm{\Sigma}_{\bm{Y}} \triangleq \textbf{C}_{\bm{p}} \bm{\Sigma}[\textbf{A}, \textbf{B}, \bm{p}] \textbf{C}_{\bm{p}} ^ \top = \textbf{A} \textbf{P} ^ {-1} + \textbf{B} = (\textbf{A} + \textbf{B} \textbf{P}) \textbf{P} ^ {- 1} = \bm{\Delta} \textbf{P} ^ {-1} = \left(\textbf{P} \bm{\Delta} ^ {-1}\right) ^ {-1}$.
		Thus, $\bar{\bm{Y}} ^ {(m)} = \textbf{C}_{\bm{p}} \bar{\bm{X}} ^ {(m)}$ for every $m$ and $\bar{\bm{Y}} = \textbf{C}_{\bm{p}} \bar{\bm{X}}$.
		Finally, $F_{K + 1}$ can be expressed by 
		\begin{equation*}
			\begin{split}
				F_{K + 1} 
				& = \sum_{m = 1}^{M} n_m \left(\bar{\bm{X}} ^ {(m)} - \bar{\bm{X}}\right) ^ \top \textbf{W}_{K + 1} \left(\bar{\bm{X}} ^ {(m)} - \bar{\bm{X}} \right) \\
				& = \sum_{m = 1}^{M} n_m \left(\bar{\bm{X}} ^ {(m)} - \bar{\bm{X}}\right) ^ \top 
				\left(
				\textbf{C}_{\bm{p}} ^ \top \textbf{P} \widehat{\bm{\Delta}} ^ {-1} \textbf{P} ^ {-1} \textbf{P} \textbf{C}_{\bm{p}}	
				\right) \left(\bar{\bm{X}} ^ {(m)} - \bar{\bm{X}} \right) \\
				& = \sum_{m = 1}^{M} n_m \left(\bar{\bm{Y}} ^ {(m)} - \bar{\bm{Y}} \right) ^ \top 
				\left(\textbf{P} \widehat{\bm{\Delta}} ^ {-1}\right)
				\left(\bar{\bm{Y}} ^ {(m)} - \bar{\bm{Y}} \right) \\
				& = \operatorname{tr}\left(
				\left( \widehat{\bm{\Sigma}}_{\bm{Y}} \right) ^ {- 1}
				\sum_{m = 1}^{M} n_m \left(\bar{\bm{Y}} ^ {(m)} - \bar{\bm{Y}} \right) \left(\bar{\bm{Y}} ^ {(m)} - \bar{\bm{Y}} \right) ^ \top
				\right),
			\end{split}
		\end{equation*}
		which is the Hotelling-Lawley $T_0 ^ 2$ statistic.
		Specifically, $(n - M) \widehat{\bm{\Sigma}}_{\bm{Y}} \sim W_{K} \left(\bm{\Sigma}_{\bm{Y}}, n - M\right)$ and $\sum_{m = 1}^{M} n ^ {(m)} \left(\bar{\bm{Y}} ^ {(m)} - \bar{\bm{Y}} \right) \left(\bar{\bm{Y}} ^ {(m)} - \bar{\bm{Y}} \right) ^ \top \sim W_{K} \left(\bm{\Sigma}_{\bm{Y}}, M - 1; \bm{\Omega}\right)$,
		where the noncentrality parameter $\bm{\Omega} \triangleq \sum_{m = 1}^{M} n ^ {(m)} \left(\bm{\nu} ^ {(m)} - \bar{\bm{\nu}} \right) \left(\bm{\nu} ^ {(m)} - \bar{\bm{\nu}} \right) ^ \top$ satisfying that $\bar{\bm{\nu}} = n ^ {-1} \sum_{m = 1}^{M} n ^ {(m)} \bm{\nu} ^ {(m)}$ and 
		$\operatorname{E}\left\{\sum_{m = 1}^{M} n ^ {(m)} \left(\bar{\bm{Y}} ^ {(m)} - \bar{\bm{Y}} \right) \left(\bar{\bm{Y}} ^ {(m)} - \bar{\bm{Y}} \right) ^ \top\right\} = (M - 1) \bm{\Sigma}_{\bm{Y}} + \bm{\Omega}$.
		Additionally, 
		Wishart variate
		$\sum_{m = 1}^{M} n ^ {(m)} \left(\bar{\bm{Y}} ^ {(m)} - \bar{\bm{Y}} \right) \left(\bar{\bm{Y}} ^ {(m)} - \bar{\bm{Y}} \right) ^ \top$ may be singular;  
		$\widehat{\bm{\Sigma}}_{\bm{Y}} > 0$ holds with probability $1$ if $n - M > K$.
	\end{proof}
	
	\begin{proof}[Proof of Theorem 4 (approximation bounds for the FDP estimator)]
		We verify the conditions of Theorem 2 and Theorem 3 in \citet{FanHan2017}.
		Specifically,
		Due to the presence of $\delta_1$, $\delta_2$, and $\delta_3$, 
		(condition 1), (condition 2), (condition 5) in \citet{FanHan2017} are satisfied.
		Thus, it remains to verify (condition 3) and (condition 4) for Theorem 2 in \citet{FanHan2017}, 
		which is equivalent to establishing the existence of $\delta_4 > 0$ such that $\left \Vert \widehat{\bm{\Psi}}\left[\widehat{\textbf{A}}_{\bm{\Psi}}, \widehat{\textbf{B}}_{\bm{\Psi}}, \bm{p}\right] - \bm{\Psi}\left[\textbf{A}_{\bm{\Psi}}, \textbf{B}_{\bm{\Psi}}, \bm{p}\right] \right \Vert = O\left(g_p p ^ {- \delta_4}\right)$.
		
		Using the modified hard-threshold estimators from \citet{YangChenChen2024}, 
		the proposed correlation matrix estimator satisfies 
		\begin{equation*}
			\Pr \left\{
				\left \Vert \widehat{\bm{\Psi}}\left[\widehat{\textbf{A}}_{\bm{\Psi}}, \widehat{\textbf{B}}_{\bm{\Psi}}, \bm{p}\right] - \bm{\Psi}\left[\textbf{A}_{\bm{\Psi}}, \textbf{B}_{\bm{\Psi}}, \bm{p}\right] \right \Vert \leq C_q (2 \lambda) ^ {1 - q}
			\right\} \to 1, \quad K, p \to \infty, \log(K) / n \to 0, 
		\end{equation*}
		where $\lambda \triangleq \eta \sqrt{\log(K) / p}$, 
		and $0 < q < 1$, $C_q$, and $\eta$ are constants.
		Thus, we can rewrite this as
		\begin{equation*}
			\left \Vert \widehat{\bm{\Psi}}\left[\widehat{\textbf{A}}_{\bm{\Psi}}, \widehat{\textbf{B}}_{\bm{\Psi}}, \bm{p}\right] - \bm{\Psi}\left[\textbf{A}_{\bm{\Psi}}, \textbf{B}_{\bm{\Psi}}, \bm{p}\right] \right \Vert = O_P\left[C_q(K)\left\{\frac{\log(K)}{p}\right\} ^ {(1 - q) / 2}\right].
		\end{equation*}
		This result aligns with the sparse matrix case studied in \citet{BickelLevina2008b}, and the existence of $\delta_4$ has been established in \citet{FanHan2017}.
		We complete the proof.
	\end{proof}
		
	\bibliographystyle{biom}
	
	\bibliography{paper_ref_UB}